\documentclass[journal,10pt,twocolumn,twoside]{IEEEtran}
\usepackage{multirow}
\usepackage{colortbl}
 \usepackage{graphicx}
\usepackage{pifont}
\usepackage{url}
\usepackage{graphics}
\usepackage{amsfonts}
\usepackage{color}
\usepackage{stfloats}
\usepackage{color,soul}
\usepackage{amsmath}
\usepackage{multirow}
\usepackage{hhline}
\usepackage{url}
\usepackage{float}
\usepackage{enumitem}
\usepackage{cite}
\usepackage{multicol}
\usepackage{color, colortbl}
\usepackage{blindtext}
\usepackage{rotating}
\usepackage{array}
\usepackage{tikz}
\usepackage{hyperref}
\usepackage{amsxtra}
\begin{document}

\title{Autonomous Vehicles in 5G and Beyond: A Survey}

\author{Saqib Hakak, Thippa Reddy Gadekallu, Swarna Priya Ramu, Parimala M, Praveen Kumar Reddy Maddikunta, Chamitha de Alwis, Madhusanka Liyanage
\thanks{Corresponding Author: Thippa Reddy Gadekallu}
\thanks{Saqib Hakak is with the Canadian Institute for Cybersecurity, Faculty of Computer Science, University of New Brunswick, Canada (email: saqib.hakak@unb.ca)}
\thanks {Thippa Reddy Gadekallu, Swarna Priya Ramu, Parimala M and Praveen Kumar Reddy Maddikunta are with the School of Information Technology and Engineering, Vellore Institute of Technology, India (Email: \{thippareddy.g, swarnapriya.rm, parimala.m, praveenkumarreddy\}@vit.ac.in).}

\thanks{Chamitha de Alwis is with the Department of Electrical and Electronic Engineering, University of Sri Jayewardeneprua, Sri Lanka, email: chamitha@sjp.ac.lk}
\thanks{Madhusanka Liyanage is with the School of Computer Science, University Collage Dublin, Ireland and Centre for Wireless Communications, University of Oulu, Finland (email: madhusanka@ucd.ie)}
}

\IEEEtitleabstractindextext{%
\begin{abstract}
\textcolor{black}{Fifth Generation (5G) technology is an emerging and fast adopting technology which is being utilized in most of the novel applications that require highly reliable low-latency communications. It has the capability to provide greater coverage, better access, and best suited for high density networks. Having all these benefits, it clearly implies that 5G could be used to satisfy the requirements of Autonomous vehicles. Automated driving Vehicles and systems are developed with a promise to provide comfort, safe and efficient drive reducing the risk of life. But, recently there are fatalities due to these autonomous vehicles and systems. This is due to the lack of robust state-of-art which has to be improved further.}
With the advent of 5G technology and rise of autonomous vehicles (AVs), road safety is going to get more secure with less human errors. However, \textcolor{black}{integration of 5G and AV} is still at its infant stage with several  research challenges that needs to be addressed.  \textcolor{black}{This survey first starts with a discussion on} the current advancements in \textcolor{black}{AVs, automation levels, enabling technologies and 5G requirements. Then, we focus on the emerging techniques required for }integrating 5G technology with AVs, impact of 5G and B5G technologies on AVs along with security concerns in AVs. \textcolor{black}{The paper also provides a comprehensive survey of} recent developments in terms of standardisation activities on 5G autonomous vehicle technology \textcolor{black}{and current projects. The article is finally concluded with lessons learnt, future research directions and challenges.}
\end{abstract}


}

\maketitle

\IEEEdisplaynontitleabstractindextext

\IEEEpeerreviewmaketitle

\begin{table}[h!]
\centering
\caption{Summary of Important Acronyms}
\label{table:acronyms}
\resizebox{\columnwidth}{!}{%
\begin{tabular}{|p{1.5cm} p{7.5cm}|}
\hline
\rowcolor{gray!30}
\textbf{Acronym} & \textbf{Definition}\\
\hline
\hline

2G & Second Generation \\
3G & Third Generation \\
4G & Fourth Generation \\
5G & Fifth Generation \\
5GNR & 5G New Radio \\
ACVs & Autonomous and Connected Vehicles \\
AD & Autonomous Driving\\
BS & Base Station \\
B5G & Beyond 5G \\
CAN & Controller Area Network\\
CL & Cooperative Localization\\
CR & Cognitive Radio \\
CVs & Connected Vehicles \\
DMM & Distributed Mobility Management\\
EC & Edge Computing \\

eMBB & enhanced mobile broadband \\
I2I & Infrastructure-to-Infrastructure \\
ITS & Intelligent Transportation System \\
ITU & International Telecommunication Union \\
MEC & Multi-access Edge Computing \\
MIMO & Massive Multiple-Input-Multiple-Output \\
mMTC & massive machine-type communications (mMTC)\\
mmWAVE & millimeter-Wave \\
NFV & Network Functions Virtualization \\
NS & Network Slicing \\
PLS & Physical Layer Security \\
ProSe & Proximity Service \\
QoE & Quality of Everything \\
RL & Reinforcement Learning \\
SAE & Society of Automotive Engineers \\
SDN & Software-Defined Networking\\
THz & Terahertz \\
TIS & Traffic Information Systems \\
V2I & Vehicle-to-Infrastructure \\
V2V & Vehicle-to-Vehicle \\

V2X & Vehicle-to-Everything \\
UAV & Unmmaned Aerial Vehicles \\

URLLC & Ultra-reliable Low-Latency Communications \\

V2P & Vehicle-to-People \\

\hline
\end{tabular}
}
\end{table}

\section{Introduction}

With the emergence of new technologies such as high-speed networks, decentralised storage platforms, edge computing and others, driving a car without human intervention or very less intervention has become possible. Autonomous vehicles (AVs), also known as driverless car or autonomous cars, are the vehicles where the operation of the vehicle occurs without the direct driver input and driver is not expected to monitor the roadway constantly \cite{manfreda2019autonomous,ravi2020driver}. With better safety measures \cite{jadaan2017connected} and improved energy efficiency resulting in lower environmental impact \cite{wadud2016help}, the AVs seems to be promising technology. For the same very reason, major car manufacturers are expanding their fleets with AVs \cite{manfreda2019autonomous}. For example, one of the major car-manufacturer company Mercedes-Benz\footnote{\url{https://www.bbc.com/news/business-56332388}} recently announced its intention to launch autonomous driving technology in its S-class model. Similalry, Tesla\footnote{\url{https://www.tesla.com/en_CA/autopilot, Accessed:Nov.09, 2021.}} has already come up with a advanced hardware and software technology to make driving completely autonomous. However, to make driving completely autonomous, high-speed networks such as 5G or beyond 5G technologies will play a key role.

With the emergence of fifth generation (5G) technology that can offer speed upto 10 Gbps with incredibly low latency of 1 ms (for everyday cellular users), the realisation of AVs has been made possible. Initiated by ITU in the year 2015, 5G new radio (NR) is a new cellular communication standard with the potential to support massive delay-sensitive applications through three key use-cases in 5G. These three use-cases include enhanced mobile broadband (eMBB), massive machine-type communications (mMTC) and URLLC. Ultra-reliable low-latency communication (URLLC)\cite{sachs20185g} is designed to support services for the delay-sensitive applications such as remote surgery and autonomous driving requiring very less error-bit rates. Similarly, the focus of eMBB is to improve the latency requirements by providing more bandwidth and achieve the speed in Gigabits to supprt applications such as virtual reality (VR). Finally, the focus of mMTC is to improve connectivity between billions of devices transmitting very short packets (not done adequately in cellular communications due to human nature type of communication) \cite{silva20205g}.

As the integration of 5G technology with other existing technologies will open doors to thousands of other use-cases, it will be interesting to understand how its integration with autonomous driving will be beneficial. To the best of our knowledge, this is the first comprehensive review paper that addresses several aspects of integrating 5G technology in AVs. Table \ref{Tab:Survey} presents a comparative analysis of this work with the existing studies.  The contributions of this work are summarised as follows: 
\begin{itemize}
    \item We have provided a comprehensive overview on AVs.
    \item Different technical aspects for the successful integration of AVs with 5G technology are enumerated.
    \item State-of-the-art on integrating AVs with 5G is conducted.
    \item Security concerns in AVs are identified and explored.
    \item Key projects and standardisation activities on 5G autonomous vehicles are highlighted.
    \item Future research challenges and directions are identified and explored.
\end{itemize}

\begin{table*}[t!]
\scriptsize
\caption{Summary of Related Surveys on 5G-based Autonomous Vehicles}
\begin{center}
\label{Tab:Survey}
\vspace{-0.5cm}
\hspace{0.5cm}
		\centering
\begin{tabular}{c p{5cm}p{2 cm}p{2cm}p{1.5cm}p{1.5cm}p{1.5cm}}
\hline
\hline
\textbf{Reference} & \textbf{Theme of the survey} & \textbf{Integration of 5G in AVs} & \textbf{Security Concerns in AVs} &\textbf{Overview of 5G} & \textbf{Technical aspects in AVs} & \textbf{Projects and Standardisation activities} \\
\hline
\hline
 \cite{rasouli2019autonomous}, 2019 & Pedestrian behaviour using AVs & Not Covered & Not Covered & Not Covered & Not Covered & Not Covered
 \\\hline
 \cite{ma2020artificial}, 2020 & Role of AI in supporting the primary applications in AVs & Not Covered & Not Covered & Not Covered & Partially Covered & Not Covered
 \\\hline
  \cite{ahangar2021survey}, 2021 & Comprehensive study of AV technologies & Partially Covered & Not Covered & Not Covered & Partially Covered & Not Covered
 \\\hline
   \cite{storck2020survey}, 2020 & Study on interaction among: IoV, 5G, and V2X & Partially Covered & Not Covered & Partially Covered & Partially Covered & Not Covered
 \\\hline
   \cite{navarro2020survey}, 2020 & Study on significant use cases expected for 5G  & Partially Covered & Not Covered & Fully covered & Partially Covered & Not Covered 
 \\\hline
   \cite{pham2021survey},2021  & State-of-the art CAV attacks and countermeasures  & Not Covered & Fully Covered & Partially Covered & Not Covered & Not Covered
 \\\hline
  This study,2022 & Study on integrating 5G and AV technology & Fully Covered & Fully Covered & Fully Covered & Fully Covered & Fully Covered
 \\\hline
\end{tabular}
\end{center}
\end{table*}

The remaining paper is organised as follows: Section I highlights the features of AVs and provides theoretical framework to achieve inter-vehicular connectivity using 5G technology. Section II discusses on the technical requirements for AVs to realise their full-potential and how 5G fulfils those requirements. Section IV explores the impact of other technologies on AVs alongwith 5G. Security concerns in AVs are discussed in Section V. Section VI highlights the key research projects and standardisation activities. Future research directions are discussed in Section VII. Section VIII finally concludes the article.

\section{Introduction to Autonomous and Connected Vehicles}
The recent era is witnessing a rapid growth in the urban life and this growth has brought the need for high mobility and drastic changes in the transportation technologies. The Autonomous and Connected Vehicles (ACVs) plays one among the major components in the innovative solutions for promoting Intelligent Transportation System (ITS) \cite{khazraeian2019intelligent}. ITS is one in which numerous vehicles communicate with one another with the utilization of a communication infrastructure for exchanging critical information like traffic, congestion and road conditions. This section discusses about the background of ACVs, architecture, key enabling technologies and various requirements for Beyond 5G (B5G).

\subsection{Evolution of Autonomous and Connected Vehicles}
ACVs are characterised by two major properties namely capability of automation and connectivity \cite{shladover2018connected}. This property portrays them as unique when compared to the conventional vehicles that are connected with each other. The concept of self-driving cars are not completely new. The experiments on the self-driving cars started early during 1930s and the first modern control system played a major role. The modern cruise control system which was developed in 1948 paved the motivation for the development and evolution of autonomous vehicle. In 1966, mechanical anti-lock braking was installed in a car which was under standard production. Next came the electronic cruise control in the year 1968. In 1987, BMW and Bosch invented the electronic stability system. Later in 1995, Mitsubishi Diamante introduced the adaptive cruise control which was laser based. Nissan introduced lane departure warning system in 2001. In 2003, a pre-crash mitigation system was tested by Toyota Harrier. Later came many significant developments like traffic sign recognition (2009), Google car (2010), Mercedes S-class (2013), autopilot (2015) in Tesla cars, and intelligent speed adaptation (2022). A comparative analysis of ACVs with all other vehicles starting from conventional vehicles are illustrated in the Table~\ref{tab:comparison}. These ACVs are expected to have a high impact on the urban society and in future smart city applications. The connectivity among the autonomous vehicles are realized using the technology called vehicular networks or vehicular ad-hoc networks \cite{hussain2018autonomous}. Before we explore the features of ACVs, it is worth while to discuss briefly the concept of CVs and their limitations which motivated the researchers to focus on ACVs. CVs are designed to support better connectivity by utilizing various levels of communication which include Vehicle-to-Vehicle (V2V) and Vehicle-to-Infrastructure (V2I) \cite{contreras2017internet}. Since the communication levels are heterogeneous, various issues arise on scalability aspects and coverage area. An ACV is an autonomous vehicle which has connectivity with all the levels of communication channels and retrieves information from vehicle to everything (V2X) communication level \cite{zhang2018vehicular} to provide a complete field view and hence optimize the autonomous driving of other road users. Due to this characteristic, apart from providing better safety and improved performance than autonomous vehicle, ACVs also improve the throughput of traffic and economy of fuel by optimizing the route and cooperative driving. 

\subsection{Features of Autonomous and Connected Vehicles}
The salient features of ACVs which paves a way for fully autonomous transportation system are as discussed below.

\textbf{Management and Organization:} The fundamental feature of ACV is the organization and management. The basic intention is to free the human from keeping track of the schedule of maintenance related operations and to provide a platform which is available 24/7 with peak performance. ACVs would assist the vehicles in adjusting the operations when software failures occur and adapt to the dynamic environment. ACVs will help in monitoring the self usage and updates of software. Apart from this, ACVs are expected to detect the malicious activity of a specific autonomous vehicle and isolate the same away from the network. The ACVs also provide an alternative solution in case of malicious activity to the passengers in a very least time.

\textbf{Optimal Configuration:} For optimal functioning of the ITS, the complex system has to be configured error free. But since the system is a large scale configuration, it is error prone and time consuming. The feature expects the ACVs to adapt and accommodate all the third party components like transport policies, traffic control authorities and road users' policies.

\textbf{Optimized Resource Allocation:} The ACVs have the capability to tune the performance and behaviour of the system using numerous factors as illustrated in the Table \ref{tab: Factors_Parameters }. For instance, this feature can support the system to allocate the resource based on the mobility of the vehicle. This also assists in tuning of performance parameters. 

\textbf{Self-Protection:} This feature is oriented towards self-protection and assists in providing security to the ACVs from malicious attacks. The hardware as well as software are protected from various attacks. Once a failure occurs, the system should be having the capability to mitigate the rigorous effect and hence avoid the failure of the entire network of systems.
\begin{table*}[h!]
  \centering
        \caption{Comparative Analysis \textemdash From Conventional Vehicles to Autonomous Connected Vehicles.}
        \label{tab:comparison}
        \resizebox{\textwidth}{!}{%
  \begin{tabular}{|p{3.8cm}||c|c|c|c|c|c|c|c|c|c|c|c|c|c|c||c|c|c|c|}
  \hline 
  \rowcolor{gray!25}
  & \multicolumn{15}{|c||}{\textbf{Features}} & \multicolumn{4}{c|}{\textbf{Technologies}} \\
  \hline
   \rowcolor{gray!25}
      	
         &{\rotatebox[origin=c]{90}{~Data Transmission~}}
         &{\rotatebox[origin=c]{90}{~Data Analysis~}}
         &{\rotatebox[origin=c]{90}{~Safety~}}
         &{\rotatebox[origin=c]{90}{~Energy Consumption~}}
         &{\rotatebox[origin=c]{90}{~Comfort for Driver~}}
         &{\rotatebox[origin=c]{90}{~Comfort for Passenger~}}
         &{\rotatebox[origin=c]{90}{~Independent Decisions~}}
         &{\rotatebox[origin=c]{90}{~Sensors~}}
        &{\rotatebox[origin=c]{90}{~Computer Vision~}}
        &{\rotatebox[origin=c]{90}{~Wireless Communication~}}
        &{\rotatebox[origin=c]{90}{~Tracking~}}
         &{\rotatebox[origin=c]{90}{~Environment Awareness~}}
         &{\rotatebox[origin=c]{90}{~Resource Consumption~}}
         &{\rotatebox[origin=c]{90}{~Process Optimization~}}
         &{\rotatebox[origin=c]{90}{~Self- Protection~}}
          
         &{\rotatebox[origin=c]{90}{~Co-operative Driving~}}
         
         &{\rotatebox[origin=c]{90}{~Physical Layer Security~}}
         
        &{\rotatebox[origin=c]{90}{~Cloud/Edge/Fog/Roof Computing~}}
         &{\rotatebox[origin=c]{90}{~mmWave~}}
                  \\

  \hline
\hline

Conventional Vehicles    & \cellcolor{blue!15}N  & \cellcolor{blue!15}N  & \cellcolor{green!15}L  & \cellcolor{red!15}H & \cellcolor{green!15}L & \cellcolor{yellow!15}M  & \cellcolor{blue!15}N  & \cellcolor{blue!15}N & \cellcolor{blue!15}N& \cellcolor{blue!15}N& \cellcolor{blue!15}N & \cellcolor{blue!15}N  & \cellcolor{red!15}H & \cellcolor{blue!15}N & \cellcolor{blue!15}N& \cellcolor{violet!25}NA& \cellcolor{violet!25}NA& \cellcolor{violet!25}NA& \cellcolor{violet!25}NA\\
\hline

Connected Vehicles   & \cellcolor{yellow!15}M  & \cellcolor{blue!15}N  & \cellcolor{green!15}L  & \cellcolor{yellow!15}M & \cellcolor{green!15}L & \cellcolor{yellow!15}M  & \cellcolor{blue!15}N  & \cellcolor{green!15}L & \cellcolor{blue!15}N & \cellcolor{yellow!15}M & \cellcolor{yellow!15}M & \cellcolor{green!15}L  & \cellcolor{red!15}H & \cellcolor{blue!15}N & \cellcolor{blue!15}N& \cellcolor{orange!35}A& \cellcolor{violet!25}NA& \cellcolor{orange!35}A& \cellcolor{violet!25}NA\\
\hline
 Advanced Connected Vehicles  & \cellcolor{yellow!15}M  & \cellcolor{green!15}L  & \cellcolor{yellow!15}M  & \cellcolor{green!15}L & \cellcolor{green!15}L & \cellcolor{yellow!15}M  & \cellcolor{blue!15}N  & \cellcolor{green!15}L & \cellcolor{green!15}L & \cellcolor{yellow!15}M  & \cellcolor{yellow!15}M & \cellcolor{yellow!15}M & \cellcolor{red!15}H  & \cellcolor{blue!15}N & \cellcolor{blue!15}N& \cellcolor{orange!35}A& \cellcolor{violet!25}NA& \cellcolor{orange!35}A& \cellcolor{violet!25}NA\\
\hline
Self-driving Vehicles  & \cellcolor{green!15}L  & \cellcolor{green!15}L  & \cellcolor{yellow!15}M  & \cellcolor{yellow!15}M & \cellcolor{yellow!15}M & \cellcolor{red!15}H  & \cellcolor{blue!15}N  & \cellcolor{yellow!15}M & \cellcolor{yellow!15}M & \cellcolor{yellow!15}M  & \cellcolor{green!15}L & \cellcolor{yellow!15}M & \cellcolor{red!15}H  & \cellcolor{blue!15}N & \cellcolor{green!15}L & \cellcolor{violet!25}NA& \cellcolor{violet!25}NA& \cellcolor{orange!35}A& \cellcolor{orange!35}A\\
\hline
Autonomous Vehicles  & \cellcolor{yellow!15}M  & \cellcolor{yellow!15}M  & \cellcolor{red!15}H  & \cellcolor{green!15}L & \cellcolor{red!15}H & \cellcolor{red!15}H  & \cellcolor{yellow!15}M  & \cellcolor{red!15}H & \cellcolor{red!15}H & \cellcolor{yellow!15}M  & \cellcolor{yellow!15}M & \cellcolor{yellow!15}M & \cellcolor{yellow!15}M  & \cellcolor{yellow!15}M & \cellcolor{yellow!15}M& \cellcolor{orange!35}A& \cellcolor{orange!35}A& \cellcolor{orange!35}A& \cellcolor{orange!35}A\\
\hline
Connected Autonomous Vehicles  & \cellcolor{red!15}H  & \cellcolor{red!15}H  & \cellcolor{red!15}H  & \cellcolor{green!15}L & \cellcolor{red!15}H & \cellcolor{red!15}H  & \cellcolor{red!15}H  & \cellcolor{red!15}H & \cellcolor{red!15}H & \cellcolor{red!15}H  & \cellcolor{red!15}H & \cellcolor{red!15}H & \cellcolor{green!15}L  & \cellcolor{yellow!15}M & \cellcolor{red!15}H& \cellcolor{orange!35}A& \cellcolor{orange!35}A& \cellcolor{orange!35}A& \cellcolor{orange!35}A\\
\hline

\hline

  \end{tabular}
  }
  \begin{flushleft}
\begin{center}
    
\begin{tikzpicture}

\node (rect) at (6,2) [draw,thick,minimum width=1cm,minimum height=0.4cm, fill= red!15, label=0:High Level: \color{black}This feature plays a vital role and is given higher importance while manufacturing. \color{black}] {H};
\node (rect) at (6,2.5) [draw,thick,minimum width=1cm,minimum height=0.4cm, fill= yellow!20, label=0:Medium Level: \color{black}This feature is taken into account while manufacturing but not wide coverage.\color{black}] {M};
\node (rect) at (6,3) [draw,thick,minimum width=1cm,minimum height=0.4cm, fill= green!15, label=0:Lower Level: \color{black}This feature does not play a vital role and is not given importance while manufacturing.\color{black}] {L};
\node (rect) at (6,1.5) [draw,thick,minimum width=1cm,minimum height=0.4cm, fill= blue!15, label=0:Not Available.] {N};
\node (rect) at (10,1.5) [draw,thick,minimum width=1cm,minimum height=0.4cm, fill= violet!25, label=0:Not Applicable.] {NA};
\node (rect) at (14,1.5) [draw,thick,minimum width=1cm,minimum height=0.4cm, fill= orange!35, label=0: Applicable.] {A};
\end{tikzpicture}
\end{center}

\end{flushleft}
  
  \end{table*}

\subsection{Levels of Automation}
\begin{figure*}[h!]
\centering
\includegraphics[width=.85\linewidth]{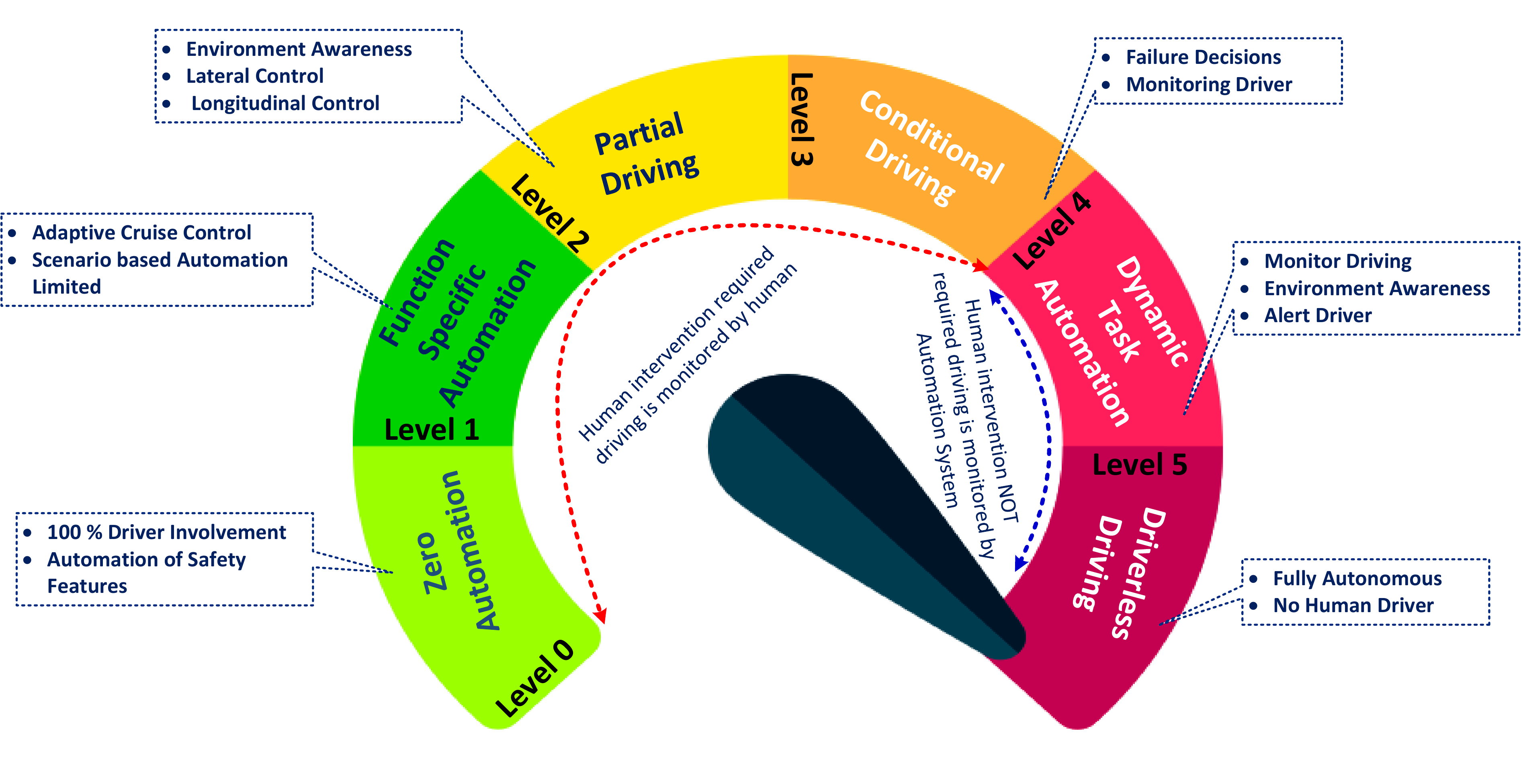}
\caption{Levels of Automation in Autonomous Vehicles.}
\label{Fig:levels}
\end{figure*}

As per the guidelines provided by Society of Automotive Engineers (SAE) \cite{taxonomy2016definitions}, there are six levels of automation for driving the vehicles autonomously as illustrated in the Fig. \ref{Fig:levels}. The levels are depicted based on the depth of automation utilized.
\begin{itemize}
	\item \textbf{Level 0:} This level is considered as no automation level since almost all the tasks and operations with respect to driving a vehicle is taken care and controlled by human. Decisions regarding driving and road usage is done by human intervention.
    \item \textbf{Level 1:} The level is focused towards automating only selective functionalities of driving and provides assistance to human in driving to a limited range. Few examples of level 1 automation are lateral and longitudinal control of motion.
    \item \textbf{Level 2:} This is oriented towards automating the driving task by combining two or more controls. Examples include adaptive cruise control, lane maintaining assistance.
    \item \textbf{Level 3:} This is a kind of automation which facilitates conditional driving. The driver is assisted while driving the vehicle for sometime in between the whole drive for performing some other activity and then he resumes his driving.  
    \item \textbf{Level 4:} The automated system takes care of the entire driving mechanism including monitoring the current environment for detection of dynamic changes and controlling of motion as per the changes. The system also allows the driver to take control of the entire system at critical situations.
    \item \textbf{Level 5:} This is the highest level of automation where the system controls the driving mechanism as well as monitors the current environment and the surroundings nearby by concept called cooperative driving. There is no intervention of human at any point of time. All the failures and dynamic decisions are also controlled by the system throughout the trip. 
\end{itemize}

\begin{table}[h!]
	\centering
	\caption{Factors for Optimized Resource Allocation \cite{jameel2019internet}. }
	\setlength{\tabcolsep}{1pt}
	\centering
\arrayrulecolor{black}
\resizebox{\columnwidth}{!}{%
\begin{tabular}{|l|l|l|} 
\hline
\rowcolor{gray!30}
\textbf{ Factors}         & \textbf{ Parameters}                                        & \textbf{ Examples of Optimization}                                                                            \\ 
\hline
\multirow{4}{*}{Technical} & \multirow{2}{*}{System Oriented}                             & Ability to migrate the load                                                                                     \\ 
\cline{3-3}
                           &                                                              & Improve System Security                                                                                         \\ 
\cline{2-3}
                           & Network Oriented                                             & Availability of Signal to Noise Ratio                                                                           \\ 
\cline{2-3}
                           & Vehicle Oriented                                             & Better Power utilization                                                                                        \\ 
\hline
\multirow{2}{*}{Physical}  & \multicolumn{1}{l!{\color{black}\vrule}}{Speed and Velocity} & \multicolumn{1}{l!{\color{black}\vrule}}{The vehicles' speed of motion}                                         \\ 
\arrayrulecolor{black}\cline{2-3}
                           & Pattern of Motion                                            & \begin{tabular}[c]{@{}l@{}}Whether the vehicle is moving in urban,\\ suburban, village or highway\end{tabular}  \\ 
\arrayrulecolor{black}\hline
\multirow{3}{*}{Temporal}  & Total Duration                                               & \begin{tabular}[c]{@{}l@{}}The total duration for which a service is\\ available\end{tabular}                   \\ 
\cline{2-3}
                           & \multicolumn{1}{l!{\color{black}\vrule}}{Time}               & \multicolumn{1}{l!{\color{black}\vrule}}{Certain services are available only during day}                        \\ 
\arrayrulecolor{black}\cline{2-3}
                           & Frequency                                                    & Frequency at which user need the application                                                                    \\ 
\arrayrulecolor{black}\hline
\multirow{2}{*}{Economy}   & Price                                                        & The cost of transmission as well as reception                                                                   \\ 
\cline{2-3}
                           & Subscribe                                                    & Subscription costs for a specific service                                                                       \\ 
\hline
\multirow{2}{*}{Others}    & Utilization                                                  & \begin{tabular}[c]{@{}l@{}}Information exchange, broadcasting information\\ due to emergency\end{tabular}       \\ 
\cline{2-3}
                           & Preference                                                   & Service need and preferred application                                                                          \\
\hline
\end{tabular}
}
	\label{tab: Factors_Parameters }
\end{table}

\subsection{Architecture}

\begin{figure*}[t]
\centering
\includegraphics[width=\linewidth]{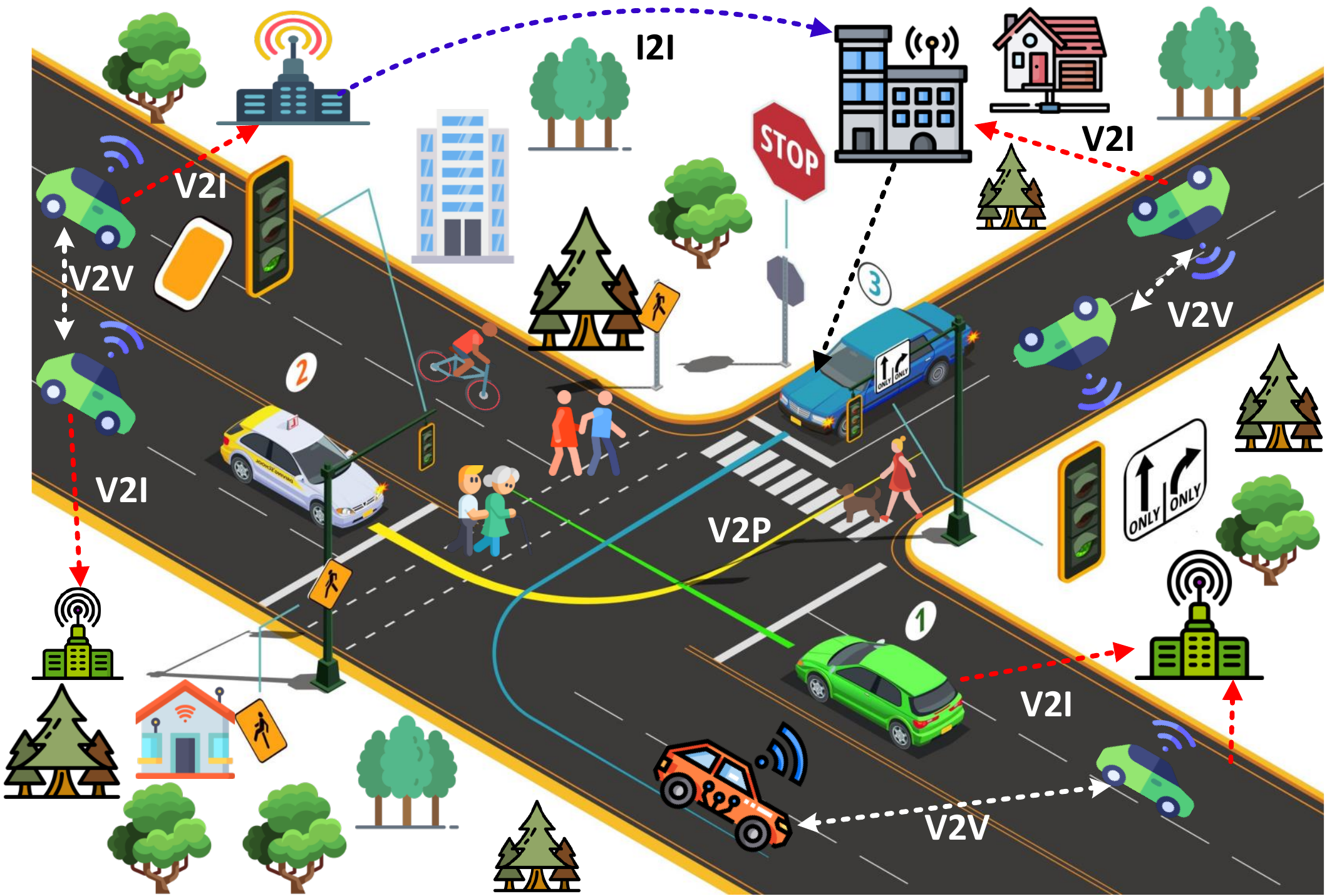}
\caption{Illustration of Autonomous Connected Vehicles, Infrastructure, Environment and Communication.}
\label{Fig:Infrastructure}
\end{figure*}

ACVs refer to an integrated domain of various technologies which help in achieving the inter-vehicular connectivity for providing various services in applications like traffic safety, assistance on roadsides, driving efficiency, remote monitoring, traffic congestion avoidance, maintenance, and system failures. To achieve these, the ACVs are configured with wide range of sensors deployed onboard which communicate among them through Controller Area Network (CAN) bus, infrastructures related to communication and other vehicles. ACV is a blooming technology that is focused by both academicians and industrialists. The motto for moving towards this technology is to provide better safety for road users, better flow of traffic, reduction in fuel consumption and cost of travel. The basic infrastructure required for implementation and deployment of the ACVs to provide ITS is illustrated in the Fig. \ref{Fig:Infrastructure}. As per the illustration, the entire network could be grouped into two major type of nodes namely vehicles which have on-board sensor units and the other communication based infrastructures along the roadsides. The channels which help in the communication within the networks are Vehicle-to-Vehicle (V2V), Infrastructure-to-Infrastructure (I2I), Vehicle-to-Infrastructure (V2I), Vehicle-to-People (V2P) and Vehicle-to-Everything (V2X). The vehicles' on-board units comprise of sensors which help in detecting the objects as well as obstacles within their dedicated range. The frequently used sensors, their range of sensing and possible applications in which they are used are tabulated in the Table \ref{tab:onboardsensors_ACV }.

\begin{table}[h!]
	\centering
	\caption{On-Board Sensor Types, Range and Usage \cite{yurtsever2020survey}}
	\setlength{\tabcolsep}{2pt}
	\begin{tabular}[t]{|p{1.5cm}| p{1.1cm}|p{2.5cm}|p{3cm}|}
		\hline
		\rowcolor{gray!30}
		\textbf{Sensor Type}  & \textbf{Range} & \textbf{Example} & \textbf{Usage}  \\ \hline
		Proximity sensors  & 5 m & Ultrasonic sensor  & Detects nearby obstacles
Parking assistance \\ \hline
Short range sensors & 30 m & Forward camera

Backward camera

Short range radars & Recognition of traffic signs

Detection of blind spots

Alerting cross traffic

Lane detection \\ \hline

Medium range sensors  & 80-160 m & LiDAR

Medium range radars & Detection of pedestrians

Collision avoidance \\ \hline
Long range sensors  & 250 m & Long range radars & Support adaptive cruise control

Information collection at high speed \\ \hline

\end{tabular}
	\label{tab:onboardsensors_ACV }
\end{table}

\begin{figure*}[t]
\centering
\includegraphics[width=\linewidth]{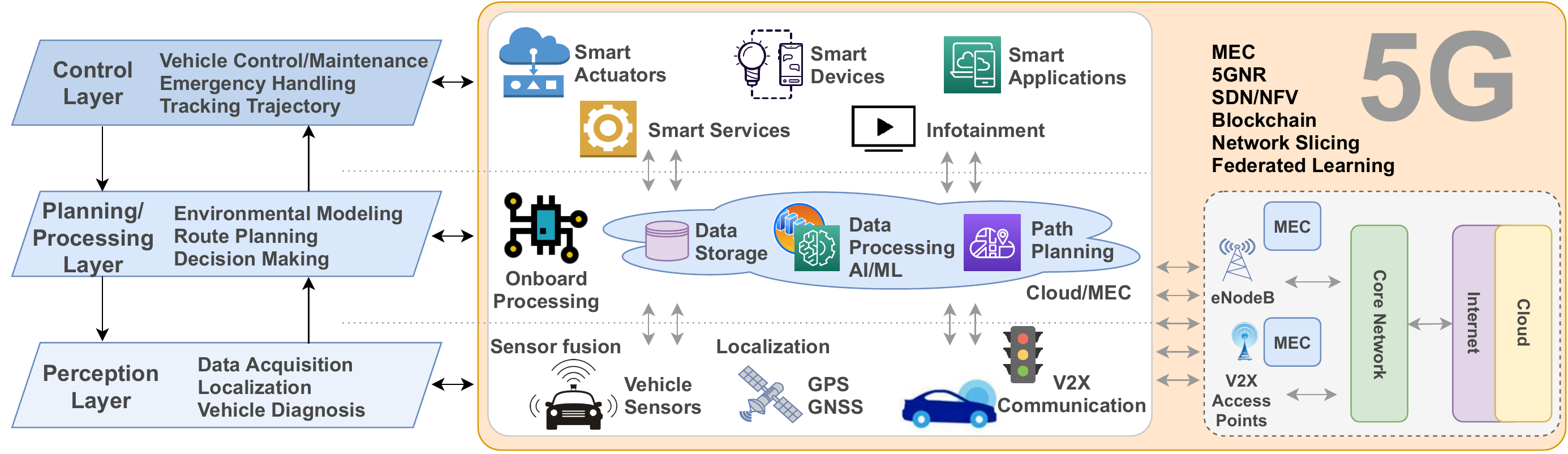}
\caption{General architecture of autonomous connected vehicles}
\label{Fig:frame}
\end{figure*}

The general architecture of ACVs can be illustrated as in Fig.~\ref{Fig:frame}. The architecture consists of three layers which are grouped based on the functionalities performed by the ACV. The first and foremost layer is the perception layer which is responsible for gathering the raw information from the environment with the help of sensors mounted over the vehicle. The sensor data collected is utilized and utilizing sensor fusion techniques the local and global location parameters are calculated by the perception layer and a map of the environment is generated\cite{van2018autonomous}. The next layer is the planning/processing layer which plays a major role in determining the global route that suits best as per the current position and the requested destination. For determining the best possible route, the layer utilizes the remote data of road and traffic. Based on the environment map generated by the perception layer, the trajectory planning and tracking is computed\cite{gonzalez2015review}. The final layer is the control layer which concentrates on providing appropriate commands to control the various actuators of the ACV like steering wheel, gas pedal and brake pedal \cite{amer2017modelling}. Apart form these major functionalities, the perception layer also shares the perceived information of the environment with the other users of the road and hence the planning layer supports co-operative driving \cite{guanetti2018control} with the help of the inter-connectivity along with the other road users. Decision making is the major functionality of the planning layer where decisions with respect to control of servo motor as well as control of the actuator is taken. For such high level and critical decisions, the connectivity among the vehicles, among the infrastructure and the road users plays a major role which increases the complexity in deployment of the ACVs.      
\subsection{Key enabling technologies}
The current traditional methodologies which are proved in real-time may not be suitable for implementing and deploying the various salient features of ACVs. But, recently there are many innovative technologies and approaches in the field of sensors, cloud computing and artificial intelligence which can help in developing and designing intelligent ACVs that can show various benefits. This subsection discusses about the recent technologies that could support the deployment of ACVs in real-world.
\subsubsection{Sensing Environment}

\begin{figure}[b]
    \centering
    \includegraphics[width=0.5\textwidth]{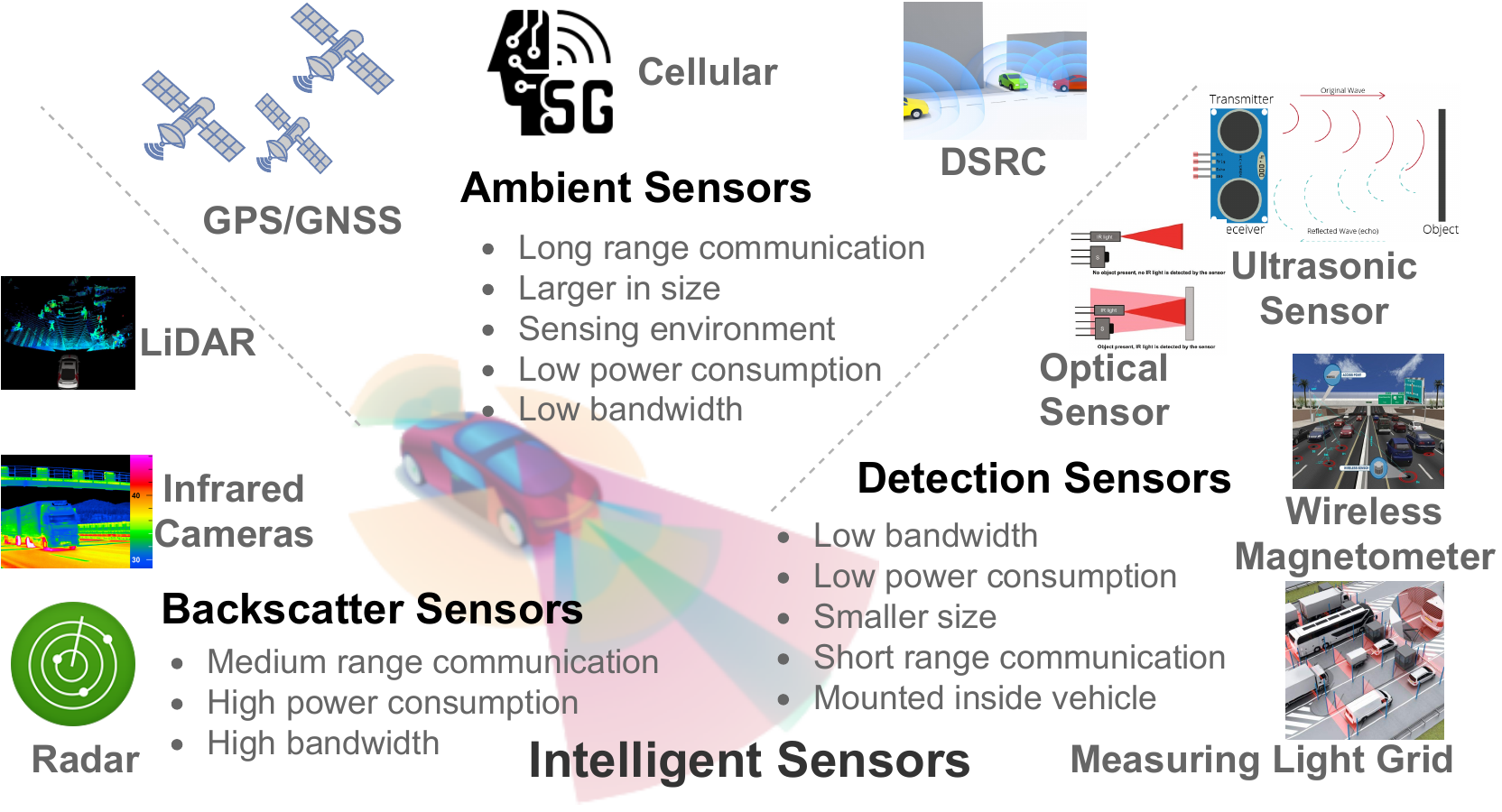}
    \caption{Intelligent sensors in autonomous and connected vehicles}
    
\label{Fig:intelligent sensors}
\end{figure}

Vehicular networks are those which focus mainly on sensor networks of either single type or homogeneous sensors whereas ACVs focus on various sensors which are heterogeneous in nature. The frequently used sensors for ACVs are broadly classified into three types\cite{jameel2019internet} as illustrated in the Fig. \ref{Fig:intelligent sensors}. The detection sensors are those which are usually mounted on the vehicle for identifying the various features in and around the environment. These sensors can also help in the process of monitoring the working as well as inner condition of the vehicle. Ambient sensors are those which generally monitor the environment and gather all the sensitive legacy data. The gathered data is transferred to the corresponding authorities by these sensors. Back-scatter sensors are specially built for usage in all kinds of objects and helps in providing better perception of the outer world including trespassers, bicyclists, etc.

\subsubsection{Accessing Data}
The ACVs generate a large amount of data in the form of signals which are emitted by the sensors mounted in, on and around the vehicles. The generated data need to be accessed by the authorised authorities, other vehicles in the network, other infrastructures in the vehicular environment for the purpose of taking decisions at the right time. For the purpose of storing temporarily or archiving and hence accessing, very high end resources as well as servers are required. Recently, the cloud computing technology has solved the data access challenge by providing virtual resources whenever required and is provided as a service \cite{joy2018internet}. This would allow the ACVs to communicate among themselves, infrastructures, things, and everything in the environment with better performance. Apart from cloud based technology, recently there are other key technologies like fog, edge and roof which also support in accessing the huge vehicular data. Also, few vehicular fog nodes are developed recently which function based on fog computing technology. These nodes provide better information about the environment as well as helps in inter as well as intra-vehicular communication. These gathered data can be processed further by the fog nodes and analytic can be performed on the crowd sourced data from the ACVs. There are various algorithms for optimizing the performance of the fog nodes further. 
\subsubsection{Vehicular Communication}
Almost all functionalities of the ACVs are dependent on the data perceived by the various heterogeneous sensors and their reception. For the purpose of improving the performance, the communication bandwidth has to be higher to an extent of giga-bits-per-second. To achieve high performance and better communication, millimeter-wave (mmWave) is being utilized for ACVs. These can be used for V2V, V2I and for intra-vehicular communications as well. The mmWave V2V links can help the vehicles to gather and share the raw information from and to the neighbouring vehicles in a particular environment. The mmWave V2I links are utilized by the ACVs in road safety based applications. These links help in gathering the data sensed from thee vehicles and forwarding further to the cloud resource for archiving or decision making. The higher data rate mmWave links are utilized by the ACVs for downloading real-time maps and streams of dynamic environment. Various communication bands can be utilized by ACVs for mmWave. To list out a few, 5G banks like 28 GHz and 28 GHz, unlicensed band like 60 GHz and automotive radar bands like 24 GHz and 76 GHz. 
\subsubsection{Security}
Security plays a major role in case of any type of vehicular communications. This is a very critical challenge which is to be handled when the environment includes both legacy as well as autonomous vehicles. In such scenarios, if the security is not handled properly, there can be a chaos \cite{zhang2018vehicular}. To handle the vehicular communications and also to maintain the security and privacy, recently physical layer security (PLS) has been designed as a replacement for cryptography. Numerous studies have shown that PLS can be utilized for improving the performance of secrecy and for employing jamming techniques among the source and destination \cite{jameel2018interference}. Also PLS can be employed for restricting the vehicles from broadcasting any false data. Recently block-chain techniques are being used in most of the applications for building a trust model for ACVs. These trust models are usually built by providing some additional policies and certificates.  

\subsection{5G and Beyond Requirements for Autonomous Vehicular Communication}

5G is a recent emerging platform which is developed for not only supporting the existing applications but also to support and leverage the upcoming novel applications that require communication at a very low latency. This technology provides building blocks for supporting the existing traditional platforms like 2G, 3G, 4G, and WiFi. Apart from this, it also leverages greater connectivity and coverage for handling higher network density. It is best suited for vehicular communication. This section discusses about the various building blocks required for using the 5G and beyond techniques for ACVs. 
\subsubsection{Proximity Service}
The most important feature of the 5G communication is the Proximity Service (ProSe). The ProSe is a service which aims in providing awareness to the vehicles about the devices, infrastructure and other objects in the environment along with the locality information. The most significant feature of ProSe is that it provides spontaneous communication and interactions within a certain locality. For example, ProSe is best suited for discovering the moving vehicles on the road. Apart from this, it can be used as a communication service for scenarios like public safety. 
\subsubsection{Multi-access Edge Computing}
The key feature of 5G enabled vehicular communication is the low latency (upto 100 ms) utilized for safety measures and ultra-low latency (upto 1 ms) utilized in ACVs. One of the methodologies for achieving such low latency is to move the basic and core functionalities towards the user or the customer. This is considered to be the edge. Multi-access Edge Computing (MEC) plays a significant role in bringing all kinds of services to the concerned network locations. Every edge includes a multi-vendor environment which helps in hosting most of the mobile edge applications. These services can be achieved through a technology called network functions virtualization (NFV). There are various kinds of access technologies like WiFi, 802.11p and 5G which could be used by the vehicles for their communication. Though MEC is suitable for communication among the ACVs, it might be crucial in case of Traffic Information Systems (TIS) where the requirement of latency is flexible. The reason behind is that TIS do not demand any hard and fast rule on the latency. Hence even a small improvement in latency by using MEC can give a better user experience. 
\subsubsection{Network Slicing}
As discussed in the previous sections, 5G supports all the existing access technologies by providing a wider range. Since there are numerous networks utilizing different access technologies, managing all of them under a single roof named 5G is the key challenge. Network slicing helps in solving this challenge by separating the networks logically. For instance, in case of ACVs the networks can be sliced based on the applications and requirements. To name a few, safety applications, infotainment applications, mission critical system applications. Each application can be logically separated as a network slice. Safety applications require low latency but reliable message transmission. Infotainment applications aim at high bandwidth. Mission critical systems require very fast and spontaneous information exchange especially during disasters. Network slicing also helps in maintaining the integrity as well as security in vehicular networks.

\section{Technical Aspects of Autonomous Vehicles}

\begin{table}[h!]
  \centering
        \caption{Technical Aspects of 5G} 
        \label{tab:TA5G}
\resizebox{\columnwidth}{!}{%
  \begin{tabular}{|p{1.5cm}|p{0.5cm}|p{0.5cm}|p{0.5cm}|p{0.5cm}| p{0.5cm}| p{0.5cm}| p{0.5cm}| p{0.5cm}|}
    \rowcolor{gray!25}
   \hline
   \cellcolor{gray!15}AV Application &
{\rotatebox[origin=c]{90}{\cellcolor{gray!15}~Latency~}}&
{\rotatebox[origin=c]{90}{\cellcolor{gray!15}~Security level~}}&
{\rotatebox[origin=c]{90}{\cellcolor{gray!15}~Privacy~}}&
{\rotatebox[origin=c]{90}{\cellcolor{gray!15}~Bandwidth~}}&
   {\rotatebox[origin=c]{90}{\cellcolor{gray!15}~Mobility~}}&
{\rotatebox[origin=c]{90}{\cellcolor{gray!15}~Scalability~}}&
{\rotatebox[origin=c]{90}{\cellcolor{gray!15}~Availability~}}&
{\rotatebox[origin=c]{90}{\cellcolor{gray!15}~Reliability~}}\\
\hline   
\hline
   
Navigation and Path Planning & \cellcolor{yellow!15}M & \cellcolor{red!15}L & \cellcolor{yellow!15}M
& \cellcolor{yellow!15}M &\cellcolor{red!15}L&\cellcolor{red!15}L & \cellcolor{green!15}H & \cellcolor{green!15}H \\
   \hline
\hline
Object Detection & \cellcolor{red!15}L& \cellcolor{red!15}L & \cellcolor{yellow!15}M
& \cellcolor{green!15}H & \cellcolor{green!15}H &\cellcolor{green!15}H & \cellcolor{green!15}H & \cellcolor{green!15}H \\
  \hline
 \hline
URLLC & \cellcolor{red!15}L & \cellcolor{yellow!15}M & \cellcolor{yellow!15}M
& \cellcolor{red!15}L & \cellcolor{green!15}H &\cellcolor{yellow!15}M & \cellcolor{green!15}H & \cellcolor{green!15}H \\
  \hline
 \hline
mMTC & \cellcolor{yellow!15}M & \cellcolor{green!15}H & \cellcolor{green!15}H & \cellcolor{red!15}L & \cellcolor{yellow!15}M & \cellcolor{green!15}H & \cellcolor{red!15}L & \cellcolor{red!15}L \\
  \hline
  \hline
   eMBB & \cellcolor{red!15}L & \cellcolor{green!15}H & \cellcolor{green!15}H
& \cellcolor{green!15}H & \cellcolor{green!15}H & \cellcolor{green!15}H & \cellcolor{yellow!15}M & \cellcolor{yellow!15}M \\
  \hline
  \end{tabular}
}
  \begin{flushleft}
\begin{center}
    
\begin{tikzpicture}
\node (rect) at (6,2) [draw,thick,minimum width=.1cm,minimum height=0.4cm, fill= green!15, label=0:High: \color{black} Most important in performing the functionality. \color{black}] {H};
\node (rect) at (6,2.5) [draw,thick,minimum width=.1cm,minimum height=0.4cm, fill= yellow!20, label=0:Medium: \color{black}More important in performing the functionality.\color{black}] {M};
\node (rect) at (6,3) [draw,thick,minimum width=.1cm,minimum height=0.4cm, fill= red!15, label=0:Low: \color{black}Less important in performing the functionality.\color{black}] {L};
\end{tikzpicture}
%
\end{center}

\end{flushleft}
  
  \end{table}


In recent years, both the academia and industry have shown great interest in the development of autonomous driving, which will liberate drivers physically and mentally, greatly improve traffic safety and energy efficiency, as well as make better use of public resources. Universities and research groups are actively involved in autonomous driving competitions and technical challenges. Some of the universities, automobile companies and Internet auto companies have joined the ranks of autonomous vehicle manufacturing and research. The development of light detection and ranging, Radar, camera, and other advanced sensor technologies inaugurated a new era in autonomous driving. However, due to the intrinsic limitations of these sensors, autonomous vehicles are prone to making erroneous decisions and causing serious disasters. Hence, this section, provides  an overview  of  technical aspects of autonomous vehicles as illustrated in Table \ref{tab:TA5G} and also discusses the networking and communication technologies that can improve autonomous vehicle's perception and planning capabilities as well as realizing better vehicle control. The summary of this section is shown in Table \ref{tab:Tech aspects of AV}.

\subsection{Navigation and Path Planning} 
\subsubsection{Introduction}

\begin{figure}[t]
    \centering
    \includegraphics[width=0.5\textwidth]{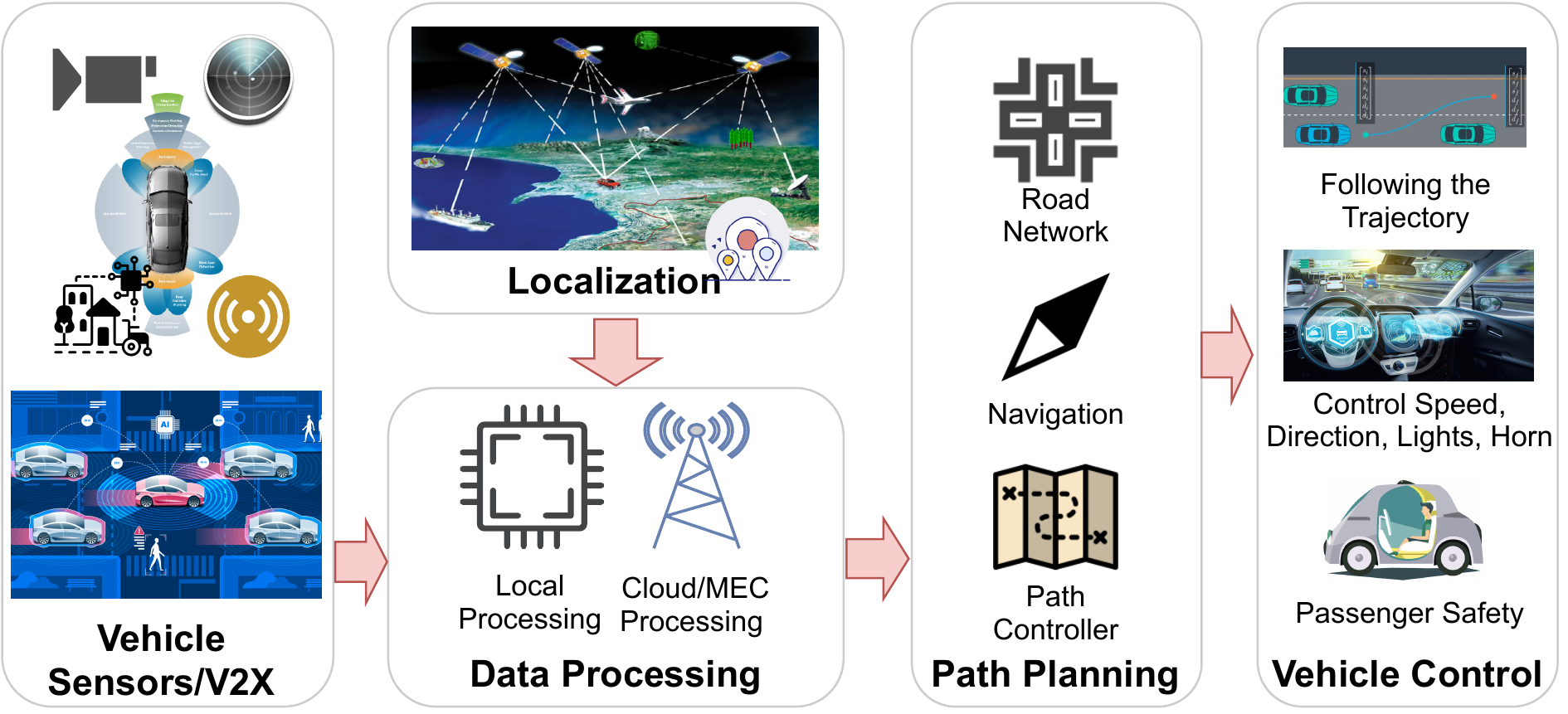}
    \caption{AV data acquisition, data processing, path planing and vehicle control}
    \label{Fig:Navigation and path planning}
\end{figure}

AV are going to revolutionize the future of road transportation. In future, ordinary vehicles would be replaced by smart vehicles which are capable of making decision, choosing the shortest path and planning the optimal travel route. Inorder to achieve these objectives, latest advancement in communication technologies like 5G can be used in local perception for controlling the short range vehicles with respect to the safety, traffic control and energy management parameters. There are more functions developed to take the control of individual driving function to fully autonomous system. Navigation is one of the key tasks in AV which automatically calculates the route from source to destination. The data of the road network should be available to the vehicle to design the route for destination point in advance. Autonomous navigation includes path planning, obstacle detection, obstacle avoidance and finding the optimal path for safe as shown in Fig. \ref{Fig:Navigation and path planning}.

\subsubsection{Existing Challenges/Limitations}
Autonomous navigation depends on these technologies like localization, planning and control. Maintaining a reliable localization along the planned path is a basic requirement in AV. Autonomous robots in UAV and UGV has gained its importance in AV which provides more stable and robust system. Specifically in UAV applications, the initial trajectory path will be provided to the system for initial guidance and further the navigation will be guided based on the waypoints of the trajectory.

Localization can be classified based on GPS and Laser Range Finder. Even though GPS based system have proved good for outdoor environments, still signals are unreliable in indoor or dense urban environments. Whereas LRF performs better in both indoor and outdoor applications. In unstructured and complex environments, the LRF readings is the key factor for planning the path to minimize the localization errors or exposing the vehicle at risk of failure. In \cite{irani2018localizability}, authors have proposed the path planning for AV based on Localizability Constraint (LC) using the Laser Range Finder (LRF) sensor model of the vehicle. LC is maintained throughout the path planning and also it reduces the overall localization error. Finally, path planned with and without LC and the influence of LRF in the model is compared and also simulated.  The proposed method ensures effective localization performance and rich environmental information when compared to existing techniques.

Navigation and path planning under uncertainty is a very tedious task for AV such as UAV, drones and self-driving cars \cite{ jiang2019probabilistic}. Probabilistic methods can be used for optimization and modeling problems to capture the uncertainty prevailing in the dynamic environment. Rather than processing with traditional time-consuming methods like Monte-Carlo methods, probabilistic decision engine is designed to provide the navigation of AV under uncertain dynamic environment. Probabilistic decision engine can be implemented in the real-time FPGA hardware. Navigation algorithms can be combined with motion control algorithms and other path planning algorithms. Authors in \cite{jardine2018robust} have proposed generalized and stable Robust Model-Predictive Control (R-MPC) technique to design the optimal path or trajectory by solving a convex quadratic program. This technique implements the constraints and detect the obstacles within the bounded uncertainty. It works well in two scenarios: military UAV vehicle flying over the target and assistive care robot to safely navigate through cluttered home.

Recently, commercial and military applications have gained high demand for UAV. Vision based technique using optical flow is proposed \cite{lin2021autonomous} for obstacle detection and Collison avoidance for autonomous navigation in quadrotor UAVs. For initial path, map based offline path planning is developed followed by trajectory points for the flight guidance. Onboard camera is used during the navigation to analyze the monocular images for environment perception.  Basically vision based technique can be divided into three categories such as monocular visual cues that depends on the properties of the sequence of image acquired from which the image is classified as obstacle or non-obstacle. Second method, stereo vision based approach uses multiple cameras for depth computation and stereo matching. Third approach is using motion parallax combined with the camera motion. Autonomous navigation for marine environment is highly required as they often fail in their mission due to harsh environments. Deep reinforcement learning \cite{yan2021reinforcement} can be used for autonomous navigation in Unmanned Surface Vehicles (USV). Long short-term memory networks are used to remember the ocean environments values.

The technologies are developed by the researchers to transform from semi-autonomous driving to full autonomous driving. These AV technologies should support the realistic road scenarios including the complex driving situations. It also should have an effective warning system to avoid collisions and accidents. So, high speed navigation path must be planned for both structured and unstructured roads. Many studies are constructed for structured roads whereas for unstructured roads free-form navigation method like motion planning algorithm is required to produce global path towards the destination. While planning for structured road, it implicitly provides the preferred path and thus planning in unstructured road is less constrained, complex and computationally expensive problem. To find a primitive the authors in \cite{chu2015real} have used graph based search algorithm and discrete kinematic vehicle model to avoid collision and accidents.  

The decision in AV are taken based on the planner which creates the collision free waypoints to reach the destination. This planner module is capable of finding optimal solution by reducing the computational cost and the distance travelled by the vehicle. The planner module in navigation is divided global and local planner. Global planner considers the prior knowledge of the environment and static obstacles whereas the local planner regulates the path by considering the dynamic obstacles. The authors in \cite{ marin2018global} have developed Time Elastic Bands(TEB) method for local planning and Dijkstra algorithm for global planning. They collect the information from all the task in navigation process and provide these solutions as input to the global and local planner algorithm. The main two components in planner are sensor analytics and Path finder. Former combines the uncertainty of all sensors and evaluates the positioning and performance for a given location and time. Path finder utilizes the performance of sensor and defines the optimal and shortest path \cite{hoang2015path} to achieve the safety on road. AV navigation is highly influenced by uncertainties in data which leads to a high risk of accidents and it could be life threatening situation. These uncertainties could be from static sources which do not vary over time like an AV driving through the tunnel. The other one is from the dynamic source like weather condition that varies from with location, time and severity. For example, the quality and noise of image depends on the amount and the intensity of rain. These uncertainties in sensor must be considered while designing global planner to modify the plan accordingly. The authors in \cite{alharbi2020global} have designed a fuzzy logic based uncertainty indicator and formulating the cost function for selecting the optimal routes with the lowest sensor uncertainty and thus ensuring the safe navigation on road.   

\subsubsection{{How B5G help (with Related work)}}
Devices used for receiving signals in AV are sensors, radar, lidar and camera. Some of the challenges faced in these devices are further discussed. Camera devices are not able to detect objects when the climatic conditions are poor. In radar, the main problem is to differentiate objects type because of its longer wavelength. Lidar is useful to detect the objects surrounding the vehicle but the laser beams does not provide the accurate results in bad weather conditions like fog and snow. Also, it is so expensive compared to radar and camera. Next challenge in AV is constructing the maps based on the inputs received from lidars and cameras which is a difficult process and a time intensive process. With respect the safety parameter, the main issue is that it is not able to predict the agent behaviour. It is very hard for AV to sense the behaviour of other objects and infrastructure on road specifically predicting the human error and behaviour is a tedious task. The other legal issues like who is going to take the responsibility in case of an accident? So the basic challenge in AV without using 5G are summarized as follows
\begin{itemize}

\item Detecting objects in poor climatic conditions
Differentiating the type of objects
\item Fails to provide accurate results in fog or snow
\item Constructing maps from the signals received from devices is a time consuming task.
\item Unable to predict the agent behaviour on roads
\item Legal issues related to accident
\end {itemize}

\subsubsection*{Summary}
The main challenge in autonomous navigation is detecting the localizing the obstacles at specific time and adjusting the path accordingly to avoid the accident and navigate to the destination safely. Also, it requires an efficient, fast and dynamic object detection and path planning algorithms to have safe and accident free navigation of vehicles on road.

\subsection{Object detection/ Collision Avoidance}
\begin{figure}[t]
\centering
\includegraphics[width=\linewidth]{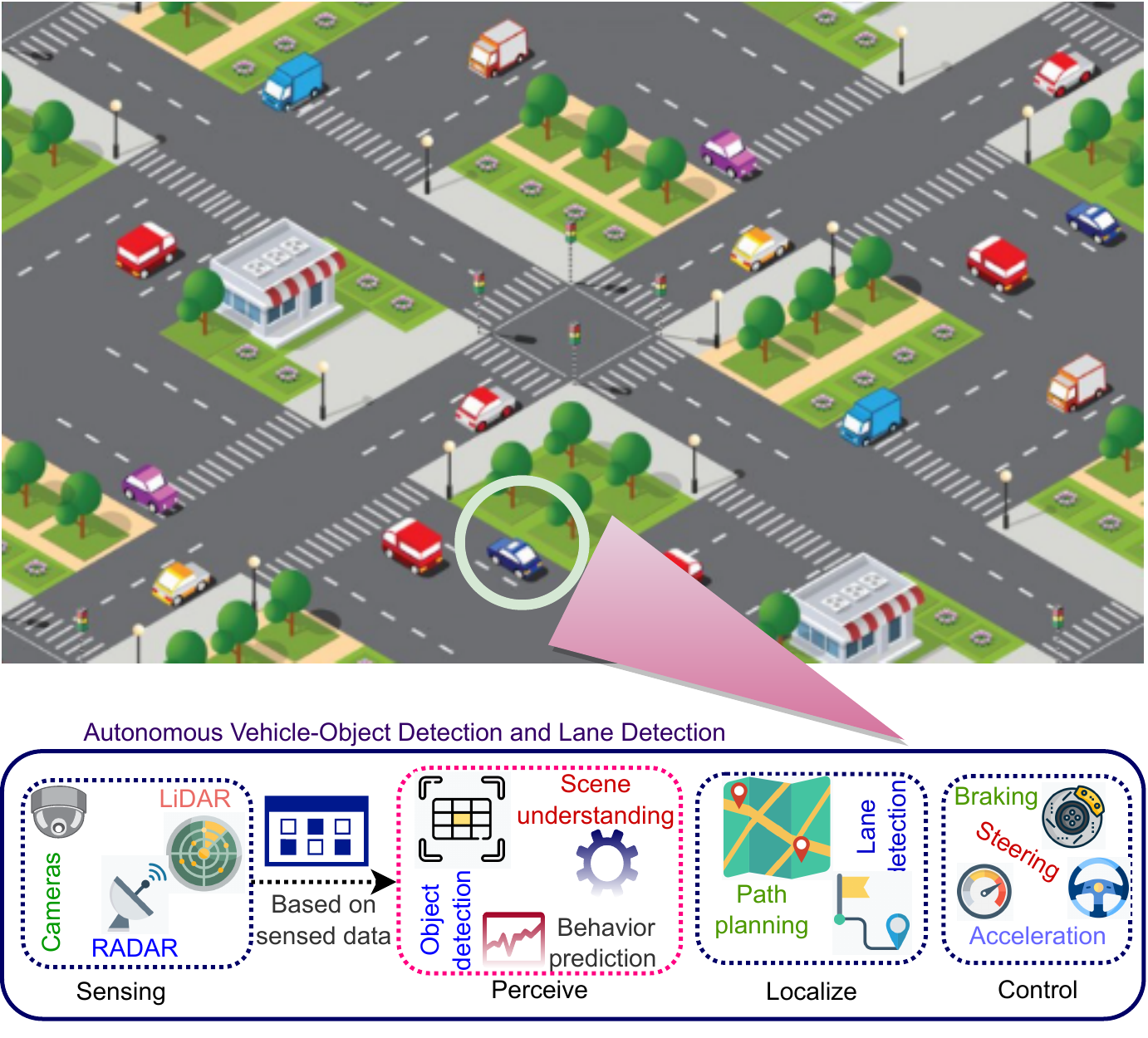}
\caption{Object Detection in Autonomous Vehicles.}
\label{Fig:object}
\end{figure}



Autonomous vehicles generate massive amounts of data, such as video streaming, sensor data, object detection, and lane condition. Providing cloud services to autonomous vehicles is considered as a major challenge in terms of latency and security. In \cite{coronado2019enabling}, the authors of proposed a software-defined networking architecture at the network edge for 5G-enabled vehicular networks in order to overcome the latency problem. Later, the study presented a computer vision application to perform object detection and lane detection for autonomous vehicles. Fig.~\ref{Fig:object} depicts object detection in autonomous vehicles.  The lane detection algorithm performs operations such as vehicle position and path detection with high accuracy.  Gaussian filter techniques are used to reduce granularity in order to estimate the correct lane with greater accuracy. The object detection algorithm detects the object; however, the authors used a limited set of traffic signals and objects in this work. The authors used Haar feature-based cascade classifiers for object classification and detection. The experimental results show that the proposed 5G-enabled vehicular network has a round trip time of 49.2 ms to transfer a message from the vehicle to the server. However, based on our observations, the authors did not demonstrate an accuracy rate in detecting the object and lane. Localization is one of the most important aspects of autonomous vehicles for avoiding collisions and ensuring safe navigation. Although GPS provides an accurate position, its accuracy is limited to 10 metres higher or lower. The precise position of autonomous vehicles must be no more than 5 metres. The use of 5G communication technology for cooperative localization (CL) provides an accurate vehicle position.  The work in \cite{piperigkos20205g} introduces a CL approach for multi-modal fusion between autonomous vehicles. The proposed model's main contribution was the introduction of a Laplacian Graph Processing framework, in which all vehicles act as vertices in a graph and communication paths act as edges. Using the Laplacian Graph Processing framework, the experimental results show that the proposed model has a faster response time and a higher GPS accuracy rate. The Intelligent Transportation System (ITS) is a combination of manual and self-driving vehicles. The ITS faces some challenges in autonomous vehicles, particularly in object detection and accurate path prediction. These obstacles raise safety and traffic concerns. Furthermore, autonomous vehicles share information with signals and act adaptively based on the situation, whereas humans in manual vehicles act more appropriately based on the situation. To address the issues associated with the mixture of manual and autonomous vehicles, the authors in \cite{yu2020deep} proposed a deep learning model in a 5G enabled ITS.  During this process, a natural-driving dataset and a driving trajectory dataset are fed into long short term memory networks. The softmax function computes the probability matrix of each lane change intention. The final lane change intention probability is then assessed in the decision layer at an accuracy rate of 85\%.

To detect nearby circumstances, AVs rely primarily on radar sensors, as well as light detection and ranging sensors. However, the reliability of such high-end sensors is limited over longer distances or when a vehicle enters an area with low visibility. One alternative solution is to enhance data exchange between vehicles by using roadside equipment. However, there are some risks associated with data sharing, such as when a malicious vehicle intentionally sends fake data in order to manipulate the receivers or when faulty sensors communicate incorrect data. The AVs will be trapped if they trust the data provided by the source vehicle, causing them to switch lanes or accelerate faster. As a result, there may be significant risks to human life. Because the decision-making process of autonomous vehicles is highly dependent on shared data and sensors, it is essential that the vehicle detect and filter out incorrect information. Motivated by such challenges, \cite{nguyen2020enhancing} introduced a novel approach to support a host vehicle in verifying the motion behavior of a target vehicle and then the truthfulness of sharing data in cooperative vehicular communications. Initially, at the host vehicle, the detection system recreates the motion behavior of the target vehicle by extracting the positioning information from the V2V received messages. Furthermore, the next states of that vehicle are predicted based on the unscented Kalman filter. Unlike prior studies, the checkpoints of the predicted trajectory in the update stage are periodically corrected with a new reliable measurement source, namely 5G V2V multi-array beamforming localization. If there is any inconsistency between the estimated position and the corresponding reported one from V2V, the target vehicle will be classified as an abnormal one.
\subsubsection*{Summary}
Autonomous vehicles generate massive amounts of data 5G-enabled vehicular networks helps to overcome the latency problem. The lane detection algorithm performs operations such as vehicle position and path detection with high accuracy.  The use of 5G communication technology for localization provides an accurate vehicle position with optimal latency and reduces the communication overhead. Some of the risks associated with data sharing, as the malicious vehicle intentionally sends fake data to manipulate the receivers. The AVs will be trapped if they trust the data provided by the source vehicle, causing them to switch lanes or accelerate faster. As a result, there may be significant risks to human life. Because the decision-making process of autonomous vehicles is highly dependent on shared data and sensors, it is critical that the vehicle detect and filter out incorrect information as quickly as possible.
\subsection{URLLC}
\subsubsection{Introduction}
URLLC is specifically designed to handle the stringent requirements on reliability and latency of critical packet transmission for autonomous driving in connected vehicles. In one-way, highway vehicular network performance is enhanced by using the joint resource allocation of the eMBB and URLLC traffics.In AV, vehicles are interconnected with each other to transfer information. To achieve these transfer of information among vehicles and infrastructure present on the road, these connections need to be dynamic with ultra-low latency and ultra high reliability.
\subsubsection{Existing Challenges/Limitations}
Ultra Reliable Low Latency Communications (URLLC) provides various services to the applications like Autonomous vehicles and Industrial IoT. It requires strict latency with 99 percentage of reliability. It is used in the applications that requires low transmission latency of 1 ms and ultra-high reliability of 1-10-5 success probability \cite{mukherjee2018energy}.

URLLC in AV needs a target latency of 1 millisecond and end-to-end security with 99 percent of reliability. Thus type of communications will more useful for autonomous driving to share and receive information both with neighboring vehicles and road infrastructure. In fully automated with no human intervention system, based on the information received the vehicles has to perform these tasks like automated overtaking of vehicles, collision avoidance, smart decisions, prioritizing the task like giving more importance to ambulance vehicles. All these task requires the high level of reliability and low latency which is provided only by URLLC. However, onboard processing or cloud computing is not sufficient to store the massive amount of data generated from high resolution sensors and cameras. Inorder to provide better safety and reliability than the human driving, processing real time traffic must be within a latency of 100ms.

Storage resources in onboard processing are very limited. For example, GPU required for low latency have high power consumption, further it requires cooling to satisfy the thermal constraints which would significantly degrades the fuel efficiency of the vehicle. Also, SSD storage device is also not a good option as it will be filled within hours to store the sensor and device data. Onboard processing can be convenient for the communication between passenger and vehicle but not for the V2V and V2I.  Cloud computing is also not sufficient to provide low latencies as in Internet of Vehicles there will be communication delay in between the server and the client.

\subsubsection{How B5G help (with Related work)}
\cite{popovski2018wireless} paper discusses about the various building blocks of URLLC in wireless communication systems for supporting in framing, access topology and use of diversity.

The authors in \cite{alsenwi2019chance}have addressed the problem of resource scheduling in URLLC using puncturing approach. The time domain is divided into equally spaced time slot with duration of 1 ms and each time slot is further divided into minislots in order to achieve the latency requirements of URLLC. The URLLC traffic received in each time slot is placed immediately to transmit in the next mini slot using the puncturing method in eMBB transmissions. The main objective of the user is to increase the eMBB users while maintaining the URLLC constraints. The Cumulative Distribution Function (CDF) is used to change the resource allocation problem into an optimization problem with a deterministic constraint.

In \cite{song2019performance}, to satisfy the strict latency constraints each time slot is divided into minislots and currently received URLLC transmission is promptly allocated in the next immediate minislot by puncturing the on-going eMBB traffic. The reliability of URLLC is ensured by deploying guard zone around the vehicle receivers and even eMBB transmission is also prohibited inside the guard zone. Association probability of the vehicle receivers for URLLC is captured and finally the coverage of vehicle-to-vehicle links and Vehicle-to- Infrastructure are analysed. 

Some of the solutions to overcome the challenges faced using URLLC in AV are discussed here. Storage and computing resources can be deployed at the wireless network edge which includes edge caching, computing and AI using Mobile Edge Computing. MEC combined with Baseband Unit (BBU) servers installed in radio access points along the roadside or base stations. Next, reliability and redundancy must be achieved in all the levels such as application, transmission, software and networking. A cloud-native BBU server is used to virtualize these levels in wireless radio network that ensures both reliability and redundancy and makes a robust system. Finally, Network slicing can provide different slice to satisfy different requirement, functions and configuration. One slice can provide low security and low reliability for mMTC services whereas another slice can provide high security and reliability like for URLLC. 

\subsubsection*{Summary}
uRLLC is a key technology to provide communication between AV and 5G network. Real-time decision has to be taken in fraction of seconds to have safe and accident-free drive. This is achieved by low latency feature of 5G network that enables vehicles to receive information at turbo speed. So, faster communication, real-time connectivity and low latency are essential factors for increasing growth in 5G enabled AV

\subsection{mMTC/ V2V communication/ V2X communication}
\subsubsection{Introduction}

\begin{figure}[t]
    \centering
    \includegraphics[width=0.5\textwidth]{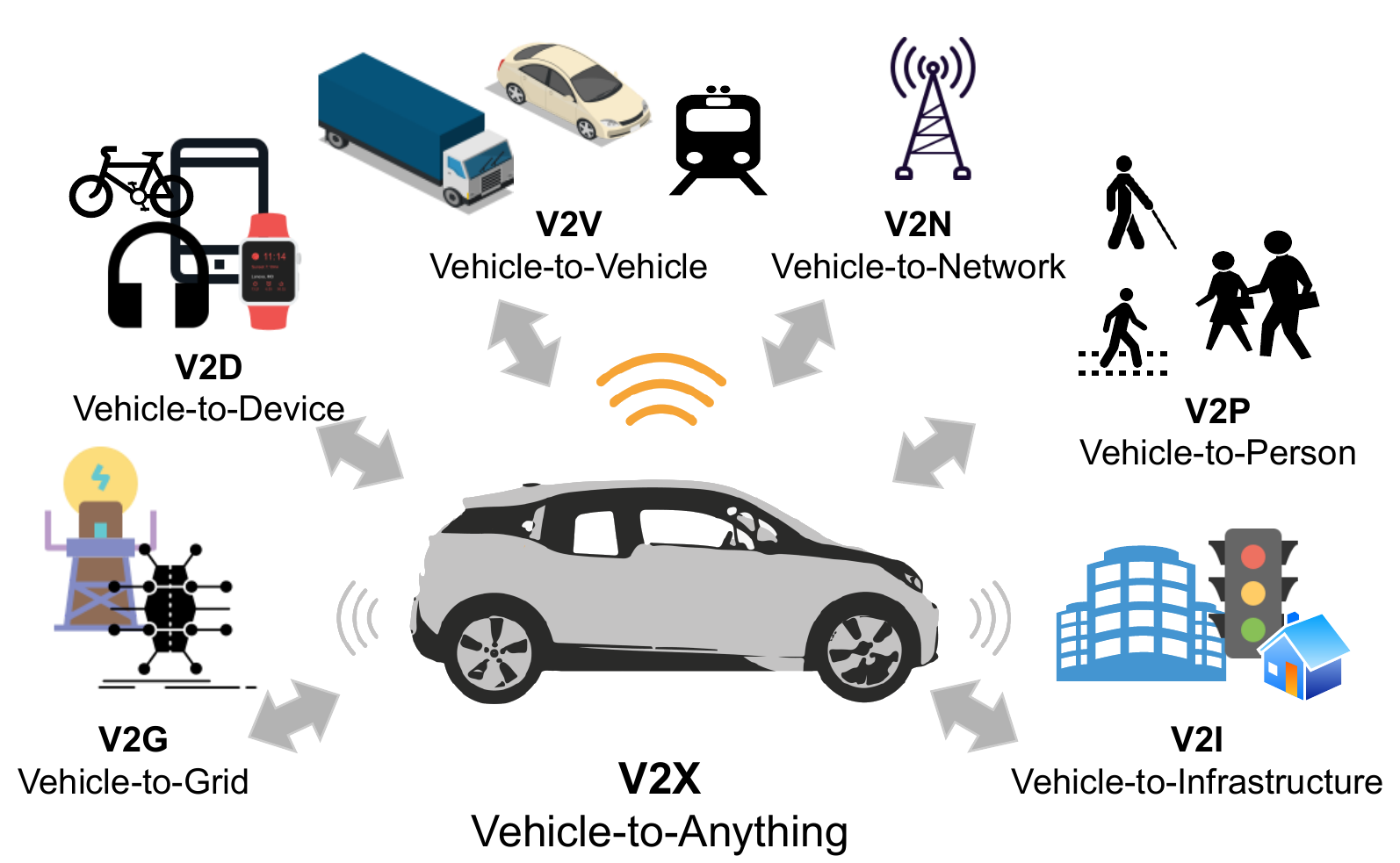}
    \caption{V2X: Vehicle-to-Anything communication}
     \label{Fig:Vehicle to Anything}
\end{figure}

Machine type communications (MTC) play a key role in 5G systems. MTC can be classified into massive Machine Type Communication (mMTC) and ultra-reliable Machine Type Communication (uMTC). mMTC provides connectivity to a large number of devices in which traffic profile is typically a minimum amount of data. It achieves minimum latency and high throughput but the main concern is about the optimal power utilisation of these connected devices. mMTC provides wireless connectivity to billions of devices with low complexity  and low-power type of devices. Some of the use cases are automated industries, remote surgeries and smart metering. On the other hand uMTC is about connecting adequate wireless link for network services which are widely used in V2X and industrial control applications.  

The main features of mMTC are as follows. Small packets are transmitted from devices with bytes and it connects large number of devices per cell. Sporadic user activity, uplink transmission and low user data rates of around 10 kb/s per user. Finally, optimal power usage and long battery life is also achieved by using mMTC. 

\subsubsection{Existing Challenges/Limitations}

The existing communication technologies \cite{bockelmann2016massive} focus only on some specific applications. They are not able to meet the latency and reliability requirements of applications such as connected cars, automated vehicles and industrial automation. The traditional technologies evolved from Long-Term Evolution (LTE) cannot be applied to IoT devices. So, these technologies are not able to optimise or handle IoT specific features like sporadic transmission, optimization of power and uplink centric transmission and so on. These factors motivated to shift from existing technologies to mMTC. 

The application of 5G technology in Various communication like V2V, V2I and V2X communication are inevitable as shown in Fig. \ref{Fig:Vehicle to Anything}. V2V communication generates the network by connecting different vehicles using mesh topology\cite{d2013its}.  Based upon the number of hops, communication can be classified as single-hop (SIVC) which is used for short range application and multi-hop vehicular communication (MIVC) \cite{zheng2020cooperative} used in traffic monitoring. Vehicle interacting with infrastructure helps to detect traffic lights lane markers and parking meters \cite{kaiwartya2016internet}. V2X communication interact with all the entities such as roadside, grid, devices and pedestrians\cite{lin2015potential}. This type of communication prevents road accidents and to alert the passengers before accident \cite{lee2017latency}

\subsubsection{How B5G help (with Related work)}

In \cite{zhu2018energy},authors have proposed the Hybrid Hovering Position Selection (HHPS) algorithm to determine the hovering positions of Unmanned Aerial Vehicles(UAV) which minimizes the power consumptions of machine type communication devices. Inorder to optimize the latency of communication devices, cuckoo search algorithm was proposed for trajectory planning. Furthermore, energy consumption and throughput of UAV is optimized. Based on the priorities given, efficiency of data and computing services are also optimized.  Bockelmann et al. \cite{bockelmann2016massive} have designed a FP7 METIS project using 5G system concept in which well-suitable chase for high data rates are summarized by the term ''Extreme Mobile Broadband'' (xMBB). The overhead seen in exchange of messages required before the transmission of data payload affects the energy efficiency of MTC devices. Hence, less frequent and shorter transmissions preserves energy. Therefore MAC protocols coupled with Physical layer approaches enable devices with long battery lives.

In \cite{dougan2018ofdm}, orthogonal frequency division multiplexing with index modulation is proposed to improve inter carrier interference caused by asynchronous transmission for uncoordinated mMTC networks. Data transmission is performed by the indices of active subcarriers. Subcarrier Mapping Scheme called inner subcarrier activation scheme is proposed to further improve the interference of adjacent user in asynchronous systems. In \cite{tanwar2019tactile}, proposed a novel 5G based network architecture called Non-Orthogonal Multiple Access (NOMA) to reduce the latency and reliability requirements in V2X communication. 

One of the critical challenge for next-generation wireless communication is to satisfy the high demand for mMTC systems which performs a random transmission between base station and machine users. Therefore, the lack of coordination between the base station causes inter-carrier interference. Researchers are facing a big challenge in providing the services for asynchronous massive machine users. Activation ration and optimization of subblock size can be evaluated for clustering users and conflicting users with respect to their requirements. Autonomous vehicle in reality faces various challenges like real-time data analytics, software heterogeneity, validation, verification and latency issues. Aforementioned challenges, researchers across the globe are trying to provide solutions using 5G-based testbeds.

\subsubsection*{Summary}

V2X and 5G connections helps AV to visualize the objects and obstacles around corners and beyond. Connectivity between the cars and infrastructure provides awareness ahead among the vehicles like reduce the speed automatically in slow-moving traffic areas. By the time when it reaches the signal, traffic would have been cleared thereby reducing the waiting time on the traffic signal. 'Traffic Light Information' is a classic case study based on V2I initiated by Audi in Europe. All these scenarios requires good connectivity at the speed of light. This can be achieved by providing 5G network that enables fast processing and prior decision making.To summarize, AV would become reality with the application of 5G technologies.

\subsection{eMBB} 
\subsubsection{Introduction}
eMBB service is used particularly to enhance the Quality of Experience(QoE) in bandwidth for in-vehicle applications.
\subsubsection{Existing Challenges/Limitations}
One of the significant way in 5G to deliver wireless broadband to previously unreached areas is through the technology Fixed Wireless Access (FWA) network. Globally, FWA has gained its importance in developed and developing countries and is expected to expand exponentially from 2018-2025. This FWA creates a platform for eMBB for wide coverage using higher-spectrum bands. Some of the use cases of eMBB are as follows:
Broadband everywhere:  FWA technology can provide wide coverage globally with minimum speed of 50 Mbps.
Public transportation:  Broadband access used in high-speed trains and public transport systems
Hot spots:  Enhancing broadband coverage in densely populated areas and in high rise buildings
Large-scale events: Enabling high speed of broadband data where thousands of people are gathered in one place for any kind of big event.
Smart offices: Delivering high-bandwidth of data connections to thousands of users even in the environment with heavy data traffic.
Enhanced Multimedia: Provides high quality video streaming and real-time content over wide coverage areas.

\subsubsection{How B5G help (with Related work)}
In \cite{alsenwi2019chance}, authors have addressed the problem of resource scheduling problem in eMBB traffics. Initially the resource blocks are assigned to the beginning of each time slot based on the channel state and the previous average data rate upto the current time slot of each eMBB users. Two dimensional Hopfield Neural Network and the energy function is used to solve the resource allocation problem. Then, chance constraint problem is applied to maximize the eMBB data rates.

The authors in \cite{abuin2020complexity} targets more number of users in data transmission in AV. Non-Orthogonal Multiple Access (NOMA) is proposed to optimize the distribution in unicast/multicast scenarios. The complexity of the algorithm is measured and compared using Time Division Multiplexing (TDM). In order to reduce the complexity of the algorithm, two solutions are proposed. First, the numbers of injection levels are reduced and second one is choosing the smart algorithm that selects the optimum injection levels.

In \cite{song2019performance}, authors have considered eMBB and URLLC to be the important prerequisites of smart intelligent transportation systems. The eMBB traffic is scheduled at the boundary of each time slot for data transmissions. During the transmission interval, random arrivals of URLLC traffics are allowed.  
\subsubsection*{Summary}
URLLC and mMTC  works together with eMBB to fulfill the needs in new wireless networks to provide the facilities in applications like healthcare, manufacturing, military and emergency response.  These three features in 5G provides the solutions for the issues with respect to bandwidth, density and latency  in applications with restricted  LTE's capabilities like autonomous vehicles, smart city and augmented reality.  

\begin{table*}[h!]
\centering
\caption{Benefits and challenges of Technical Aspects of Autonomous Vehicles.}
\label{tab:Tech aspects of AV}
\resizebox{\textwidth}{!}{%
\begin{tabular}{|p{2cm}|p{5.5cm}|p{8.5cm}|p{6.5cm}|}
\hline
\multicolumn{1}{|c|}{\textbf{Application}} &
  \multicolumn{1}{c|}{\textbf{Existing Challenges}} &
  \multicolumn{1}{c|}{\textbf{Solutions}} &
   \multicolumn{1}{c|}{\textbf{Limitations}}\\ \hline
Navigation and Path Planning 
\cite{irani2018localizability, jiang2019probabilistic, jardine2018robust, lin2021autonomous, yan2021reinforcement, chu2015real, marin2018global, hoang2015path, alharbi2020global}
&
  1)  Autonomous planning and navigation  easily plunge into the local optimum in the complex scenarios and thus fall into traps, resulting in a low probability of finding a reasonable route to the target   
  
  2)  conventional algorithms cannot plan the environment completely in advance
  
  3)  path planning for unstructured roads are very complex. 
     
  &
  1) Deep learning and Deep reinforcement learning is used to achieve end-to-end learning for complex tasks
    
  2)  Gradually transformation of low-level feature representations into high-level feature representations through a multilayer process is implemented

  3)  Free-form navigation like graph based search algorithm can be used to plan for navigation in unstructured roads
  
  4) V2X shares intent, accurate and fast information to perform higher level of prediction
  
  5) V2X wireless sensor visualize 360 degree and sense non-line sight and wide range of area
  
  6) Precise positioning of location and dynamic decision making are faster and accurate using 5G technologies enabled with AI
  
  &
  
  1) Detecting and localizing the object at specific time and adjusting the path dynamically has to be still optimal and faster
  
  2) Constructing the maps from the inputs received from the sensors and devices is a time consuming task
  
  3) Fails to provide accurate predictions in poor climatic conditions
  
  4) Unable to predict the agent behaviour on roads
  
     \\ \hline
Object Detection/ Collision Avoidance \cite{coronado2019enabling,  piperigkos20205g, yu2020deep, nguyen2020enhancing} &
  1) Massive volumes of data are generated by autonomous vehicles. In terms of latency and security, delivering cloud services to autonomous vehicles is seen as a significant challenge    
  
  2) Localization is one of the most major aspects of autonomous vehicles for avoiding collisions and ensuring safe navigation
  
  3) The reliability of high-end sensors is limited over longer distances or when a vehicle enters a low-visibility region
  
  &
  1) To address the latency issue in 5G-enabled vehicular networks, a software-defined networking architecture at the network edge is being constructed.
  
  2) Gaussian filter techniques are used to reduce granularity in order to estimate the correct lane with greater accuracy
  
  3) 5G communication technology for CL provides an accurate vehicle position
  
  4) Laplacian Graph Processing provides faster response time and a higher GPS accuracy rate
  
  5) Roadside equipment is used to improve data sharing between automobiles
  &
  1) The lane detection algorithm performs tasks such as vehicle location and path detection with a moderate accuracy rate
  
  2) AV share information with signals and act adaptively according to the situation, whereas humans in manual vehicles act more appropriately based on the scenario
  
  3) Certain risks are inherent in data sharing, such as when a malicious vehicle sends fake data in order to manipulate the receivers, or when faulty sensors communicate incorrect data
  
  4) If the AV trust the data provided by the malicious vehicle, they will be trapped, causing them to switch lanes or accelerate faster. As a result, there may be serious threats to human life
  \\ \hline
URLLC \cite{mukherjee2018energy, popovski2018wireless, alsenwi2019chance, song2019performance, bockelmann2016massive, d2013its, zheng2020cooperative, lin2015potential, lee2017latency, zhu2018energy, dougan2018ofdm, tanwar2019tactile}&
  1) Onboard processing is not sufficient to store massive amount of data generated from sensors and other devices.   
  
  2)  Connecting the vehicles based on cloud computing provides high latency which is not suitable AV 
  
  3)  Providing communication with low latency and high reliability is a complex task in dynamic real time process which is connected to different devices.
  &
  1)   Resource scheduling  problem is solved using puncturing approach  
  
  2)  To maintain the latency constraint, URLLC are considered inside the minislot of each eMBB time slot
  3) To ensure the reliability, guard zones are deployed around the vehicle.
  
  4) 5G can make instantaneous communication with URLLC
  
  5) The time delay using 5G is 1 to 5 millisecond when compared to 20 milliseconds in 4G
  
  6) Reduced time delay in AV provides users with safety information
  
  7) One slice in network slicing can provide high reliability and security for URLLC
  
  &
  1) Resource scheduling should be considered for both uplink and downlink scenario for URLLC
  
  2) Guard zone based URLLC scheduling policy is compared with the baseline policy where the guard zone receiver is not applied

  \\ \hline
mMTC/ V2V communication/ V2X communication \cite{bockelmann2016massive, d2013its, zheng2020cooperative, kaiwartya2016internet, lin2015potential, lee2017latency, zhu2018energy, bockelmann2016massive, dougan2018ofdm, tanwar2019tactile} &
  1)    A common framework for Machine type communication must be available
  
  2) Current link adaption mechanisms are not suitable for MTC 
  
  3)   Resource allocation and channel coding schemes are unsuitable for small packets
  
  &
  
  1)     optimization algorithms are used for trajectory planning which in turn minimizes the power consumption of MTC
  
  2)   Extreme mobile broadband is used for high data rates
  
  3) less frequent and shorter transmissions preserves energy. So MAC protocols with physical layer approaches ensure device with long battery life
  
  4) NOMA architecture is used to reduce the latency and reliability requirements in V2X
  
  &
  
  1) As the number of devices increases rapidly, lack of coordination between base station and the machine causes inter-carrier interference.
  
  2) Real-time data analytics is complex
  
  3) Handling software heterogeneity and latency issues are challenging in mMTC
  
  \\ \hline
eMBB \cite{alsenwi2019chance, abuin2020complexity, song2019performance} &
  1) Wide coverage with minimum speed is expected in AV for exchange of information    
  
  2) Broadband access in moving objects, densely populated areas must be available
  
  3) Should deliver high band width data even in heavy data traffic environment. 
  
  &
  
  1) Resource Blocks allocation problem are modeled using 2-Dimensional Hopfield Neural Network
  
  2) Optimization problem is designed  using the Generalized Proportional Fair(GPF)and then it is converted in the form of energy function to neural network
  
  3) Time Division Multiplexing is used to measure the complexity of algorithms
  
  4) One way vehicular network with location of the vehicles and road side unit is modeled using  poisson  point processes
  
  5) The eMBB traffic is buffered at the boundary of each time slot
  
  6) Using 5G, it is very fast and is able to support media data like 3D videos, augmented and virtual reality
  
  &
  1) Due to the hard latency rate of URLLC, it affects the data rate of eMBB traffic
  
  2) Even though the various algorithms reduces the complexity   by reducing the injection levels, the tradeoff between the complexity and performance should be maintained
  
  3) eMBB transmission is punctured due to the arrival of URLLC data transmission. This degrades the flexibility of eMBB users

  \\ \hline

\end{tabular}%
}
\end{table*}

\section{Impact of 5G and B5G technologies on AV} 

In this section we discuss the impact of some of the prominent technologies in 5G and B5G on AV.

\subsection{Multi-access Edge Computing (MEC)}

\begin{figure}[htb]
    \centering
    \includegraphics[width=0.5\textwidth]{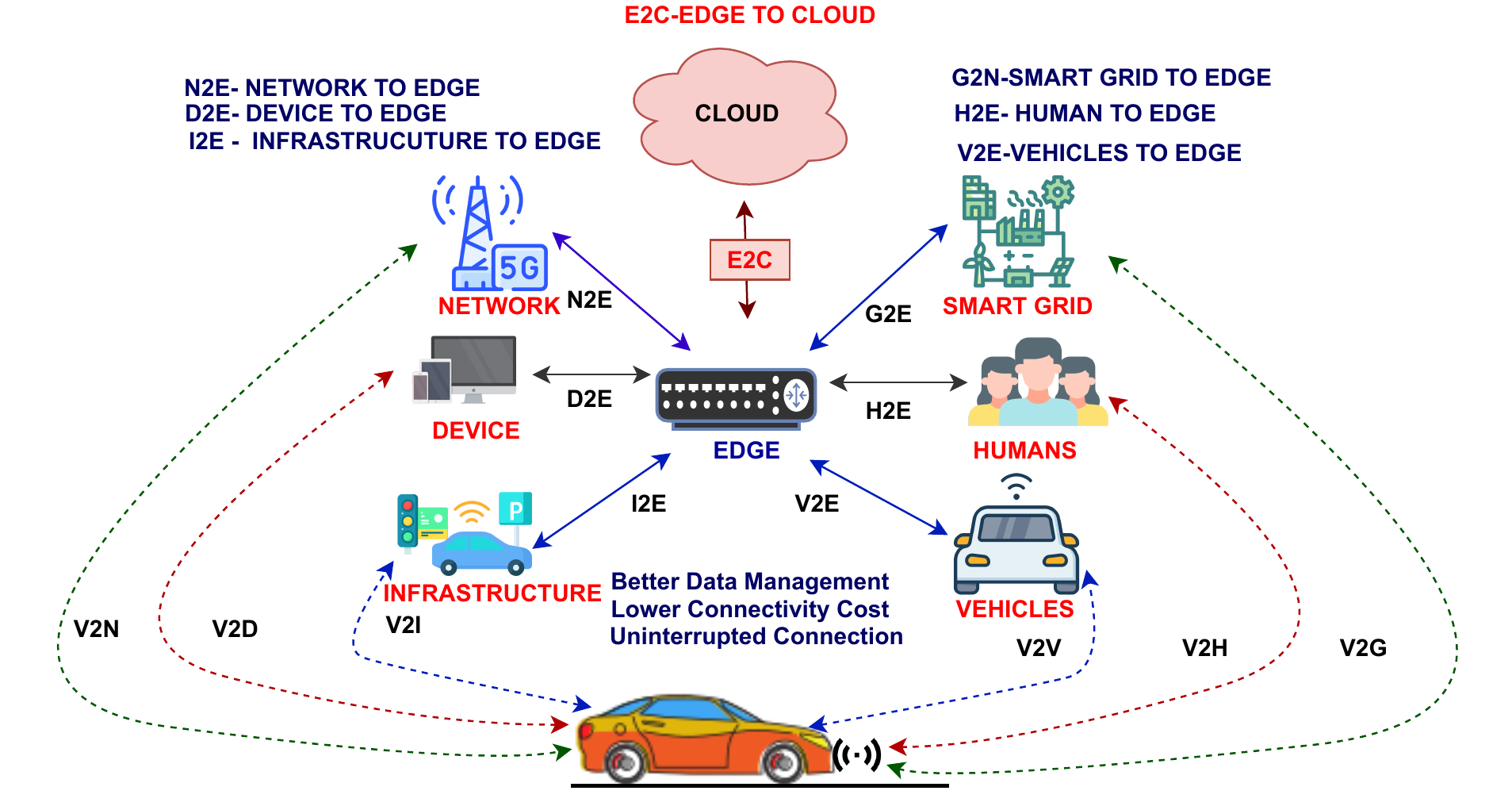}
    \caption{MEC enabled 5G and Beyond for AV}
    \label{fig:mec}
   \end{figure}
MEC extends the capabilities of cloud computing and Information Technology environment by bringing them closer to the network or to the edge of the network, i.e. to the end users. By bringing in these capabilities to the network edge, the congestion, latency are reduced and also the applications run faster. Real-time analytics for applications such as traffic analysis, big data analytics in social media, smart cities, etc. can be realized by  MEC \cite{prabadevi2021toward,li2022computation}. MEC is a key enabling technology in 5G and B5G technology as MEC enabled B5G networks offer high-bandwidth, low latency and real-time access to the resources of the radio network. MEC can help the network operators to support wide range of innovative services and also integration of IoT based applications in B5G \cite{liyanage2021driving}.  

Enormous amounts of data will be generated from AV. This data can be used for recognition of several external and road features such as, pedestrians, traffic signs, lanes, road condition, etc. When multiple AVs cooperate for tasks such as collision avoidance and lane merging, data processing from the vehicles at global perspective is required. Enormous computing power is required to process several computer vision operations that are essential for AD like extraction  of features from images to perform the analytics in real-time. Hence, MEC enabled 5G services plays a major role in fast communication, fast processing of complex data in real-time analytics which allows the offloading latency-sensitive and computing-intensive tasks to the edge devices as depicted in Fig. \ref{fig:mec} \cite{coronado2020enabling,coronado2019enabling,zhdanenko2019demonstration}. Rest of this section discusses some of the recent literature on MEC enabled B5G for AV.  

\subsubsection{Related work)}
Zhou~et al. \cite{zhou2018mec} proposed a novel MEC enabled 5G architecture for vehicular networks. They discuss how the proposed architecture can accommodate V2I and V2V communication with guaranteed low packet delay and high scalability. The proposed architecture addressed the issue of mobility management on MEC for AVs. The IP-handoff procedure is made transparent and seamless by the application of distributed mobility management. The authors used NS3 tool to conduct proof of concept simulation. Fabio~et al. \cite{giust2018multi} discussed the importance of MEC in 5G-enabled connected cars. They have provided several technical aspects and practical use cases on MEC-enabled 5G for V2X. In a similar work, Claudia Campolo~et al.  \cite{campolo2019mec} proposed an architecture MEC enabled 5G for V2X application. The authors mainly focus on service migration for V2X applications to reduce the duration of non-availability of services provided by the virtualized migrating instance (service downtime). Sokratis Barmpounakis~ et al. \cite{barmpounakis2020collision} proposed a an architecture for 5G systems that uses MEC and NFV to select the computing resources for V2X applications. A novel algorithm, VRU-Safe, is proposed by the authors that operates on  top of the proposed architecture to identify and predict road hazards like vehicle collisions. Coronado~et al. \cite{coronado2019addressing} proposed a novel MEC and NS enabled 5G for ACV. The proposed architecture enables features such as object detection and lane tracking to be offloaded to the MEC without affecting the effectiveness of 5G. The authors have developed an application named as road data processing to analyze the videos streams from the AVs. An ML model is used for lane tracking, detecting traffic and road signs. The MEC server will instruct the AV with an appropriate command to take required action. 

Lee~et al. \cite{lee2020multiaccess} have proposed a simulation as a service (SIMaaS) to offload the simulation by making use of cloud infrastructure which is based on computational offloading for the MEC enabled 5G platform for AVs. As an use case, the authors conducted Monte-Carlo simulations using SIMaaS for optimal tollgate selection in a highway for AVs. Shie~et al. \cite{shi20185g} proposed an algorithm based on distributed heuristics for 5G-V2X to solve the problem of motion planning of Avs that travel in industrial parks. The proposed algorithm focuses on avoiding the collision of AVs at intersections, ensuring safety of the pedestrians. To achieve these goals, mutual cooperation and seamless communication among the AVs are desired. To ensure highly reliable and fast communications, the authors employed 5G uRLLC slice and for mutual cooperation of the AVs, a distributed heuristic algorithm is employed. The proposed solution will pass on the information related to the pedestrians and the vehicles passing through the intersections that are downloaded downloaded from the MEC to the vehicle that is closest to the intersection. Rasheed ~et al. \cite{rasheed2020application} proposed a hierarchical offloading scheme which is application-aware for MEC-5G enabled AV architecture. In the proposed architecture, the network is divided into three layers based on the requirements of the applications that result in efficient network and quick response that meets the requirements of the application. To handle each application at appropriate layer to cater the individual computation and latency requirements, every application is treated independently based on their complexity, computation requirements and data rate for deciding the computation layer. Huisheng Ma~et al. \cite{ma2020cooperative} proposed a prototype for cooperative AD oriented MEC enabled 5G-V2X. The proposed prototype is based on the experimental platform for the next generation RAN, a cooperative AD vehicle platoon, and to provide high-definition 3D map service dynamically, MEC server is used. Lian~et al. \cite{lian2020semantic} proposed a scalable MEC-enabled 5G infrastructure for unmanned vehicular systems. The proposed system provides high-precision maps through the offloading capacity of MEC-enabled 5G, through which the unmanned vehicles can sense the environment that enables AD. Sabella~et al. \cite{sabella2020mec} have proposed and implemented an infotainment service based on MEC for AVs in smart road environment in 5G network. MEC has enabled the reduction in improving the usage of the  backbone bandwidth.  It also reduced the latency by accelerating the contents which are frequently accessed. The also the speed of content transfer between the AVs has been significantly increased due to 5G network. 

\subsubsection*{Summary}
Many researchers have used the MEC enabled 5G to successfully solve many issues such as latency, computational offloading successfully in several applications such as object detection, infotainment services, lane tracking etc. However, there are several issues like standardization of protocols, heterogeneity in computing and communication, privacy and security, that have to be addressed to fully realize the potential of MEC enabled 5G in AV.

\subsection{Network Slicing} 
\subsubsection{Introduction}
\begin{figure}[t]
    \centering
    \includegraphics[width=0.5\textwidth]{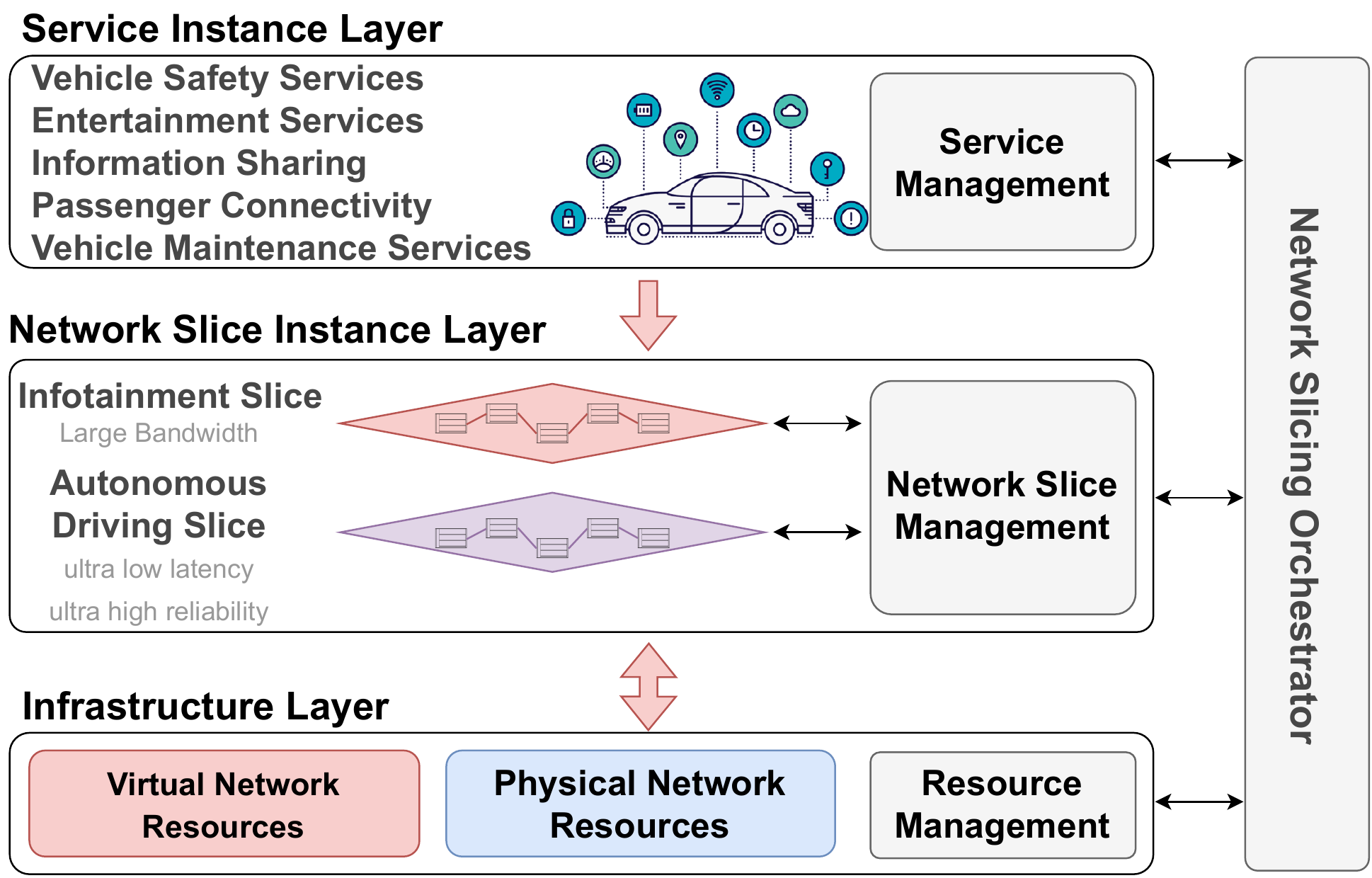}
    \caption{Network slicing for ACVs}
    \label{Fig:nsacv}
\end{figure}

Network slicing is considered to be an important enabler of softwarized 5G and beyond communication networks~\cite{3gpp2017study}. Network slicing enables multiple end-to-end logical networks on the same physical and virtual network resources~\cite{alliance2016Description}. This supports diverse communication requirements of emerging applications, such as, ACVs, which cannot be satisfied by a generic ''one-fits-all'' type of network architecture in pre-5G communication networks~\cite{wijethilaka2021survey}.  The network slicing architecture consists of three functional layers. They are service instance layer, network slice instance layer, and resource layer, as presented in Fig.~\ref{Fig:nsacv} ~\cite{alliance2016Description}. The functionality of these three layers are coordinated by the network slice orchestrator~\cite{huang20205g}. Each network slice consists of network functions that are customized to satisfy the communication requirements of different verticals, applications and use-cases. For instance, in case of ACVs, a network slice with large bandwidth can be allocated for vehicle infotainment while a network slice with ultra low latency and high reliability can be assigned for autonomous driving, as depicted in Fig.~\ref{Fig:nsacv}.

\subsubsection{Related Work}

Many research work consideres network slicing as an enabler of ACVs. For instance, Campolo~\emph{et. al.}~\cite{campolo20175g} presented a vision for 5G network slicing to facilitate V2X services by partitioning the resources in vehicular devices, radio access network, and the core network. This work emphasizes the importance of network slicing towards guaranteeing the performance of V2X services through dedicated V2X slices. In addition, a model for network slicing based vehicular networks is presented in ~\cite{khan2021network} where dedicated network slices are allocated for autonomous driving and infotainment. The autonomous driving slice is concerned on communicating safety messages while the infotainment slice is utilized for video streaming. Furthermore, vehicles are assigned to different clusters and slice leaders are allocated for each cluster to facilitate V2V communication and safety information with high reliability and low latency. In similar vein, a network slicing based communication model for diverse traffic requirements in vehicular networks is proposed in~\cite{khan2018application}. This work also proposes two slices for autonomous driving and infotainment. The quality of infotainment slices are improved through V2X links to improve the packet reception rate. Moreover,~\cite{sanchez2019empowering} proposes a framework for 5G network slicing to control bandwidth with full-flow-isolation in network slicing for vehicular systems. This method incorporates heterogeneous radio access technologies, such as cellular and IoT communication (e.g. LoRaWAN) to generate end-to-end network slices. The proposed method also relies on distributed MEC and cloud computing to facilitate the dynamic deployment of virtual network functions required for network slices. In addition, a MEC-based online approach towards performing tasks such as power allocation, coverage selection and network slice selection of vehicular networks using deep reinforcement learning is proposed in ~\cite{mlika2021network}. This algorithm is also proved to be resilient against the high mobility in vehiclular networks. Furthermore,~\cite{xiong2019smart} proposes a method to tailor network slicing for vehicular fog-RANs. A smart slice scheduling mechanism formed as a Markov decision process is considered for dynamic resource allocation for network slices while overcoming the limitations caused by the dynamic nature of vehicular fog resources and vehicular movement.

\subsubsection*{Summary}
Network slicing is yet to be matured as a technology to realize fully functional ACVs in the future. The dynamic communication requirements of ACVs ranging from time critical communication for safety and high volume data requirements for vehicle infotainment can be satisfied through network slicing enabled 5G networks. However, the dyanmic nature of vehicular movement and vehicular resources make it challenging to deploy network slicing for vehicular networks. Several research work have attempted to address this problem through smart resource allocation algorithms. Furthermore, utilizing generic slices in various combinations to satisfy the demands of vehicular networks can also be explored. These methods should also evolve from simulations to fully-fledged and functional technologies. In addition, the security and privacy aspects of network slicing based vehicular networks should also be considered in order to ensure passenger and road safety, data security and privacy of ACV users.

\subsection{Software-Defined Networking/Network Functions Virtualization}
\subsubsection{Introduction}
\begin{figure*}[t]
\centering
\includegraphics[width=1\linewidth]{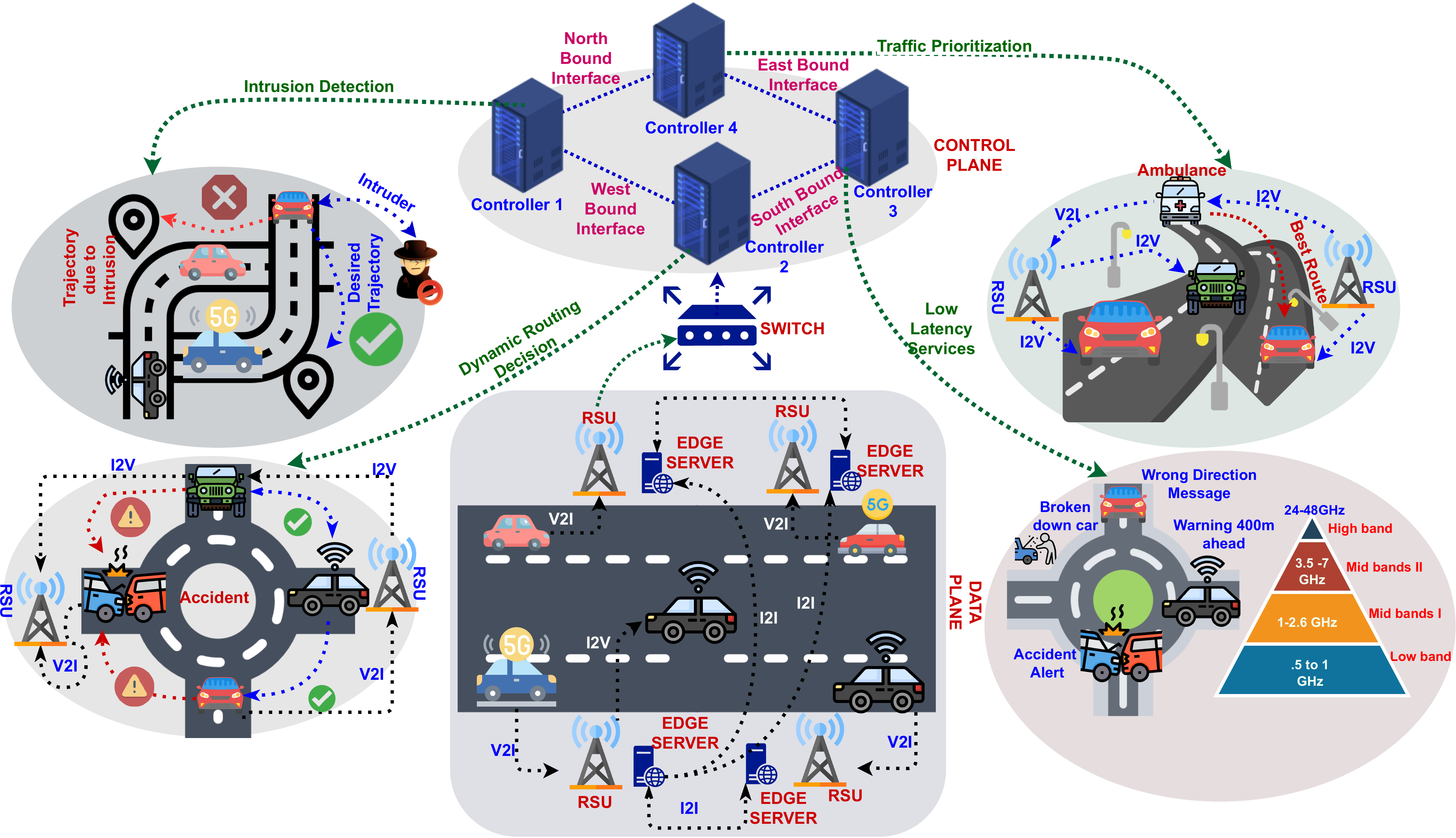}
\caption{Impact of 5G enabled SDN on ACVs.}
\label{Fig:sdnacv}
\end{figure*}

Recent advancements in telecommunication and automotive industries have paved a way for innovative communication and intelligent sensing. This has further led to the innovation of next generation ACVs. Nowadays the autonomous vehicles try to communicate with other vehicles and pedestrians via wireless for sharing the critical information for collision mitigation. Also the vehicles try to coordinate with the TIS entities using the various network infrastructures for gaining extra awareness about the road hazards and guidance on speed while travelling. This obviously provides an efficient traffic flow. To achieve these discussed goals, a secure as well as low latency communication is the need of the hour. The major challenge faced while deploying are the heterogeneity in radio access technologies, deployment inflexibility, network fragmentation issues and inefficient utilization of resources. These bottlenecks need to be handled efficiently to realize a best vehicular networking architecture for ACVs. This subsection discusses about the recent advancements in 5G enabled SDN vehicular networking architecture which ensures a very high agile infrastructure for rapid and secure communication among ACVs as illustrated in the Fig. \ref{Fig:sdnacv}.

\subsubsection{Related Work}
In \cite{mahmood2019software}, Mahmood~\emph{et. al.} proposed an architecture named SDHVNet for handling the heterogeneous vehicular network using SDN. The architecture primarily consists of two planes namely infrastructure plane and control plane. The infrastructure plane handles all the vehicles, vehicle users, road users like pedestrians, infrastructure on the roads like access points, traffic signals, etc. The control plane is facilitated by the various application programming interfaces like OpenFlow. This plane is responsible for all sorts of network related functionalities like virtualization of vehicular environment, gathering status of SDN switches, dynamic topology and frequency of service request response. The control plane includes a cache manager and handover decision manager that is responsible for providing a seamless interconnection among the ACVs and satisfy the safety as well as critical requirements. They also preserve the resource utilization by preventing handover failures. In \cite{goudarzi2020dynamic}, Goudarzi~\emph{et. al.} has formulated a three-tier framework consisting of cloud layer, edge computation layer and device layer for calculating the capacity of elastic processing and dynamic route for the purpose of monitoring the ACVs in their respective environment. They have also proposed a resource allocation policy combining SDN and edge computation based on reinforcement learning (RL). The proposed methodology improves the performance of the network layer. SDN is specifically used for managing the connectivity as a whole and to dynamically handle the computation in the edge devices. In \cite{garg2019sdn}, Garg \emph{et. al.} utilized SDN and formulated a framework for providing end-to-end security as well as privacy in 5G enabled ACVs. The major contribution of this framework is that it acts as an optimized network management layer. The intrusions in case of high speed ACVs are also identified potentially by this framework. The intrusion detection module is designed using a multi-objective evolutionary algorithm and a dimensionality reduction scheme. This integrated scheme improves the performance of the intrusion detection module. Also the framework consists of a module which authenticates the environment before the commencement of communication. SDN is utilized for reducing the latency in a high density traffic \cite{duan2017sdn}. The ACVs are grouped into a cluster based on the real-time conditions of the road. The clustering scheme follows a double head concept for each cluster. This enhances the quality of the trunk link thereby increasing the throughput. Recently, Garg \emph{et. al.} utilized SDN and edge computing for solving the primary challenges of ITS like high level of mobility, minimum latency, high quality of service and other real time services \cite{garg2019mobqos}. The author has proposed a framework named SDN-DMM based on EC, SDN and distributed mobility management (DMM). DMM helps in solving the routing issues and hence improves the quality of service in ACVs. The framework proposed proved to have an improvement in end-to-end delay of around 15.9 percent. This improvement in performance was achieved by utilization of multi-objective evolutionary algorithm while communication between the edge and the cloud. This enhanced the overall communication latency and also reduced overheads by better utilization of bandwidths.                 
\subsubsection*{Summary}
The rapid advancements in autonomous driving would result in adoption of the ACVs in near future. This would surely challenge the current ITS and its existing solutions. The major focus for facing these challenges should be on providing support for mobility, reducing the latency, increasing connectivity and providing higher intelligence. 5G enabled SDN could be utilized for handling the communication overheads, reducing the latency, reducing the overall throughput and also reducing the delay hence improving the mobility. 5G enabled SDN architectures are recently designed by researchers for handling the security concerns like intrusion detection. Though researchers have focused on utilizing SDN for finding few solutions for the above mentioned challenges, there are still significant open issues like analysis of energy utilization with respect to communication and data, real-time verification and validation, quality of everything (QoE), reduction in overheads and end-to-end delay.
\subsection{5GNR/Physical layer stuff}

To support the potential applications of future intelligent and autonomous vehicles in the coming B5G era, more demanding requirements have to be supported. Transmission of extremely large quantity of the data is one of the major requirements in vehicular communications in B5G\cite{ahangar2021survey}. To meet the growing traffic demands, B5G is supposed to provide a transmission rate up to 1 terabit/second. There exist many technologies that can improve the efficiency of spectrum for B5G. One of the most used technologies among them is massive multiple-input-multiple-output (MIMO), that can have more than 100 antennas. To increase the bandwidth of the spectrum, the millimetre wave (MMWAVE) is considered in B5G \cite{cheng2020vehicular,duan2020emerging}.

\begin{table}[h!]
  \centering
        \caption{Suitability of various vehicular communication technologies for vehicular communication use-cases~\cite{boban2017use}}
        \label{tab:5GNR}
  \begin{tabular}{|p{4.5cm}|p{0.5cm}|p{0.5cm}|p{0.5cm}|p{0.5cm}|}
  
  \rowcolor{gray!25}
 
  \hline
   \cellcolor{gray!15}Use-Case &{\rotatebox[origin=c]{90}{\cellcolor{gray!15}~LTE-V2X ~}}&{\rotatebox[origin=c]{90}{\cellcolor{gray!15}~DSRC 802.11p~}}&{\rotatebox[origin=c]{90}{\cellcolor{gray!15}~Vehicular VLC~}}&{\rotatebox[origin=c]{90}{\cellcolor{gray!15}~5GNR C-V2X~}}   \\
\hline   
\hline
   Remote Driving & \cellcolor{yellow!15}M & \cellcolor{red!15}L & \cellcolor{red!15}L & \cellcolor{green!15}H \\
   
\hline
\hline
   Pedestrian Communication & \cellcolor{yellow!15}M & \cellcolor{yellow!15}M & \cellcolor{red!15}L & \cellcolor{green!15}H \\
  \hline
 \hline
   Traffic Management & \cellcolor{green!15}H & \cellcolor{yellow!15}M & \cellcolor{red!15}L & \cellcolor{green!15}H \\
  \hline
 \hline
   Cooperative Sensing & \cellcolor{yellow!15}M & \cellcolor{yellow!15}M & \cellcolor{yellow!15}M & \cellcolor{green!15}H \\
  \hline
  \hline
   Cooperative Awareness & \cellcolor{green!15}H & \cellcolor{green!15}H & \cellcolor{red!15}L & \cellcolor{green!15}H \\
  \hline
  \hline
   Cooperative Maneuvering & \cellcolor{yellow!15}M & \cellcolor{yellow!15}M & \cellcolor{red!15}L & \cellcolor{green!15}H \\
  \hline
  \end{tabular}
  \begin{flushleft}
\begin{center}
    
\begin{tikzpicture}
\node (rect) at (0,2) [draw,thick,minimum width=.1cm,minimum height=0.4cm, fill= green!15, label=0:High: \color{black}This communication technology suits best \color{black}] {H};
\node (rect) at (0,2.5) [draw,thick,minimum width=.1cm,minimum height=0.4cm, fill= yellow!20, label=0:Medium: \color{black}This communication technology can be used.\color{black}] {M};
\node (rect) at (0,3) [draw,thick,minimum width=.1cm,minimum height=0.4cm, fill= red!15, label=0:Low: \color{black}This communication technology does not suit.\color{black}] {L};
\end{tikzpicture}
\end{center}

\end{flushleft}
  
  \end{table}


Vehicles in the next generation will be equipped with radar, cameras, wireless technologies, GNSS, and other sensors that can support autonomous driving. The requirement of line-of sight propagation limits the functionality of the aforementioned sensors. By adding the cellular V2X technology in the vehicles, this issue can be overcome. The functionalities of sensors can be complemented by the cellular V2X technology by exchanging of the sensory data among the vehicles that can provide situational awareness to the drivers at a higher level. Cellular V2X has attracted significant interest from both academia and the industry recently. One of the promising technologies that can be used to realize autonomous driving and communications in V2X is 5G New Radio (5GNR) \cite{dahlman20205g}. From a communication point of view, 5G should support the following three broad categories of services: URLCC, eMBB, and mMTC \cite{bagheri20215g}. eMBB, that aims to provide an uplink data rate of 10 Gbps and for downlink channels, data rates of at least 20 Gbps, plays a pivotal role for several functionalities in autonomous cars such as several multimedia services, video gaming/conferencing within the car, downloading of high-precision maps, etc \cite{albonda2019efficient}. For constantly sensing and learning the changes in the environment through the sensors that are equipped within the cars/infrastructure by the future driver-less vehicles, mMTC will play a very important role \cite{pokhrel2020towards}. High reliability and over the air round trip time, which are crucial for autonomous driving can be realized by URLCC \cite{ashraf2018dynamic}.

The 5G NR physical layer has to deal with diverse data requirements and tough V2X channel conditions like : 1) Extremely dynamic mobility, in which slow moving vehicles with speeds of upto 60 kilometers/hour and high speed trains/cars travelling with a speed of greater than 500 kilometers/hour have to be dealt with. Resources with more time-frequency are needed for high mobility communication in the design of air interface for dealing with the disabilities resulting from multi-path channels and Doppler spread. 2) Conflicting requirements like data services with wide range such as downloading of high precision maps,video conferencing, multimedia entertainment with the car,, etc. with diverse quality of requirements with respect to latency, reliability and data rates are difficult to be supported simultaneously \cite{anwar2019physical}. To address this issue, 5GNR frame structure ensures flexible configurations to support wide variety of use cases for cellular V2X. Orthogonal frequency-division multiplexing is used by 5GNR. A maximum of 400 MHz bandwidth/new radio carrier is the limit for the channels.   

In cellular V2X physical layer, channel coding plays a very important role in accommodating several requirements such as decoding latency, data throughput, mobility, packet length, support for hybrid automatic repeat request, and compatibility of rates. A common physical infrastructure is expected to be used for several applications for system resources such as computing, storage, bandwidth, etc. in cellular V2X services in 5GNR. NS can be used for providing E2E data services by slicing the resources of physical layer in radio access network \cite{mei2019intelligent,husain2019ultra}. The information related to the status of every vehicle like intended route, trajectory, speed, and location, can be exchanged with the road infrastructure,  pedestrians, other vehicles, in the network by Sidelink protocols \cite{ashraf2020supporting, ganesan2020nr,lien20203gpp}. Some of the major enhancements of NR Sidelink to enable use cases of 5G for enhancing AD are i) For low latency and high reliability, Sidelink feedback channel is proposed, ii) Support of up to 16 carriers for the aggregation of carriers, iii) To increase the throughput for every carrier upto 256-QAM, a modulation scheme is proposed, and iv) To reduce the time for selecting the resources, a modified resource scheduling is proposed. NR-V2X supports uncast and groupcast communications along with traditional broadcast communication, in which a vehicle can transmit variety of messages with several QoS requirements. As an example, some periodic messages can be transmitted by a vehicle through broadcasting and some aperiodic messages can be transmitted by groupcast or unicast \cite{gyawali2020challenges}.

The available resources in NR-V2X can be either exclusive or shared by the cellular users for direct communication among the vehicles. Mode-1 and Mode-2, which are two Sidelink nodes are defined for managing the resources in NR-V2X. It is expected that vehicles are covered completely by the base stations (BS) in Sidelink Mode-1. The vehicles are allocated with the resources depending on the dynamic and configured scheduling by the BSs. Resources are allocated based on a pre-defined bitmap in configured scheduling, whereas the resources are either reallocated or allocated every millisecond depending of the dynamic coverage of the cellular network in dynamic scheduling. Resources are allocated distributedly without the coverage of cellular network in Sidelink Mode-2. In Mode-2, there are four sub-modes; sub-node 2(a) - 2(d). Every vehicle may select the required resources in an autonomous manner by using a mechanism based on semi-persistent transmission in 2(a). In 2(b), most suitable resources for transmission are selected in a cooperative approach, where the vehicles will be assisting each other. The resources are selected by the vehicles based on pre-determined scheduling in 2(c). The sidelink transmissions are scheduled by the vehicles for their neighboring vehicles in 2(d) \cite{chen2017vehicle}. Several vehicular communication technologies for vehicular communication use-cases are summarized in Table~\ref{tab:5GNR}.

Some of the recent works related to the physical layer communication in 5G for AVs is discussed below.

A location aware handover for enabling self-organized communication for AVs using a robotic platform enabled by mmWave in multi-radio environment is presented by Lu~et al. \cite{lu2020feasibility}. The location aware handover mechanism studied by the authors enhances the robustness of the wireless link for IIoT based applications. A geometry based positioning algorithm is proposed by the authors for acquiring the location awareness in this work.

The obstacles for achieving high transmission rates in V2X networks are the increased demands for data bandwidths and the dynamic nature in the traffic. Rasheed~et al. \cite{rasheed2021intelligent} presented a predictive routing approach for V2X based on SDN controlled CR to overcome these obstacles. The proposed routing  supports the smart switching between mmWave and THz. The V2X network is segregated into several clusters for efficient network management. To form the clusters, a stability-aware clustering method is used by the authors. Te authors achieved minimum transmission delay and high transmission rate fir THz anf mmWave communications by using the following design approaches. (1) For real-time and high-resolution prediction, a deep neural network with extended kalman filter approach is used to predict future positions of the vehicles in 3D. (2) CR-enabled road side units perform ThZ band detection. For optimal route selection in THz communications, a hybrid fruitfly-genetic algorithm algorithm is used. (3) A multi-type2 fuzzy inference system is used for selecting optimal beam in V2X communications based on mmWave.

Eric~et al. \cite{kampert2020millimeter} investigated commuication links based on 5G mmWave for a low-speed AV. They have studied the effects of the positions of antenna on the performance of the links and also the quality of the received signal. The authors have observed that the communication losses in front of the vehicles for the road side infrastructure is almost half-power beam independent, also a major role in the quality of the signal is played by the increased root mean square delay spread.

\subsubsection*{Summary}

5GNR plays a very important role in several use cases for AVs such as providing high data rate, throughput and URLCC for coordinated driving, real-time updates on local conditions, trajectory sharing and sharing of raw data gathered by the vehicles.

\subsection{Federated Learning } 
\subsubsection{Introduction}

The ITS can be fully realized when the issues such as high availability, data privacy, and scalability are addressed. The traditional ML models are not efficient in dynamic environments such as ITS and Internet of Vehicles (IoV). They face challenges such as complexity of the system, performance of the model, privacy preservation and data management. In IoV and ITS, even though roadside infrastructure remains constant, vehicles enter and leave the system constantly, which makes the system volatile and complex. Traditional ML algorithms find it difficult to handle such volatility and complexity. Traditional ML models use static local intelligence making it difficult for them to adapt to the dynamically changing environments, that can hamper their performance. Any degradation in the performance of the ML algorithms can affect he human lives. The next issue faced by the traditional ML models is the latency and privacy preservation involved when the sensitive data generated from the autonomous vehicles has to be transferred to the cloud for model training. Also, data management is very difficult as new vehicles with processing capabilities will be continuously added to the IoV/ITS. As the roadside infrastructure has limited resources, efficient storage of the data is a challenge, due to which, the traditional ML methods may not have access to sufficient data for training phase \cite{manias2021making}. The aforementioned issues of ITS/IoV can be effectively handled by  FL coupled with B5G. B5G connectivity can enable flexible V2X interactions with other vehicles (V2V) and road infrastructure (V2I) in vehicular URLCC. FL can solve the issues inITS/IoV such as privacy preservation, real time analytics in complex and dynamic environments, and latency.  FL enabled B5G can ensure seamless, reliable communication between the autonomous vehicles in ITS/IoV that will enable collaborative driving through sharing of maps, sharing the information about the conditions of the roads, weather, accidents, landslides, etc., without exposing the sensitive information of the vehicles \cite{savazzi2021opportunities,du2020federated,pokhrel2020improving}. Some of the recent works on the FL enabled B5G for ITS/IoV and autonomous vehicles are presented in the rest of this subsection.

\begin{table}[h!]
  \centering
        \caption{Benefits of AI/FL/Edge AI in Automated Vehicles \cite{lu2020blockchain, lu2020federated, kong2021fedvcp}  
        }
        \label{tab:FL_AV}
\resizebox{\columnwidth}{!}{%
  \begin{tabular}{|p{2.8cm}|p{0.5cm}|p{0.5cm}|p{0.5cm} |p{6.5cm}|}
  \rowcolor{gray!25}
  \hline
   \cellcolor{gray!15}Benefits &{\rotatebox[origin=c]{90}{\cellcolor{gray!15}~AI and ML  ~}}&{\rotatebox[origin=c]{90}{\cellcolor{gray!15}~Edge AI  ~}}&{\rotatebox[origin=c]{90}{\cellcolor{gray!15}~Federated learning
   ~}}&{{\cellcolor{gray!15}~Challenges of FL~}} \\
\hline
   Data security & \cellcolor{red!15}L &\cellcolor{yellow!15}M& \cellcolor{green!15}H& Poisoning and backdoor attacks are difficult to handle. \\
\hline
   Real-time learning& \cellcolor{red!15}L & \cellcolor{green!15}H &\cellcolor{yellow!15}M&Generation of labels for the data from AVs in real-time is a challenge.\\
  \hline
   Privacy & \cellcolor{red!15}L & \cellcolor{yellow!15}M &\cellcolor{green!15}H& The attributes can be reconstructed by the attackers from the model updates sent from the local devices, hence compromising the privacy preservation.\\
 \hline
   Latency & \cellcolor{green!15}H & \cellcolor{red!15}L&\cellcolor{yellow!15}M& The frequent local updates sent from AVs to the central cloud will increase latency. \\
  \hline
   Resource optimization & \cellcolor{red!15}L & \cellcolor{yellow!15}M&\cellcolor{green!15}H& Optimization of resources for training the data and processing at individual AVs is a challenge.  \\
  \hline
   Data offloading & \cellcolor{red!15}L & \cellcolor{green!15}H&\cellcolor{yellow!15}M& Due to limited computational capabilities of the AVs, offloading the data in real-time for training the local FL model is a significant challenge. \\
  \hline
  \end{tabular}
}
  \begin{flushleft}
\begin{center}
    
\begin{tikzpicture}
\node (rect) at (0,2) [draw,thick,minimum width=.1cm,minimum height=0.4cm, fill= green!15, label=0:High Capability: \color{black}Technique is highly utilized. \color{black}] {H};
\node (rect) at (0,2.5) [draw,thick,minimum width=.1cm,minimum height=0.4cm, fill= yellow!20, label=0:Medium Capability: \color{black}Technique can be utilized but not wider.\color{black}] {M};
\node (rect) at (0,3) [draw,thick,minimum width=.1cm,minimum height=0.4cm, fill= red!15, label=0:Low Capability: \color{black}Technique is not utilized.\color{black}] {L};
\end{tikzpicture}
\end{center}

\end{flushleft}
  
  \end{table}

To improve the service quality and the driving experience in IoV, vehicles can share the data among themselves for collaborative analysis. The data providers hesitate to participate in the process of data sharing due to several issues such as privacy, security and bandwidth issues. The efficiency and reliability of the data sharing has to be enhanced further as the communication in IoV is unreliable and intermittent. Lu~et al. \cite{lu2020blockchain} proposed a hybrid blockchain architecture that consists of local directed acyclic graph and permissioned blockchain that is run by the vehicles for reliable and secured data sharing in 5G network. To improve the efficiency, the authors have proposed a scheme based on asynchronous FL that uses deep reinforcement learning for learning models on the data at the edge nodes.  In a similar work, Lu~et al. \cite{lu2020federated} have proposed a FL enabled 5G networks for vehicular cyber-physical systems to reduce the leakage of private and sensitive data while sharing the data among the vehicles that leads to preserving the privacy of the passengers and improve their safety.

The key applications of autonomous vehicles such as collision avoidance and autonomous driving require that the positioning of the vehicles is precise. Due to the advancements in V2I communications and sensing techniques, the vehicles can communicate with the nearby landmarks to enable precise positioning. This involves sharing data related to the the positioning of the vehicles loaded with sensors. Collecting and training the data centrally is difficult due to the sensitive nature of the trajectory data. The continuous sharing updating the location data of the vehicles wastes network bandwidth and other resources. To address these issues, Kong~et al. \cite{kong2021fedvcp} proposed a vehicular cooperative positioning system based on FL, namely, FedVCP, that utilizes collaborative edge computing to provide accurate positioning information in a privacy preserving manner.  The comparative analysis of AI, Fl and Edge AI for AVs in 5GB networks are summarized in Table \ref{tab:FL_AV}.

\subsubsection*{Summary}
FL has a huge potential in solving some of the pressing issues in connected AVs in B5G era such as improved latency, privacy issues, optimal resource utilization, giving customized recommendations/predictions in heterogeneous networks \cite{gadekallu2021federated}. However, FL faces several challenges such as poisoning attacks, high false alarms, energy efficiency in low powered devices used in AVs, etc. These issues are to be addressed to realize the full potential of FL enabled B5G networks for AVs.

\subsection{Blockchain}

 \begin{table}[!t]
	\centering
        \caption{Impact of Blockchain on Technical Aspects for AVs in 5g and beyond \cite{gao2019blockchain, xie2019blockchain, bera2020blockchain, aloqaily2021design, gupta2021blockchain, jian2021blockchain, feng2021efficient, ghimire2021sharding, gumaei2021deep,nguyen2021federated} 
        }
        \label{tab:Blockchain}
\resizebox{\columnwidth}{!}{%
  \begin{tabular}
  {|p{1.8cm}|c|c|c|c|c|c|c|p{5.5cm}|}  
  \rowcolor{gray!25}
  \hline
   \cellcolor{gray!15} Technical Aspect &{\rotatebox[origin=c]{90}{\cellcolor{gray!15}~Increased Capacity~}}
   &{\rotatebox[origin=c]{90}{\cellcolor{gray!15}~Security~}}
   &{\rotatebox[origin=c]{90}{\cellcolor{gray!15}~Immutability~}}
   &{\rotatebox[origin=c]{90}{\cellcolor{gray!15}~Transparency~}}
   &{\rotatebox[origin=c]{90}{\cellcolor{gray!15}~Traceability~}}
   &{\rotatebox[origin=c]{90}{\cellcolor{gray!15}~Authenticity  ~}}
   &{\rotatebox[origin=c]{90}{\cellcolor{gray!15}~Reliability~}}   
    &{{\cellcolor{gray!15}~Challenges~}}\\
\hline
   MEC  & \cellcolor{green!15}H & \cellcolor{green!15}H & \cellcolor{red!15}L & \cellcolor{yellow!15}M & \cellcolor{red!15}L & \cellcolor{green!15}H &\cellcolor{green!15}H& Due to big data generated by AVs, data storage and resource optimization through blockchain are significant challenges.\\
\hline
   Network Slicing& \cellcolor{red!15}L & \cellcolor{red!15}L & \cellcolor{red!15}L & \cellcolor{red!15}L & \cellcolor{red!15}L & \cellcolor{red!15}L &\cellcolor{red!15}L& Interoperability and scalability of blockchain integrated NS for heterogeneous AVs are major challenges.\\
 \hline
   SDN/NFV & \cellcolor{red!15}L & \cellcolor{red!15}L & \cellcolor{red!15}L & \cellcolor{red!15}L & \cellcolor{red!15}L & \cellcolor{red!15}L &\cellcolor{red!15}L & On-demand allocation of resources and efficient
utilization of network resources to optimize the energy efficiency of the sensors in AVs is a chalenge.\\
\hline
   5GNR & \cellcolor{green!15}H & \cellcolor{yellow!15}M & \cellcolor{yellow!15}M & \cellcolor{green!15}H & \cellcolor{green!15}H & \cellcolor{green!15}H &\cellcolor{green!15}H & Resource allocation, interoperability during mobility in real-time with less latency are significant challenges in blockchain enabled 5GNR.   \\
\hline
   FL  & \cellcolor{green!15}H & \cellcolor{green!15}H & \cellcolor{green!15}H & \cellcolor{red!15}L & \cellcolor{red!15}L & \cellcolor{green!15}H &\cellcolor{green!15}H & Privacy and security of AVs, increased communication latency in transferring of the data to and from AVS and training cost of the data from AVs are significant challenges when using blockchain for federated learning. \\
\hline
 
  \end{tabular}
}
  \begin{flushleft}
\begin{center}

\begin{tikzpicture}
\node (rect) at (0,2) [draw,thick,minimum width=.1cm,minimum height=0.4cm, fill= red!15, label=0:High Impact: \color{black}Blockchain Technology has highest impact. \color{black}] {H};
\node (rect) at (0,2.5) [draw,thick,minimum width=.1cm,minimum height=0.4cm, fill= yellow!20, label=0:Medium Impact: \color{black}Blockchain Technology has medium impact.\color{black}] {M};
\node (rect) at (0,3) [draw,thick,minimum width=.1cm,minimum height=0.4cm, fill= green!15, label=0:Low Impact: \color{black}Blockchain Technology has low impact.\color{black}] {L};
\end{tikzpicture}
\end{center}

\end{flushleft}
  
  \end{table}

\subsubsection{Introduction}

Blockchain is a distributed ledger technology that provides several features like immutability, transparency, and trustability.  The records in the blockchain are distributed and duplicated throughout the nodes in the network. Blockchain consists of millions of blocks; each block in the blockchain consists of information related to transactions, its own hash and also previous block's hash \cite{kumar2021ppsf}. Whenever a new transaction is executed in the blockchain, that transaction's record is added in the ledger of every participant. Consensus algorithm is an important concept in blockchain, where majority of the blocks/nodes in the blockchain have to agree to the present state of the distributed ledger. Consensus algorithms make sure that, in order to update/modify the information stored in the blockchain, more than  50\% of the millions of nodes in the blockchain have to approve it. This inherent property of the blockchain makes it nearly impossible to tamper or alter the data, that make the data in the blockchain immutable. Consensus algorithms also ensure that if a new node has to be added into the network, majority nodes in the blockchain have to approve it \cite{bouraga2021taxonomy}. 

\subsubsection{How it used in AV (with Related work)}

AVs that are interconnected through 5G can share crucial information about the environments, road blockages, traffic conditions, etc. with each other. However, if a malicious AV enters into the network, it can pose serious security issues to the AVs. A malicious AV can post wrong information about the environment, can get access to private information about the vehicles such as its location, path, modify the information shared by other AVs, etc. To address security and privacy issues, blockchain can play a major role. When a blockchain is integrated with 5g for connected AVs, it can make sure that only trusted vehicles can enter into the network. Also, the blockchain ensures that the information stored in it is immutable \cite{ortega2018trusted,shrestha2020evolution,rahmadika2019blockchain}. These features of blockchain make it a very important enabling technologies of 5G and beyond for AVs. Rest of this subsection discusses about the state-of-the-art on blockchain enabled 5GB for AVs. Reebadia~et al. \cite{reebadiya2021blockchain} presented a blockchain based approach to assure the privacy and security of the AVs that automatically sense events and act with necessary responses. The events such as accident detection, home transfer, congestion in traffic, etc. through nearby AVs or RSUs and will respond by actions such as calling an ambulance, calling the logistics, or informing the traffic department by sharing the location of the event. In this work, for communication edge-enabled 5G network is proposed to support huge bandwidth requirements and to process the data in real time. The blockchain is used in this work to ensure that the privacy and security of the messages that are shared in the network regarding the incidents/people/AV are preserved and by ensuring the trustworthiness of the AVs that join the network.

AVs generate event driven messages (EDM) when they detect events such as road sidings or accidents. The AVs have to send videos or audio or pictures for proof of the occurrence of those events. To ensure the reliability and authenticity of EDMs, Lewis~et al. \cite{nkenyereye2020secure} have proposed a private blockchain enabled edge nodes for 5G for storing the records of EDMs. Gao~et al. \cite{gao2019blockchain} proposed SDN and blockchain enabled 5G network for VANETs. Due to its distributed nature, the blockchain is used in this work to avoid single point of failure in VANETs. Blockchain ensures trust-based message propagation that can evaluate the data passed by the peers in the connected environment and can also control the doctored or false data shared in VANETs for communication systems based on fog computing and 5G. For effective management of the network, SDN is used. In a similar work, Xie~et al. \cite{xie2019blockchain} proposed a blockchain based system for addressing trustworthiness, security and privacy issues in 5G integrated with SDN for VANETs. In this work, the authors used the proposed framework for securing and preserving the privacy of the videos shared in VANETs by allowing only the trusted vehicles to be entered into the VANET. SDN is used o control and network and also enables global information gathering.

The traditional approach of using costly base stations that are fixed in a particular locations needs a revamp to meet the rise in number of interconnected IoT devices, irregular service and data requests. UAVs provide on-demand mobile access points for realization of several scenarios such as smart cities, smart manufacturing, etc. MEC enabled B5G networks can provide low latency, high data rates that can meet several high-end QoS requirements of UAVs. Integrating B5G with blockchain ensures secured and decentralized service delivery \cite{bera2020blockchain}. To address several issues such as requirement of high-seed communication, security of the messages passed between the UAVs/drones, trustworthiness of the UAvs joining the network, Aloqaily~et al. \cite{aloqaily2021design} proposed a blockchain based 5G network. To identify the intrusive traffic in the network, ML methods are proposed by the authors. In a similar work, Gupta~ et al. \cite{gupta2021blockchain} proposed an architecture based on blockchain enabled softwarized 5G for secured and easy management of UAVs. In this work, SDN along with blockchain can help in securing the 5G enabled UAV network against intruders, DoS and DDoS attacks, identifying anomalies, man in  the middle attacks, etc. Similarly, Jian~et al. \cite{jian2021blockchain} proposed a blockchain empowered UAV ad hoc network for B5G era. The authors have designed HELLO packet, consensus packet, and topology control packet for secured and reliable communication in the 5G enabled UAV ad hoc network. For ensuring trust in the network, the authors have used a consensus mechanism based on Byzantine fault tolerance for secured multi-point relaying of messages. Feng~et al. \cite{feng2021efficient} proposed a blockchain enabled 5G networks for secured and efficient sharing of data in flying drones. In the proposed approach,  attribute based encryption and blockchain are used for securing the data sharing and instructions in the drones.

Ghimire~et al. \cite{ghimire2021sharding} proposed a blockchain and software defined Internet of UAV for defence applications. To provide visibility of the network and to dynamically configure the network parameters for better manageability and better security of the Internet of UAVs. To prevent the tempering of the control and command operations exchanged by the UAVs, blockchain technology is used in this work. UAVs in the network are responsible to create the blocks in the blockchain and also act as miners to validate each transaction. To address the scalability issue of blockchain technology, the authors have proposed sharding-enabled blockchain. In this work, shards  of light weight UAVs are used for maintaining the required number of miners whenever a miner is damaged or destroyed in the battlefield. Some sensitive data of UAVs like flight modes and identification of drones can be communicated with other UAVs using radio frequency (RF) signals in 5G networks. Gumaei~et al. \cite{gumaei2021deep} proposed a blockchain enabled 5G network for secured and decentralized transmission of RF signals among the drones/UAVs. The authors in this work have proposed a framework that integrated blockchain with deep  recurrent neural network for 5G enabled flight mode detection and identifications of drones. Blockchain can be used for privacy preservation and securing the data during transmission among the UAVs connected through 5G in medical applications \cite{chen2021exploiting}. The impact of blockchain technology on several technical aspects for AVs in 5GB is summarized in Table \ref{tab:Blockchain}.

\subsubsection*{Summary}

UAVs/drones are being used for several applications like crop/soil analysis, road surveillance, monitoring of natural disasters, delivering the product to the customers, etc. apart from traditional military and defense applications. But there are several challenges that need to be addressed to integrate blockchain in B5G for Avs such as lack of standardization, scalability (handling the ever increasing AVs), etc.

\begin{table*}[h!]
\centering
\caption{Benefits and challenges of B5G Technologies in Autonomous Vehicles.}
\label{tab:FL_BD_Applications}
\resizebox{\textwidth}{!}{%
\begin{tabular}{|p{3cm}|p{6.5cm}|p{7.5cm}|}
\hline
{\textbf{Technology}} &
  {\textbf{Challenges Faced by Autonomous Vehicles}} &
 {\textbf{Benefits of using the B5G  Technology}}  
  \\ \hline
  
Multi-access Edge Computing (MEC) \cite{liyanage2021driving, zhou2018mec, 
giust2018multi, campolo2019mec, barmpounakis2020collision, coronado2019addressing, lee2020multiaccess, shi20185g, rasheed2020application, ma2020cooperative, lian2020semantic, sabella2020mec}
&
  1)    Increased latency
  
  2)   Communication cost
  
  3)  Real-time analytics

  &
  Edge computing extends the capabilities of cloud computing by bringing the resources closer to the vehicles. Hence, the latency and communication cost will be reduced, that in turn will help in providing the analytics in real time.  
   
     \\ \hline
Network Slicing 
\cite{campolo20175g, khan2021network, khan2018application, sanchez2019empowering, mlika2021network, xiong2019smart}

&
  1) Ultra reliable connectivity for autonomous driving
  
  2)  High data rates required for vehicle infotainment
  
  3) Support for high mobility
  
  4) Dynamic network connectivity
  
  &
  1) Guaranteed performance  through  dedicated  V2X  slices
  
  2)  Dedicated  slices  for  autonomous  driving and infotainment
  
  3)  Dynamic resource allocation for network slices
  
  4) Support for heterogeneous radio  access  technologies

  \\ \hline
SDN/NFV \cite{kakkavas2021network,mahmood2019software, goudarzi2020dynamic, garg2019sdn, duan2017sdn, garg2019mobqos} &
  1) Static routing decisions    
  
  2) Requirement of traffic prioritization   
  
  3) Requirement of reduced end-to-end delay
  
  4) Requirement of secured communication of data
  &
  1) Minimal latency     
  
  2) Reduction in communication overheads
  
  3) Dynamic routing decisions
  
  4) Enhanced congestion control 
  
  5) Reduction in end-to-end delay
  
  6) Optimized resource utilization
 
  \\ \hline
  
  5GNR  \cite{lu2020feasibility, rasheed2021intelligent, kampert2020millimeter} &
  1)   Requirement of increased bandwidth, increased data speeds
  
  2)   Improved links between the users
  
  3)  Network loss and more lag time
 
  4) Quality of connection in broad geographical locations
  &
  Adaptive bandwidth is provided by 5G NR, through which larger capacity for wireless devices, ease of maintenance  of quality in connections covering large geographical locations,  reduced lag time, and increased data speeds can be realized.

  \\ \hline
  
Blockchain \cite{gao2019blockchain, xie2019blockchain, bera2020blockchain, aloqaily2021design, gupta2021blockchain, jian2021blockchain, feng2021efficient, ghimire2021sharding, gumaei2021deep}
&
  1)    Possibility of malicious vehicle entering into the network
  
  2)  Privacy and security of the AVs

  &
  
  Blockchain ensures that only trusted vehicles enter the vehicular networks. Blockchain also ensures privacy and security of the sensitive information of the vehicles through immutable property.  
   \\ \hline
Federated Learning \cite{lu2020blockchain, lu2020federated, kong2021fedvcp} &
  1) Privacy preservation    
  
  2) Requirement of large number of computational resources and communication infrastructure
  
  &
  
  FL brings the computation to the devices, which will reduce the possibility of sensitive information of the vehicles being exposed and also the ensures the reduction in optimization of computational and communication resources.

  \\ \hline
  
\end{tabular}%
}
\end{table*}

\subsection{ZSM(AI/ML)}

AVs in B5G era need extreme range of requirements, such as personalized services with dramatic improvements in customer-experience, ultra-high reliability, massive seemingly infinite capacity, global web-scale reach, support for massive machine-to-machine communication, and imperceptible latency. Fully automated network and service management is the need of the hour for delivering the aforementioned services for AVs in B5G era. The fully automated futuristic network will be driven by high-level rules and policies. This automation in the B5G networks for AVs will be capable of self-optimization, self-monitoring, and elf-configuration of the B5G networks without human intervention. Zero touch network \& Service Management (ZSM) architecture is proposed by European Conference of Postal and Telecommunications Administrations (ETSI) to achieve full automation of the networks \cite{benzaid2020ai,liyanage2022survey}. ZSM can play a major role in providing seamless, reliable B5G network for AVs.



\section{Security Concerns in AVs}
Autonomous-based vehicles are vulnerable to several cyber attacks. Based on safety features, AVs can be categorised into two categories \cite{cui2019review} i.e. AV components (including vehicles hardware/software components) and the infrastructure/environment within which vehicle is operating (such as pedestrian movement, traffic signs, road conditions, network-type etc). The attackers can jeopardize the safety of passengers by targeting these two components via In-Vehicle communication (through control network bus (CAN)) and Vehicle-to -everything communication (V2X). Hence, in this section, we will provide overview of potential attacks in AVs based on CAN and 5G-enabled V2X communication. We will also discuss the emerging security and privacy concerns in AVs based on B5G connectivity. 
\subsection{Security Concerns via In-Vehicle Communication}
From the security perspective, there are four key requirements for any intelligent transport system (ITS) \cite{malla2013security,manvi2017survey,othmane2015survey,zhang2017secure} such as AVs to achieve the security goals. One of the requirement is \textit{authentication} which requires proper mechanism to identify an authorised user (such as driver) and grants access to features of the vehicle based on the privilege. Authentication might also include source authentication (making sure data is generated from legitimate sources) and location authentication (ensuring integrity of the received information). The second requirement includes \textit{availability} which ensures that the exchanged/shared data is always available for processing in a real-time. The third requirement is data integrity/trust which ensures that the received data is free from any kind of modifications, manipulation etc. Finally, the last requirement involves \textit{confidentiality/privacy} to ensure that the exchanged data is not exposed to unauthorised or malicious users. Based on the above-mentioned requirements, the attackers can exploit a specific weakness and compromise a specific security goal. Attacker can be anyone such as active/passive or external/insider \cite{raya2007securing} with different malicious intentions. With respect to In-Vehicle communication, the following components are vulnerable to different cyberattacks:  

\subsubsection{Sensors and Actuators} Although, AVs come up with several sensors, from the security point of view, the sensors such as Light detection and ranging (LiDAR), image sensors (Camera's) and radio detection and ranging (RADAR) are vulnerable to cyberattacks \cite{pham2021survey}. LiDAR sensors are mostly used for obstacle detection by using light as a measure to probe surrounding environment. In case of an obstacle detection, emergency autonomous brakes are applied. LiDAR sensors are vulnerable to spoofing attacks where an attack can create a counterfeit signal representing some near-by object making AVs to apply emergency braking system, thereby reducing the speed of vehicles which can prove fatal. Spoofing-based attacks were successfully demonstrated by \cite{petit2015remote}. There is also a possibility of performing denial of service attacks by jamming the functionality of LiDAR sensors. This can be done by preventing LiDAR sensors to acquire the legitimate light wave through sending out the light of same wavelength. One such attack was demonstrated by the work of \cite{stottelaar2015practical}. RADAR-based sensors are also vulnerable to spoofing \cite{chauhan2014platform} and jamming attacks \cite{buehler2014airborne} as their working principle is same as of LiDAR i.e. object detection. The main difference is RADAR sensors use radio waves to detect objects instead of laser/LED and has comparatively long object detection range. On the otherhand, Camera's consisting of image-based sensors are attached at many places of the AVs to acquire 360 degree view for different purposes such as detection of traffic signs \cite{fairfield2011traffic}, lane-detection \cite{hillel2014recent}, etc. Although, these sensors can be used to replace LiDAR sensors, however performance of image-based sensors is not good under certain conditions such as rain, fog, etc \cite{wang2019pseudo}. Image-based sensors are also vulnerable to DOS attacks as in case of LiDAR sensors. Camera's can be blinded by supplying extra-light as successfully demonstrated by \cite{petit2015remote}. Besides, the developments in adversarial machine learning where a perturbed image can be used to trick the underlying artificial-intelligence model to cause incorrect predictions is one more serious concern. The attackers just need to place perturbed input such as stickers \cite{brown2017adversarial}in front of the observing camera to confuse the intelligent system within AVs.
\subsubsection {Controller Area Network}  Controller area network is a central network within AVs with the main purpose of connecting different electronic control unit (ECUs) to each other. ECUs are embedded electronic systems used to control different sub-systems or electrical systems in a vehicle (such as brake-control system, engine control system etc), which are necessary requirements for autonomous driving. The detailed explanation with respect to working of these modules can be found in the work of \cite{kassakian1996automotive}. Different ECUs within AVs communicate through CAN and provides necessary inputs for running the safe autonomous driving system \cite{jitpakdee2008neural}. The whole CAN is vulnerable to different cyberattacks due to the fact that CAN-based communication protocols have no authentication mechanism with no support for encrypting messages to fulfill the security goal of confidentiality \cite{matsumoto2012method}. The strong authentication mechanism requires large processing power, memory, and bandwidth which is lacking in CAN. On an average, CAN supports a data-rate between 33kbps and 500kbps) and has a limited bandwidth \cite{lin2012cyber}. Following these limitations, the CAN is vulnreable to DOS attack \cite{matsumoto2012method}, replay attack\cite{lin2012cyber} and Eavesdrop \cite{matsumoto2012method}. 
\subsubsection{On-board Computer(OBD)} To get different information from the vehicle such as its emission, speed, and other data, there is a connection known as OBD which can be used for this purpose. Modern vehicles do come up with OBD-II standard. OBD can be also used to perform firmware update and modify software embedded in control units \cite{checkoway2011comprehensive,pham2021survey}. OBD can serve as a gateway to several attacks as OBDs do not encrypt data and have no authentication mechanism. There are several third-party devices such as Telia Sense, AutoPi and car-manufacturer based devices that can be used to connect smart-phones or computers to OBD ports for self-diagnostic purposes. Through OBD ports, attacks such as man-in-the-middle, code-injection are feasible to compromise certain critical components of the vehicle \cite{koscher2010experimental}.
\begin{table*}[!h]
	\scriptsize
	\begin{center}
			\vspace{-0.5cm}
		\centering
		\caption{Cyberattacks on Autonomous Vehicles}
		\label{tab:cyberattack}
\begin{tabular}{p{1.7cm}p{6 cm}p{1.7cm}p{1.7cm}p{1.7cm}p{1.7cm}}
			\hline
			\hline
Attacks	& Description & Sensors and Actuators & Controller Area Network & On-board Computer  & V2X \\ \hline
Spoofing\cite{kosmanos2020novel} & A type of attack in which an attacker impersonates the authorised device/program for different malicious activities &\checkmark &\checkmark	&\checkmark  &\checkmark \\ \hline
Denial of service \cite{hossain2021observer} & In DoS attack, illegitimate requests are sent to the network to void legitimate users in accessing the required sources &\checkmark & \checkmark & $\times$ &\checkmark  \\ \hline
Jamming \cite{fraiji2018cyber} & It is a type of DoS attack where radio signals are used to disrupt the communication between the legitimate nodes &\checkmark &$\times$ &$\times$ &\checkmark  	\\ \hline
Adversarial \cite{ilahi2021challenges} & In this type of attack, deceptive inputs are supplied to the systems relying on artificial-intelligence technology to make incorrect predictions  &\checkmark  &$\times$ &$\times$ &\checkmark \\ \hline

Injection \cite{pham2021detecting} & In this type of attack, a malicious input is supplied by the attacker which gets executed by an interpreter as part of a command to alter the execution of a program  &\checkmark &\checkmark &\checkmark &\checkmark \\ \hline
Physical \cite{fraiji2018cyber} & In this type of attack, the attacker gains physical access to cause damage to the asset & \checkmark & \checkmark & \checkmark &\checkmark  \\ \hline
     \hline

		\end{tabular}
	\end{center}
\end{table*}

\subsection{Security Concerns Via V2X technologies}
For AVs to perform different functions such as information exchange between different sensors, vehicles, its surrounding environment (i.e. infrastructure, pedestrians etc), a high-bandwidth, low latency and highly reliable communication network is required. 
The ultimate goal of V2X technology is increase traffic efficiency, reduce accidents by enhancing road safety measures and energy savings. As progress in this direction is still at its infant stages, there are two key standards i.e. IEEE 802.11P\cite{strom2011medium,van2012impact} and Cellular C2X that has unique characteristics such as standard performance under bad weather conditions and low latency makes them idea to achieve Vehicle-to-Vehicle and Vehicle-to-Infrastructure communication. Based on the literature, to achieve the goal of V2X paradigm, communication technologies have been categorised into three main categories i.e. short-range, medium-range and long-range \cite{ahangar2021survey}. For short-range communication, bluetooth, ZigBee, Ultra-wide band (UWB) technologies are being considered to achieve the following short-range functionalities within V2X paradigm i.e. vehicle localisation, real-time driving assistance, vehicle identification, collision warning and other functions such as car-owner identification, detection passenger position in the vehicle. Similarly,in medium-range 
technologies Wi-Fi and DSRC (IEEE 802.11p) are being explored to achieve the functionalities related to traffic safety, traffic management system and vehicle management. Finally, C-V2X and 5G-NR standard are being explored to achieve long-range based objectives within the scope of V2X paradigm. To get detailed information in 5G-NR and C-V2X, please refer to the works of \cite{parkvall2017nr} and \cite{abboud2016interworking} respectively.
Based on all the communication technologies discussed above, irrespective of which technology will be used to achieve the V2X communication, the security concerns will remain there. Based on the literature, we have summarised the potential attacks in Table \ref{tab:cyberattack}.

\section{Projects and Standardization} \label{sec:projects}
 
This section highlights the key projects and standardization activities related to the 5G autonomous vehicles. 

\subsection{Research Projects}
Research projects plays vital role in realizing the deployment of 5G autonomous vehicles. This section summarises the several key global level research and development projects related to the 5G autonomous vehicles.

\subsubsection{5G HarmoniseD Research and TrIals for serVice Evolution between EU and China
 (5G-DRIVE)}
5G-DRIVE \cite{5gdrive} is an European Union (EU) funded project under the Horizon (H2020) framework. This project is EU and China collaboration project which is mainly focusing on operating 5G 3.5 GHz bands for eMBB scenarios as well as 3.5 GHz and 5.9 GHz bands for V2X use cases.  Moreover, 5G-DRIVE project researching on use of key 5G technologies such as networking slicing, NFV, MEC and 5G New Radio (5GNR) for real-world V2X deployments. The 5G-DRIVE is also focusing on increasing  EU-China collaboration on 5G activities related to V2X research activities. 5G-DRIVE involves 17 European partners from 11 countries. It carried out trials as three locations i.e. Espoo (Finland), Surrey (United Kingdom) and JRC Ispra (Italy)

\subsubsection{5G for Connected and Automated Road Mobility in the European UnioN (5G-CARMEN)}
5G-CARMEN \cite{5gcarmen} is also an European Union (EU) funded project under the Horizon (H2020) framework. It focuses on building  trial 5G networks along Bologna-Munich corridor to support the European  research activities related connected and automated mobility. 5G-CARMEN builds a 600 km long trial site which spread across  three countries. 5G-CARMEN project mainly focuses on 5G NR, C-V2X (Cellular vehicle to everything), and secure service orchestration. This trial  platform  supports V2I, direct short range V2V and  long-range V2N (Vehicle to Network) communication modes. Moreover, 5G-CARMEN focuses on four main use cases i.e. cooperative manouvering, situation awareness, video streaming and green driving.

\subsubsection{Fifth Generation Cross-Border Control (5GCroCo)}
5GCroCo \cite{5GCroco} is also an European Union (EU) funded project under the Horizon (H2020) framework. It focuses on building  trial 5G networks along France, Germany and Luxembourg to support the European-level  research activities by integrating telecommunication and automotive industries. 5GCroCo builds two main 5G trial sites across Germany - Luxembourg border and France - Germany  border. In addition, 5GCroCo develops five small scale test sites at Montlhéry (France), Motorway A9 (Germany), Munich (Germany), AstaZero (Sweden) and Barcelona (Spain)    5GCroCo project  mainly focuses on emerging 5G technologies such as MEC, 5G NR V2X and secure service orchestration. It also focuses on defining new business models related to 5G autonomous vehicles. Finally, 5GCroCo focuses on three main use cases i.e. tele-operated driving, High-Definition (HD) mapping and anticipated cooperative collision avoidance.


\subsubsection{5G for cooperative and connected automated MOBIility on X-border corridors (5G-MOBIX)}
5G-MOBIX \cite{5G-MOBIX} is another European Union (EU) funded project under the Horizon (H2020) framework. It focuses on building  trial 5G networks to enable sustainable future for connected and automated vehicle.  5G-MOBIX contains two cross-boarder corridors at Greece – Turkey and Spain – Portugal boarders.  In addition, 5G-MOBIX develops several small scale test sites at Espoo (Finland), Jinan (China), Versailles Satory (France),  Paris (France), Berlin (Germany), Stuttgart (Germany), Yeonggwang (South Korea) and Eindhoven-Helmond (Amsterdam). In addition, 5G-MOBIX builds the International research cooperation between Europe, China and Korea in  5G autonomous vehicles related research activities. Finally, 5G-MOBIX focuses on several 5G autonomous vehicles use cases such as cooperative overtake, highway lane merging, truck platooning, valet parking, urban environment driving, road user detection, vehicle remote control, see through, HD map update, media and entertainment.


\subsubsection{ICT Infrastructure for Connected and Automated Road Transport (ICT4CART)}
ICT4CART \cite{ict4cart} is an H2020 project got funded by EU. It focuses on blending the technological developments in  telecommunication, automotive and IT industries  to realize the transition towards connected and automated vehicles. It focuses on research areas such as hybrid connectivity, flexible network slicing, data management and privacy, network security and localisation.  ICT4CART project has a cross-boarder corridors at Italy – Austrian boarder and  three small scale trail sites at Graz (Austria),  Ulm (Germany), and Verona (Italy). 


\subsubsection{Terahertz sensors and networks for next generation smart automotive electronic systems (car2TERA)}
Car2TERA \cite{car2tera} project is an EU H2020 funded project which focuses on in-cabin radar and  high speed onboard data communications for autonomous automobiles. On this regards, the Car2TERA project focuses on adapting sub-terahertz (150-330 GHz) communication to offer efficient and sufficient  bandwidth for high resolution demands of in-vehicle radar communications. In addition, short-range, sub-THz frequency radar technology can be useful to  improve in-cabin as well as   outdoor sensing.

\subsubsection{Fifth Generation Communication Automotive Research and innovation (5GCAR)}
5GCAR \cite{5GCar} project is an EU H2020 funded project which focuses on 5G  C-V2X. It has focuses on different aspects of 5G  C-V2X such as radio access network,  spectrum matters, system architectural options, network orchestration and management security and privacy issues, Edge computing enhancements, multi-connectivity cooperation  and possible business models. 5GCAR focuses on three 5G autonomous vehicles use cases i.e. lane merge coordination, cooperative perception for maneuvers of connected vehicles and vulnerable road user protection. 


\subsubsection{Other Projects}
Several other projects are listed below which have 5G autonomous vehicles as a minor focus.

\begin{itemize}
\item Enhance driver behaviour and Public Acceptance of Connected and Autonomous vehicLes (PAsCAL) Project \cite{PAsCAL} is focusing on studying the opinions and expectations of general public towards the autonomous vehicles and connected driving technologies.  

\item Artificial Intelligence based cybersecurity for connected and automated vehicles (CARAMEL) project \cite{CARAMEL} is developing AI/ML based anti-hacking intrusion detection and prevention systems for automotive industry including autonomous vehicles. It considers the security impact of several novel technological directions such as 5G, autopilots, and smart charging.

\item Next generation connectivity for enhanced, safe and efficient transport and logistics (5G-Blueprint) project \cite{5gblueprint} is focusing on 5G based  tele-operated cross-border transport and logistics.

\item Unmanned Aerial Vehicle Vertical Applications' Trials Leveraging Advanced 5G Facilities (5G!Drones) project\cite{5GDrones} is focusing on trial several UAV use-cases covering eMBB, URLLC, and mMTC 5G services.
\end{itemize}

%

\subsection{Standards Developing Organizations (SDOs)}

Standardization activities are the key to define technological requirements of autonomous vehicles and also define possible 5G technologies to realize these requirements. This section summarises the several key global level standardization activities related to the 5G autonomous vehicles.

\subsubsection{European Commission (EC)}

EC supports the development of  autonomous vehicles to realize the Connected and Automated Mobility (CAM) across the Europe\footnote{Shaping Europe's digital future \url{https://digital-strategy.ec.europa.eu/en/policies/connected-and-automated-mobility}}.  This initiative aims on developing a safer, user-friendly, greener and more-efficient European-level transport and mobility systems. EC supports the introduction and deployment of  autonomous vehicles in different levels such as 
\begin{itemize}
\item European-level policy development
\item European-level standard development
\item Providing funding for research and innovation projects
\item Developing EU-level legislation 
\end{itemize}

\begin{table}[h!]
	\renewcommand{\arraystretch}{1.1}
	\caption{Important standardization efforts by EC related to 5G and B5G Autonomous Vehicles
	}
	\label{table:specsEC}
\resizebox{\columnwidth}{!}{%
	\begin{tabular}{|p{3.5cm}|p{2.7cm}|p{2cm}|}
		\hline
 \textbf{Standards document/ Specification/Publication} & \textbf{Topics/Description} & \textbf{Update/publi-cation date}\\
				\hline \hline
 Certificate Policy for Deployment and Operation of European Cooperative Intelligent Transport Systems (C-ITS)\cite{EC1}
		        & Presents certificate policy defined by European C-ITS Trust model 
		        & 2018-06 
		        \\
		        
		         \hline
 C-ITS Point of Contact (CPOC) Protocol \cite{EC2}
		        & Describes the CPOC Protocol  of the EU C-ITS Security Credential Management System (EU CCMS).
		        & 2019-01 
		        \\
 \hline
		        
 Security Policy and Governance Framework for Deployment and Operation of European Cooperative Intelligent Transport Systems (C-ITS) \cite{EC3}
		        & Presents the governance framework and security policies for  the European C-ITS system 
		        & 2017-12
		        \\

		         \hline
		     COM (2016) 766: A European strategy on Cooperative Intelligent Transport Systems, a milestone towards cooperative, connected and automated mobility \cite{EC4}
		        & Presents EC's strategy on C-ITS towards connected and automated mobility
		        & 2016-11
		        \\
		        \hline

	\end{tabular}
	}
\vspace{1em}

\textit{\textbf{Note:} We consider the latest update date of the document stated in the SDO repositories. There may be more recent in-progress yet unpublished versions due to ongoing SDO work group efforts.}
\end{table}

\begin{table*}[h!]
	\renewcommand{\arraystretch}{1.1}
	\caption{Recent important white papers by 5GAA related to 5G and B5G Autonomous Vehicles
	}
	\label{table:specs5GAA}
\resizebox{\textwidth}{!}{%
	\begin{tabular}{|p{6.5cm}|p{9cm}|p{2cm}|}
		\hline
		\textbf{Publication} & \textbf{Topics/Description} & \textbf{Update/publi-cation date}\\
				\hline \hline

		         Safety Treatment in V2X Applications\cite{5GAA1}
		        & Present the important safety aspect of 5G based C-V2X applications.
		        & 2021-07
		        \\
		        \hline
		        
		         Cooperation Models enabling deployment and use of 5G infrastructures for CAM in Europe \cite{5GAA2}
		        & Illustrates five different cooperation models that can be used to deploy 5G infrastructures for CAM. 
		        & 2021-03
		        \\
		        \hline
		        
		         C-V2X Use Cases Volume II: Examples and Service Level Requirements \cite{5GAA3}
		        &  Present examples of 5G enabled C-V2X use cases and their Service Level Requirements (SLR)
		        & 2020-10
		        \\
		        \hline
		        
		          A Visionary Roadmap for Advanced Driving Use Cases, Connectivity Technologies, and Radio Spectrum Needs \cite{5GAA4}
		        &  Presents the advancement of 5G to achieve  road safety,  traffic efficiency, energy efficiency, and compatibility in CAM.
		        & 2020-09
		        \\
		        \hline
		        
		          5GAA Efficient Security Provisioning System \cite{5GAA5}
		        &  Present the design specifications of  'Efficient Security Provisioning System' (ESPS) for 5G V2X systems.
		        & 2020-05
		        \\
		        \hline
		        
		         Making 5G Proactive and Predictive for the Automotive Industry \cite{5GAA6}
		        &  Describes and specifies the concept of predictive QoS in the context of 5G CAM systems
		        & 2020-08
		        \\
		        \hline
		        
		         C-V2X Use Cases: Methodology, Examples and Service Level Requirements \cite{5GAA7} 
		        &  Present examples of 5G enabled C-V2X use cases and their Service Level Requirements (SLR)
		        & 2019-07
		        \\
		        \hline
		        
		         C-V2X Conclusions based on Evaluation of Available Architectural Options \cite{5GAA8} 
		        &  Illustrates and analyses the architectural options two C-V2X communication  use cases, i.,e.  Intersection Movement Assist (IMA) and  Vulnerable User (VRU) Discovery.
		        & 2019-02
		        \\
		        \hline
		        
		          Benefits of using existing cellular networks for the delivery of C-ITS \cite{5GAA9}
		        &  Presents the benefits of using telecommunication networks for C-ITS services.
		        & 2019-01
		        \\
		        \hline
		        
		          Toward fully connected vehicles: Edge computing for advanced automotive communications \cite{5GAA10}
		        &  Provides an overview of edge Computing enabled autonomous vehicle use cases.
		        & 2017-12
		        \\
		        \hline
		        
		         The cost-benefit analysis on cellular vehicle-to-everything (C-V2X) technology and its evolution to 5G-V2X \cite{5GAA11}
		        &  Presents the cost-benefit analysis of relevant to the deployment of C-V2X and 5G-V2X in Europe.
		        & 2017-12
		        \\
		        \hline
		        
		          The Case for Cellular V2X for Safety and Cooperative Driving \cite{5GAA12}
		        &  Elaborates necessity of  C-V2X for connected autonomous driving.
		        & 2016-11
		        \\
		        \hline


	\end{tabular}
	}
\vspace{1em}

\textit{\textbf{Note:} We consider the latest update date of the document stated in the SDO repositories. There may be more recent in-progress yet unpublished versions due to ongoing SDO work group efforts.}
\end{table*}

\begin{table*}[h!]
	\renewcommand{\arraystretch}{1.3}
	\caption{Recent important standardization efforts by ESTI related to 5G and B5G Autonomous Vehicles
	}
	\label{table:specsESTI}
\resizebox{\textwidth}{!}{%
	\begin{tabular}{|p{8cm}|p{7cm}|p{2.5cm}|}
		\hline
		  \textbf{Standards document/Specification/Publication} & \textbf{Topics/Description} & \textbf{Update/publi-cation date}\\
				\hline \hline

		        TR 103 099:	Intelligent Transport Systems (ITS); Architecture of conformance validation framework \cite{ESTI1}
		        & Present the base architectural specifications  of conformance validation framework to support the LTE-V2X access technology.  & 2020-06 \\
		        
		        \hline
		        
		          TR 103 193: 	Intelligent Transport Systems (ITS); Testing; Interoperability test specifications for ITS V2X use cases; Architecture of ITS Interoperability Validation Framework \cite{ESTI2}
		        &  Present the updated architecture of ITS interoperability validation framework 
		        & 2020-06\\
		        
		        \hline
		        
		          		TR 101 607:	Intelligent Transport Systems (ITS); Cooperative ITS (C-ITS); Release 2  \cite{ESTI3}
		        &   Present the updated system requirements of Cooperative ITS (C-ITS) system.
		        & 2021-01\\
		        
		        \hline
		        
		        	TS 103 141:	Intelligent Transport Systems (ITS); Facilities layer; Communication congestion control \cite{ESTI4}
		        &  Defines the specifications of facilities layer algorithms. 
		        & 2021-06 \\
		        
		        \hline
		        
		       TR 102 638: Intelligent Transport Systems (ITS); Use cases; Description \cite{ESTI5}
		        &  Describes the ITS use cases studied by ESTI  ITS TC. 
		        & 2021-05\\
		        
		        \hline
		        
		        	TS 103 324: Intelligent Transport Systems (ITS); Cooperative Perception Services \cite{ESTI6}
		        &  Defines the syntax and semantics of  message and data formats of the Cooperative Perception Service (CPS) for ITS applications. 
		        & 2021-05\\
		        
		        \hline
		        
		        	TS 103 561:	Intelligent Transport Systems (ITS); Vehicular Communications; Basic Set of Applications; Maneuver Coordination Service \cite{ESTI7} &
		        	Specify the maneuver coordination service to support automatic driving functions. 
		        & 2018-01\\
		        
		        \hline

		       	TR 103 578:	Intelligent Transport Systems (ITS); Vehicular Communications; Informative report for the Maneuver Coordination Service \cite{ESTI8}
		        &  	Analyzes the maneuver coordination service to support automatic driving functions.
		        & 2020-10\\
		        
		        \hline

		        	TS 102 894:	Intelligent Transport Systems (ITS); Users and applications requirements;  \cite{ESTI9}
		        &  	Specify the  facilities layer Protocol Data Unit and Service Data Unit. 
		        & 2019-10\\
		        
		        \hline

		        	TS 103 724:	Intelligent Transport Systems (ITS); Facilities layer function; Interference Management Zone Message (IMZM); Release 2 \cite{ESTI10}
		        &  Specify an "Interference Management Zone" (IMZ) message and protocol.
		        & 2021-07\\
		        
		        \hline

		        		TS 103 696:	Intelligent Transport System (ITS); Communication Architecture for Multi-Channel Operation (MCO) \cite{ESTI12} 
		        &  	Specify the ITS architectural extensions to enable Multi-Channel Operation (MCO) operation. 
		        & 2021-06 \\ \hline

		         TR 102 962: Intelligent Transport Systems (ITS); Framework for Public Mobile Networks in Cooperative ITS (C-ITS); Release 2 \cite{ESTI13}
		        &  Defines framework for public mobile networks in C-ITS
		        & 2021-02\\ \hline

		        	TR 103 439: Intelligent Transport Systems (ITS); Multi Channel Operation study \cite{ESTI14}
		        &  Defines the requirements of multi channel operations to support ITS use cases.
		        & 2021-07\\ \hline
		        
	\end{tabular}
	}
\vspace{1em}

\textit{\textbf{Note:} We consider the latest update date of the document stated in the SDO repositories. There may be more recent in-progress yet unpublished versions due to ongoing SDO work group efforts.}
\end{table*}

EC also has special interest in unitizing 5G technologies for autonomous vehicles and connected mobility. In 2017, EC has initiated  the task of building designate 5G cross-border corridors with the support of  29 signatory countries. These 5G cross-border corridors can be used to test and demonstrate the EU level automated driving projects\footnote{5G cross-border corridors \url{https://digital-strategy.ec.europa.eu/en/policies/cross-border-corridors}}. EU H2020 program has funded three 5G cross-border corridor projects (i.e. 5G-CARMEN \cite{5gcarmen}, 5GCroCo \cite{5GCroco} and 5G-MOBIX \cite{5G-MOBIX}) to realize this vision.

In 2016, EC launched a new high level group for the automotive industry called GEAR 2030\footnote{Commission launches GEAR 2030 to boost competitiveness and growth in the automotive sector \url{https://ec.europa.eu/growth/content/commission-launches-gear-2030-boost-competitiveness-and-growth-}
\url{automotive-sector-0_en}}. It consist of members such as EU commissioners, representatives from member states and stakeholders from automotive, telecommunications and insurance sectors. This GEAR 2030 group is focusing on developing coherent EU level policy, legal and public support framework for autonomous  vehicles.

In addition, EC supports the development of  Cooperative Intelligent Transport Systems (C-ITS) which can intelligently and securely share the road users and traffic information \footnote{Intelligent transport systems Cooperative, connected and automated mobility (CCAM) \url{https://ec.europa.eu/transport/themes/its/c-its_en}}. These information are useful for autonomous  vehicles  to take more informed decision and  efficiently coordinate their actions. Under EU funding programs, EU is offer funding for research and innovation projects to utilize and also further developing this C-ITS framework.
Table \ref{table:specsEC} presents the important standardization efforts by EC.

\subsubsection{European Automotive - Telecom Alliance (EATA)}

As a result of EU commission's  high level round table discussion on CAM and  autonomous  vehicles, EATA has formed  to promote the the development of EU level autonomous  vehicles activities\cite{EATA}. The vision of EATA is to support the  collaboration between stakeholders in automotive and telecommunication sectors  to explore and  accelerate the deployment of connected and automated driving in across the Europe. EATA mainly focuses on addressing regulatory and legislative obstacles on deployment of connected and automated vehicles. It proactively involve in initiating dialogue with national, EU-level as well as global level policy makers to eliminate the potential obstacles and define new  technical and regulatory measures support connected and automated driving. Moreover, it supports  research and innovation projects focusing on automated vehicles by mobilizing public funding for trail sites.

\subsubsection{CAR 2 CAR Communication Consortium(C2C-CC) }

The CAR 2 CAR Communication Consortium (C2C-CC)\cite{car2car} is an global-level organization which consist of members from road operators,  automotive manufactures, IT service providers,  telecommunication operators and research organizations. It was founded in 2002. The vision of C2C-CC is to realize the goal of vision zero, i.e accident free traffic as early as possible. To realize this vision,  C2C-CC is supporting the development of ultra-reliable, robust and matured safety solutions. It also supports the innovations in 5G and wireless technologies with special focus on spectrum efficiency, ad-hoc short-range V2X communications.  It contributes to the EU and global level standardisation activities to  harmonized the development of V2X communication. Moreover, C2C-CC contributes to development of the EC's C-ITS framework.

\subsubsection{{5G Automotive Association (5GAA)}} 

5GAA\cite{5GAA} is the one of the leading global level SDOs which is mainly focus on the integration of    5G with automotive industry. It has cross-industry members from different sectors such as  automotive, IT and  telecommunication. The 5GAA is working on developing 5G as the ultimate platform to enable C-ITS and V2X. 5GAA has established seven Working Groups (WGs), i.e. 

\begin{itemize}
\item \textbf{WG 1:} Use Cases and Technical Requirements
\item \textbf{WG 2:} System Architecture and Solution Development
\item \textbf{WG 3:} Evaluation, Testbeds, and Pilots
\item \textbf{WG 4:} Standards and Spectrum
\item \textbf{WG 5:} Business Models and Go-To-Market Strategies
\item \textbf{WG 6:} Regulatory and Public Affairs
\item \textbf{WG 7:} Security and Privacy
\end{itemize}

These 5GAA WGs develop relevant standards, architectures, frameworks and business cases related to 5G based autonomous vehicles and its applications. Table \ref{table:specs5GAA} presents the important standardization efforts by 5GAA.

\subsubsection{{European Telecommunications Standards Institute (ESTI)}} 
ETSI\cite{ETSI} is one of the world largest telecom SDOs which is focusing of standardization of mobile networks and its services. ETSI initiates a Technical Committee (TC) called Intelligent Transportation Systems (ITS) which has a special focus on automated and connected vehicles\footnote{Automotive Intelligent Transport Systems (ITS)}. ITS TC is working under three main themes i.e.  Cooperative-ITS (C-ITS), automotive radar and anti-collision radar. Under these themes, they develop  standards related overall automotive communication architecture, system management, communication protocols, security and  access layer agnostic protocols. Table \ref{table:specsESTI} presents the important standardization efforts by ESTI.

\subsubsection{{3rd Generation Partnership Project (3GPP)}} 3GPP\cite{3GPP} is another one of the largest telecom SDO which is actually a consortium of  seven other telecommunication SDOs. 3GPP is mainly responsible for developing standards for C-V2X which replaces the Dedicated short-range communications (DSRC) developed in US and C-ITS originated in Europe\footnote{V2X \url{https://www.3gpp.org/v2x}}. Initial C-V2X standard was included in the Release 14 \footnote{3GPP Release 14 \url{https://www.3gpp.org/release-14}} which was published in 2017. This C-V2X standards are developed to as an decisive step towards enabling autonomous driving with the support of future telecom networks such as 5G and beyond. In 2020, NR-V2X as a part of Release 16\footnote{3GPP Release 16 \url{https://www.3gpp.org/release-16}} comes as an improvement to support automated driving. It supports  complex interaction Use cases such as cooperative automated driving.

		        
		        
		        
		        



\subsubsection{{International Telecommunication Union - Telecommunication  (ITU-T)}} 

ITU-T\cite{ITU} is a global level telecom
SDO. In 2019, ITU-T has launched a Focus Group on AI for autonomous and assisted driving (FG-AI4AD) to support the standardization activities of AI-enabled autonomous and assisted driving systems and services\footnote{Focus Group on AI for autonomous and assisted driving (FG-AI4AD) \url{https://www.itu.int/en/ITU-T/focusgroups/ai4ad/Pages/default.aspx}}. This FG is also focusing on harmonize global-level activities to define a minimal performance threshold for AI-enabled autonomous and assisted driving systems.


		        
		        



\subsubsection{{Alliance for Telecommunications Industry Solutions (ATIS)}} 

ATIS\cite{ATIS} is a leading ICT SDO  which is focusing of various technologies such as  5G/6G, intelligent networks, blockchain,  IoT, smart Cities and security. ATIS has formed a Connected Car-Cybersecurity Ad Hoc Group which studies collaboration opportunities between telecoms and automotive sectors in-terms of cybersecurity. Specially, ATIS group analyzes the different types of security threats in connected automated vehicles, and  discuss role of future mobile network to prevent these attacks\cite{AITS1}.


\subsubsection{{5G Americas}} 

5G Americas\cite{5GAmericas} is an SDO based in America which support the advancement 5G and beyond network applications throughout the American countries. 5G Americas identifies and promotes 5G V2X as an critical technology to enable connected and autonomous vehicles.  5G V2X can be used to enable critical information exchange among the autonomous vehicles to improve navigation and situation awareness to  avoid road accidents. In 2018, 5G Americas published a white papers on "Cellular V2X Communications Towards 5G"  \cite{5GAmericas1} which provides insights on the role of emerging 5G technologies to realize advance V2X
communication.


\subsubsection{{Next Generation Mobile Networks (NGMN) Alliance}} 
NGMN\cite{NGMN} is an telecom SDO which is also focusing on 5G and beyond networks. In 2016,  NGMN formed a V2X task force to evaluate V2X technologies to speedup the deployment of the  C-V2X technology. This task force has fuel the cooperation between telecom and automotive sectors to develop new policies and business models. The task force also studies spectrum management, security and privacy aspects of future ITS. 
In 2018, NGMN published a white paper \cite{NGMN1} which presents eight V2X use cases and their technological requirements.

\subsubsection{Other SDOs}
Several other SDOs are listed below which have 5G autonomous vehicles as a minor focus.

\begin{itemize}
\item \textbf{The European Council for Automotive R\&D (EUCAR)}\cite{eucar} is a consortium of vehicle manufacturers which was founded in 1994. The EUCAR is focusing on development of strategies and solutions for future challenges in car industry by defining common frameworks and supporting research and innovation activities.

\item \textbf{5G Alliance for Connected Industries and Automation (5G-ACIA)}\cite{5GACIA} is an global-level SDO mainly focusing on deployment of 5G for Industrial Internet and smart factory applications. 5G-ACIA is also working on research and trail use-cases of 5G-enabled cloud-controlled Automated Guided Vehicles (AGVs) which can be used in industrial environments\footnote{ 5G Alliance for Connected Industries and Automation endorses testbeds for evaluation of industrial 5G use cases \url{https://5g-acia.org/press-releases/5g-alliance-for-connected-industries-and-automation-endorses-testbeds-for-}
\url{evaluation-of-industrial-5g-use-cases/}}.

\end{itemize}

\section{Lessons Learned and Future Research Directions} \label{sec:lesson}

This section discusses the lessons learned and based on these lessons it synthesizes the future research directions that paves the way for the researchers to carry out their research on this domain.

 \subsection{ Technical Aspects }
\subsubsection{Lessons Learned}
AV basically relies on V2X technology and 5G communications to fulfill the requirements like high transmission rate, low latency less than 5ms, reliability, and with good response time to communicate with other vehicles and communication with infrastructure present on the road. Telecommunication and automobile industry are trying to build an innovative ecosystem by integrating 5G in AV with the recent technologies. Recently, 5G Automotive Association (5GAA) a global cross industry released a new roadmap called automotive Cellular V2X technology which received Innovation Award Honour in the ''Vehicle Intelligence and Self-Driving Technology'' category at the CES 2019.In navigation and path planning, 5G is used in local perception to control the short range vehicles with respect to safety, traffic control and energy management parameters. Accurate positioning of vehicles in AV is achieved in object detection using 5G enabled networks. URLLC is designed to handle V2V communication dynamically with ultra-low latency and ultra-high reliability. Real-time decision are made faster in 5G enabled AV. Integration of 5G with V2X helps to visualize the objects and obstacles more quickly. Inorder to provide fast processing and prior decision making, good connectivity is required which is achieved by 5G in AV. emBB provides high quality in bandwidth for vehicular communication. URLLC, mMTC and eMBB in 5G works together to provide faster connection speeds, higher device capacity and lower latency in AV applications.Some of the benefits of using 5G in AV are driving the vehicles at high speed with high bandwidth, information delivery in minimal time limits due to low latency, notification alerts ahead about any hazardous events and provides self-decision using AV integrated with AI

\subsubsection{Possible Future Directions}
Some of the challenges in using 5G in AV are basically 5G spectrum are too expensive and can be purchased by spectrum auctions.The cost incurred in extending the existing network to 5G is too high and mapping network with reference to geo locations is a tedious task. AV using 5G and beyond can be further extended to perform the following activities
\begin{itemize}
\item \textbf{Predictive maintenance:}Drivers can proactively maintain the vehicle to avoid failure during driving. In-vehicle sensors monitors the conditions of the components present inside the vehicle like battery life, fuel pump and starter motor. Data from cloud combined with AI can predict potential maintenance issues before the occurrence of failure. 
\item\textbf{Advanced Infotainment} Customers
 can be provided with additional information by integrating AI with AR, VR and sensors to have realistic experience.
 \item\textbf{Traffic safety service: }
 5G communication devices embedded inside the vehicles collects the information from the environment and pedestrians are stored in the cloud. Analytics on these data warns the drivers about the hazardous roads and traffic congestion which in turn recommends the solution for alternate path.
\end{itemize}
5G network is considered as a key technology to design a driverless vehicles which is one of the most exciting domain in near future. It has become an important technology in automobile industry integrated with the telecom industry to provide a best experience to the customer. With the advancement in the technologies like V2X and wireless communication, new generation of driverless vehicles is going to drive the automobile industry globally.

 \subsection{B5G technologies }

\subsubsection{Lessons Learned}
Apart from large bandwidth requirements, several functionalities of AVs like object detection, lane detection, collision avoidance, navigation, V2V/V2X/V2I communications require several features like reduced latency, improved physical layer infrastructure, privacy \& security for seamless AD. The supporting technologies of B5G such as MEC, network slicing, SDN/NFV, 5GNR, blockchain, FL, and ZSM can help in providing these requirements to AVs. However, several challenges such as generation of labels in real time for training the ML algorithms, lack of justification of the predictions/recommendations of ML algorithms, huge dimensionality of the data generated, etc. have to be addressed to realize the full potential of B5G for AV. Also, the autonomous vehicles may be connected to multiple networks on the road. Due to mobility of vehicles, it is likely that an AV may move out of coverage area of the access network and may have to join another network. Availability of availability of MEC for multiple access technologies is a significant challenge. Another significant challenge is the management of huge data generated from the AVs that are interconnected in real time \cite{shah20185g}.

\subsubsection{Possible Future Directions}
Explainable Artificial Intelligence (XAI) can be adopted in B5G to address for trustability/justification/explainability of the decisions/predictions from ML algorithms in applications such as object detection, lane detection, collision avoidance, etc \cite{arrieta2020explainable,wang2021explainable,srivastava2022xai}. For instance, consider that the AV is taking a human to the destination. Suddenly, the AI/ML model gives suggestion to the AV to take another route, which is different than the intended route. In these situations, the AV should be in a position to explain/justify the actions it has taken to avoid a particular route or choosing a particular route. XAI can help the humans to understand the actions of AVs in these kinds of situations.   Unsupervised ML models can work on unlabelled data, that can address the challenge of big data generated by AV in 5GB era \cite{kaur2021machine}. Several dimensionality reduction techniques can be used to extract the most important features from the large volumes of data generated in real time \cite{reddy2020analysis}. Handover problems associated with the AVs registered to one network and requiring to join another network during the mobility can be addressed by open radio access network (Open RAN) platform \cite{hewavithana2022overcoming}.  

 \subsection{Security Concerns in AVs}
\subsubsection{Lessons Learned} 
From the security point of view, as highlighted in Table \ref{tab:cyberattack}, it can observed that compared to In-Vehicle communication, the vulnerabilities brought by V2X technologies are more. V2X technologies will widen the attack surfaces thereby increasing the level of threats for the AVs. All the basic security goals i.e. confidentiality, integrity, availability and authentication needs to be addressed while incorporating V2X technologies with AVs. There is need of effective countermeasures to mitigate the threats arising from three main attack categories i.e. spoofing, denial of service, injection and physical attacks.
\subsubsection{Possible Future Directions}
The potential future directions can be categorised based on basic security goals as mentioned above especially with the main focus on V2X technologies. A strong and robust authentication mechanisms will be required for AVs to communicate with trusted entities while moving from one place to the other. An authentication mechanism with high latency and computing power requirement might provide attackers ample time to carry out spoofing or man-in-the-middle attack. Similarly, most of the protocols within AVs do not use encryption, how to secure communication within vehicle and while interacting with other entities outside the vehicles also seems interesting future research direction. Also, there will be huge amount of data generated by AVs, storing that data while preserving privacy of users and making that data available to legitimate users is yet another interesting research direction.
 \subsection{Projects}
\subsubsection{Lessons Learned}

Today, many research projects and SDO activities are focusing on the development of 5G and B5G networks to support the deployment of autonomous vehicles. These activities primarily focus on technical developments of 5G/b5G related technologies such as  MEC, NS, 5G NR, ZSM, and AI. Significantly, NS and MEC got particular focus as these technologies can play a critical role in deploying autonomous vehicle-related applications and services. In addition, there are a significant number of research projects and standardization activities contributing to relevant 5G and B5G technical aspects such as path planning, mobility, service migration, security, and privacy. However, it might take a few more years for these standardizations to be fully deployed by different stakeholders such as mobile operators, governments, regulators, manufacturers, and third-party service providers. Thus, the continuous cooperation between different working groups (i.e., academia and industry) and SDOs are highly required in the coming years. Especially, funding frameworks such as H2020 by European Commission (EC) fuel these activities.

\subsubsection{Possible Future Directions}

In the future, there are two primary questions to be addressed. First, how to address the lack of AV-specific standardization in core 5G  SDOs? Currently, AV is considered just one use case of 5G/B5G networks and still lacks a dedicated focus on AV within the core 5G  SDOs. To resolve this, more AV stakeholders should contribute to 5G SDOs, and more AV-related sub-groups would be formed within core 5G  SDOs to develop dedicated AV-related standards.  

Second, how do we integrate and cooperate between
different working groups and SDOs? The strong cooperation between AV SDOs and core 5G  SDOs would be further encouraged. This will resolve some of the issues related to the first issue. More joint research and SDOs activities should be planned in the future.

\subsection{Emerging and Future Research Directions related to AV} \label{sec:AVFD}
5G network is considered as a key technology to design a driverless vehicles which is one of the most exciting domain in near future. It has become important technology in automobile industry integrated with the telecom industry to provide a best experience to the customer. With the advancement in the technologies like V2X and wireless communication, new generation of driverless vehicles is going to drive the automobile industry globally.
\subsubsection{Quantum Computing}
Quantum computing is the next future generation of automotive technology. Electric vehicles are significant part of quantum revolution. Automobile manufactures have started taking advantage of quantum computers to solve various automotive problems. AI in AV requires large amount of data for analysing and providing optimal response in dynamic situations. For detecting real time car locations and designing the optimal path requires high computing power and speed in AI. The former feature, high computing power can achieved by quantum computers. German automobile company Volkswagen collaborated with D-Wave systems to design and develop traffic routing in Beijing based on quantum computing systems. These systems can also solve optimization problems like waiting time, deployment of fleets etc., Volkswagen has also partnered with google to predict the state of traffic to avoid accident and to simulate the behaviour of electrical component and embed AI in driverless cars. As the AV are more vulnerable to outside world, security breaches can be solved by quantum security. AV requires tremendous computing powers like optimized route planning and change the entire transport systems into smart systems. The cars will become smarter by communicating among themselves and outside world. More developments are expected in the field of AV integrated with quantum computers to achieve the benefits of computing and processing power.

\subsubsection{Cognitive Cloud}
Cognitive AI and algorithms would help us to simulate human-level performance specifically in level 5 AV.  Satisfying the level 5 AV is a tedious task as it needs accurate decision making, object detection and localization under uncertain conditions like fog, rain and extreme darkness. Cognitive computing enhances the model accuracy to achieve closeness to humanlike performance in object detection and decision making. Integration of Cognitive computing in AV leads to improved safety and accuracy. Cognitive Internet of Vehicles allows the AV to focus on what, how and where to compute dynamically closer to human brain.

\section{Conclusion}
In this study, several aspects of AVs such as its features, levels of automation, architecture, key-enabling technologies  and  requirements for autonomus vehicular communication were discussed. Key requirements in terms of latency, security level, privacy, bandwidth, mobility, scalability, availability and reliability for potential AV applications (navigation and path planning, object detection, URLLC, mMTC, eMBB) were also identified. Several emerging technologies such as MEC, SDN and others were studied in detail and impact of 5G/B2G on these technologies was discussed. We also identified key security concerns in AVs with respect to 5G/B2G technology and highlighted  recent standardization efforts by different organisation. Finally, several key research challenges and future research directions were also identified and discussed.

\bibliographystyle{IEEEtran}
\bibliography{Ref}

\begin{thebibliography}{100}
\providecommand{\url}[1]{#1}
\csname url@samestyle\endcsname
\providecommand{\newblock}{\relax}
\providecommand{\bibinfo}[2]{#2}
\providecommand{\BIBentrySTDinterwordspacing}{\spaceskip=0pt\relax}
\providecommand{\BIBentryALTinterwordstretchfactor}{4}
\providecommand{\BIBentryALTinterwordspacing}{\spaceskip=\fontdimen2\font plus
\BIBentryALTinterwordstretchfactor\fontdimen3\font minus
  \fontdimen4\font\relax}
\providecommand{\BIBforeignlanguage}[2]{{%
\expandafter\ifx\csname l@#1\endcsname\relax
\typeout{** WARNING: IEEEtran.bst: No hyphenation pattern has been}%
\typeout{** loaded for the language `#1'. Using the pattern for}%
\typeout{** the default language instead.}%
\else
\language=\csname l@#1\endcsname
\fi
#2}}
\providecommand{\BIBdecl}{\relax}
\BIBdecl

\bibitem{manfreda2019autonomous}
A.~Manfreda, K.~Ljubi, and A.~Groznik, ``Autonomous vehicles in the smart city
  era: An empirical study of adoption factors important for millennials,''
  \emph{International Journal of Information Management}, p. 102050, 2019.

\bibitem{ravi2020driver}
C.~Ravi, A.~Tigga, G.~T. Reddy, S.~Hakak, and M.~Alazab, ``Driver
  identification using optimized deep learning model in smart transportation,''
  \emph{ACM Transactions on Internet Technology}, 2020.

\bibitem{jadaan2017connected}
K.~Jadaan, S.~Zeater, and Y.~Abukhalil, ``Connected vehicles: an innovative
  transport technology,'' \emph{Procedia Engineering}, vol. 187, pp. 641--648,
  2017.

\bibitem{wadud2016help}
Z.~Wadud, D.~MacKenzie, and P.~Leiby, ``Help or hindrance? the travel, energy
  and carbon impacts of highly automated vehicles,'' \emph{Transportation
  Research Part A: Policy and Practice}, vol.~86, pp. 1--18, 2016.

\bibitem{sachs20185g}
J.~Sachs, G.~Wikstrom, T.~Dudda, R.~Baldemair, and K.~Kittichokechai, ``5g
  radio network design for ultra-reliable low-latency communication,''
  \emph{IEEE network}, vol.~32, no.~2, pp. 24--31, 2018.

\bibitem{silva20205g}
M.~M.~d. Silva and J.~Guerreiro, ``On the 5g and beyond,'' \emph{Applied
  Sciences}, vol.~10, no.~20, p. 7091, 2020.

\bibitem{rasouli2019autonomous}
A.~Rasouli and J.~K. Tsotsos, ``Autonomous vehicles that interact with
  pedestrians: A survey of theory and practice,'' \emph{IEEE transactions on
  intelligent transportation systems}, vol.~21, no.~3, pp. 900--918, 2019.

\bibitem{ma2020artificial}
Y.~Ma, Z.~Wang, H.~Yang, and L.~Yang, ``Artificial intelligence applications in
  the development of autonomous vehicles: a survey,'' \emph{IEEE/CAA Journal of
  Automatica Sinica}, vol.~7, no.~2, pp. 315--329, 2020.

\bibitem{ahangar2021survey}
M.~N. Ahangar, Q.~Z. Ahmed, F.~A. Khan, and M.~Hafeez, ``A survey of autonomous
  vehicles: Enabling communication technologies and challenges,''
  \emph{Sensors}, vol.~21, no.~3, p. 706, 2021.

\bibitem{storck2020survey}
C.~R. Storck and F.~Duarte-Figueiredo, ``A survey of 5g technology evolution,
  standards, and infrastructure associated with vehicle-to-everything
  communications by internet of vehicles,'' \emph{IEEE Access}, vol.~8, pp.
  117\,593--117\,614, 2020.

\bibitem{navarro2020survey}
J.~Navarro-Ortiz, P.~Romero-Diaz, S.~Sendra, P.~Ameigeiras, J.~J. Ramos-Munoz,
  and J.~M. Lopez-Soler, ``A survey on 5g usage scenarios and traffic models,''
  \emph{IEEE Communications Surveys \& Tutorials}, vol.~22, no.~2, pp.
  905--929, 2020.

\bibitem{pham2021survey}
M.~Pham and K.~Xiong, ``A survey on security attacks and defense techniques for
  connected and autonomous vehicles,'' \emph{Computers \& Security}, p. 102269,
  2021.

\bibitem{khazraeian2019intelligent}
S.~Khazraeian and M.~Hadi, ``Intelligent transportation systems in future smart
  cities,'' in \emph{Sustainable Interdependent Networks II}.\hskip 1em plus
  0.5em minus 0.4em\relax Springer, 2019, pp. 109--120.

\bibitem{shladover2018connected}
S.~E. Shladover, ``Connected and automated vehicle systems: Introduction and
  overview,'' \emph{Journal of Intelligent Transportation Systems}, vol.~22,
  no.~3, pp. 190--200, 2018.

\bibitem{hussain2018autonomous}
R.~Hussain and S.~Zeadally, ``Autonomous cars: Research results, issues, and
  future challenges,'' \emph{IEEE Communications Surveys \& Tutorials},
  vol.~21, no.~2, pp. 1275--1313, 2018.

\bibitem{contreras2017internet}
J.~Contreras-Castillo, S.~Zeadally, and J.~A. Guerrero-Iba{\~n}ez, ``Internet
  of vehicles: architecture, protocols, and security,'' \emph{IEEE internet of
  things Journal}, vol.~5, no.~5, pp. 3701--3709, 2018.

\bibitem{zhang2018vehicular}
S.~Zhang, J.~Chen, F.~Lyu, N.~Cheng, W.~Shi, and X.~Shen, ``Vehicular
  communication networks in the automated driving era,'' \emph{IEEE
  Communications Magazine}, vol.~56, no.~9, pp. 26--32, 2018.

\bibitem{taxonomy2016definitions}
S.~Taxonomy, ``Definitions for terms related to driving automation systems for
  on-road motor vehicles (j3016),'' Technical report, Society for Automotive
  Engineering, Tech. Rep., 2016.

\bibitem{jameel2019internet}
F.~Jameel, Z.~Chang, J.~Huang, and T.~Ristaniemi, ``Internet of autonomous
  vehicles: architecture, features, and socio-technological challenges,''
  \emph{IEEE Wireless Communications}, vol.~26, no.~4, pp. 21--29, 2019.

\bibitem{yurtsever2020survey}
E.~Yurtsever, J.~Lambert, A.~Carballo, and K.~Takeda, ``A survey of autonomous
  driving: Common practices and emerging technologies,'' \emph{IEEE access},
  vol.~8, pp. 58\,443--58\,469, 2020.

\bibitem{van2018autonomous}
J.~Van~Brummelen, M.~O’Brien, D.~Gruyer, and H.~Najjaran, ``Autonomous
  vehicle perception: The technology of today and tomorrow,''
  \emph{Transportation research part C: emerging technologies}, vol.~89, pp.
  384--406, 2018.

\bibitem{gonzalez2015review}
D.~Gonz{\'a}lez, J.~P{\'e}rez, V.~Milan{\'e}s, and F.~Nashashibi, ``A review of
  motion planning techniques for automated vehicles,'' \emph{IEEE Transactions
  on Intelligent Transportation Systems}, vol.~17, no.~4, pp. 1135--1145, 2015.

\bibitem{amer2017modelling}
N.~H. Amer, H.~Zamzuri, K.~Hudha, and Z.~A. Kadir, ``Modelling and control
  strategies in path tracking control for autonomous ground vehicles: a review
  of state of the art and challenges,'' \emph{Journal of intelligent \& robotic
  systems}, vol.~86, no.~2, pp. 225--254, 2017.

\bibitem{guanetti2018control}
J.~Guanetti, Y.~Kim, and F.~Borrelli, ``Control of connected and automated
  vehicles: State of the art and future challenges,'' \emph{Annual reviews in
  control}, vol.~45, pp. 18--40, 2018.

\bibitem{joy2018internet}
J.~Joy, V.~Rabsatt, and M.~Gerla, ``Internet of vehicles: Enabling safe,
  secure, and private vehicular crowdsourcing,'' \emph{Internet Technology
  Letters}, vol.~1, no.~1, p. e16, 2018.

\bibitem{jameel2018interference}
F.~Jameel, S.~Wyne, M.~A. Javed, and S.~Zeadally, ``Interference-aided
  vehicular networks: Future research opportunities and challenges,''
  \emph{IEEE Communications Magazine}, vol.~56, no.~10, pp. 36--42, 2018.

\bibitem{irani2018localizability}
B.~Irani, J.~Wang, and W.~Chen, ``A localizability constraint-based path
  planning method for autonomous vehicles,'' \emph{IEEE Transactions on
  Intelligent Transportation Systems}, vol.~20, no.~7, pp. 2593--2604, 2018.

\bibitem{jiang2019probabilistic}
Z.~Jiang and S.~A. Raziei, ``A probabilistic decision engine for navigation of
  autonomous vehicles under uncertainty,'' in \emph{2019 IEEE National
  Aerospace and Electronics Conference (NAECON)}.\hskip 1em plus 0.5em minus
  0.4em\relax IEEE, 2019, pp. 131--138.

\bibitem{jardine2018robust}
P.~T. Jardine and S.~N. Givigi, ``A robust model-predictive guidance system for
  autonomous vehicles in cluttered environments,'' \emph{IEEE Systems Journal},
  vol.~13, no.~2, pp. 2034--2045, 2018.

\bibitem{lin2021autonomous}
H.-Y. Lin and X.-Z. Peng, ``Autonomous quadrotor navigation with vision based
  obstacle avoidance and path planning,'' \emph{IEEE Access}, vol.~9, pp.
  102\,450--102\,459, 2021.

\bibitem{yan2021reinforcement}
N.~Yan, S.~Huang, and C.~Kong, ``Reinforcement learning-based autonomous
  navigation and obstacle avoidance for usvs under partially observable
  conditions,'' \emph{Mathematical Problems in Engineering}, vol. 2021, 2021.

\bibitem{chu2015real}
K.~Chu, J.~Kim, K.~Jo, and M.~Sunwoo, ``Real-time path planning of autonomous
  vehicles for unstructured road navigation,'' \emph{International Journal of
  Automotive Technology}, vol.~16, no.~4, pp. 653--668, 2015.

\bibitem{marin2018global}
P.~Marin-Plaza, A.~Hussein, D.~Martin, and A.~d.~l. Escalera, ``Global and
  local path planning study in a ros-based research platform for autonomous
  vehicles,'' \emph{Journal of Advanced Transportation}, vol. 2018, 2018.

\bibitem{hoang2015path}
V.-D. Hoang and K.-H. Jo, ``Path planning for autonomous vehicle based on
  heuristic searching using online images,'' \emph{Vietnam Journal of Computer
  Science}, vol.~2, no.~2, pp. 109--120, 2015.

\bibitem{alharbi2020global}
M.~Alharbi and H.~A. Karimi, ``A global path planner for safe navigation of
  autonomous vehicles in uncertain environments,'' \emph{Sensors}, vol.~20,
  no.~21, p. 6103, 2020.

\bibitem{coronado2019enabling}
E.~Coronado, G.~Cebrian-Marquez, and R.~Riggio, ``Enabling computation
  offloading for autonomous and assisted driving in 5g networks,'' in
  \emph{2019 IEEE Global Communications Conference (GLOBECOM)}.\hskip 1em plus
  0.5em minus 0.4em\relax IEEE, 2019, pp. 1--6.

\bibitem{piperigkos20205g}
N.~Piperigkos, A.~S. Lalos, K.~Berberidis, C.~Laoudias, and K.~Moustakas,
  ``{5G} enabled cooperative localization of connected and semi-autonomous
  vehicles via sparse laplacian processing,'' in \emph{2020 22nd International
  Conference on Transparent Optical Networks (ICTON)}.\hskip 1em plus 0.5em
  minus 0.4em\relax IEEE, 2020, pp. 1--4.

\bibitem{yu2020deep}
K.~Yu, L.~Lin, M.~Alazab, L.~Tan, and B.~Gu, ``Deep learning-based traffic
  safety solution for a mixture of autonomous and manual vehicles in a
  5g-enabled intelligent transportation system,'' \emph{IEEE transactions on
  intelligent transportation systems}, 2020.

\bibitem{nguyen2020enhancing}
V.-L. Nguyen, P.-C. Lin, and R.-H. Hwang, ``Enhancing misbehavior detection in
  {5G} vehicle-to-vehicle communications,'' \emph{IEEE Transactions on
  Vehicular Technology}, vol.~69, no.~9, pp. 9417--9430, 2020.

\bibitem{mukherjee2018energy}
A.~Mukherjee, ``Energy efficiency and delay in 5g ultra-reliable low-latency
  communications system architectures,'' \emph{IEEE Network}, vol.~32, no.~2,
  pp. 55--61, 2018.

\bibitem{popovski2018wireless}
P.~Popovski, J.~J. Nielsen, C.~Stefanovic, E.~De~Carvalho, E.~Strom, K.~F.
  Trillingsgaard, A.-S. Bana, D.~M. Kim, R.~Kotaba, J.~Park \emph{et~al.},
  ``Wireless access for ultra-reliable low-latency communication: Principles
  and building blocks,'' \emph{Ieee Network}, vol.~32, no.~2, pp. 16--23, 2018.

\bibitem{alsenwi2019chance}
M.~Alsenwi, S.~R. Pandey, Y.~K. Tun, K.~T. Kim, and C.~S. Hong, ``A chance
  constrained based formulation for dynamic multiplexing of embb-urllc traffics
  in 5g new radio,'' in \emph{2019 International Conference on Information
  Networking (ICOIN)}.\hskip 1em plus 0.5em minus 0.4em\relax IEEE, 2019, pp.
  108--113.

\bibitem{song2019performance}
X.~Song and M.~Yuan, ``Performance analysis of one-way highway vehicular
  networks with dynamic multiplexing of embb and urllc traffics,'' \emph{IEEE
  Access}, vol.~7, pp. 118\,020--118\,029, 2019.

\bibitem{bockelmann2016massive}
C.~Bockelmann, N.~Pratas, H.~Nikopour, K.~Au, T.~Svensson, C.~Stefanovic,
  P.~Popovski, and A.~Dekorsy, ``Massive machine-type communications in 5g:
  Physical and mac-layer solutions,'' \emph{IEEE Communications Magazine},
  vol.~54, no.~9, pp. 59--65, 2016.

\bibitem{d2013its}
P.~M. d'Orey and M.~Ferreira, ``Its for sustainable mobility: A survey on
  applications and impact assessment tools,'' \emph{IEEE Transactions on
  Intelligent Transportation Systems}, vol.~15, no.~2, pp. 477--493, 2013.

\bibitem{zheng2020cooperative}
Y.~Zheng, Y.~Zhang, B.~Ran, Y.~Xu, and X.~Qu, ``Cooperative control strategies
  to stabilise the freeway mixed traffic stability and improve traffic
  throughput in an intelligent roadside system environment,'' \emph{IET
  Intelligent Transport Systems}, vol.~14, no.~9, pp. 1108--1115, 2020.

\bibitem{kaiwartya2016internet}
O.~Kaiwartya, A.~H. Abdullah, Y.~Cao, A.~Altameem, M.~Prasad, C.-T. Lin, and
  X.~Liu, ``Internet of vehicles: Motivation, layered architecture, network
  model, challenges, and future aspects,'' \emph{IEEE Access}, vol.~4, pp.
  5356--5373, 2016.

\bibitem{lin2015potential}
J.-R. Lin, T.~Talty, and O.~K. Tonguz, ``On the potential of bluetooth low
  energy technology for vehicular applications,'' \emph{IEEE Communications
  Magazine}, vol.~53, no.~1, pp. 267--275, 2015.

\bibitem{lee2017latency}
K.~Lee, J.~Kim, Y.~Park, H.~Wang, and D.~Hong, ``Latency of cellular-based v2x:
  Perspectives on tti-proportional latency and tti-independent latency,''
  \emph{IEEE Access}, vol.~5, pp. 15\,800--15\,809, 2017.

\bibitem{zhu2018energy}
K.~Zhu, X.~Xu, and S.~Han, ``Energy-efficient uav trajectory planning for data
  collection and computation in mmtc networks,'' in \emph{2018 IEEE Globecom
  Workshops (GC Wkshps)}.\hskip 1em plus 0.5em minus 0.4em\relax IEEE, 2018,
  pp. 1--6.

\bibitem{dougan2018ofdm}
S.~Do{\u{g}}an, A.~Tusha, and H.~Arslan, ``Ofdm with index modulation for
  asynchronous mmtc networks,'' \emph{Sensors}, vol.~18, no.~4, p. 1280, 2018.

\bibitem{tanwar2019tactile}
S.~Tanwar, S.~Tyagi, I.~Budhiraja, and N.~Kumar, ``Tactile internet for
  autonomous vehicles: Latency and reliability analysis,'' \emph{IEEE Wireless
  Communications}, vol.~26, no.~4, pp. 66--72, 2019.

\bibitem{abuin2020complexity}
A.~Abuin, E.~Iradier, L.~Fanari, J.~Montalban, and P.~Angueira, ``Complexity
  reduction techniques for noma-based rrm algorithms in 5g networks,'' in
  \emph{2020 IEEE International Conference on Electrical Engineering and
  Photonics (EExPolytech)}.\hskip 1em plus 0.5em minus 0.4em\relax IEEE, 2020,
  pp. 86--89.

\bibitem{prabadevi2021toward}
B.~Prabadevi, N.~Deepa, Q.-V. Pham, D.~C. Nguyen, T.~Reddy, P.~N. Pathirana,
  O.~Dobre \emph{et~al.}, ``Toward blockchain for edge-of-things: A new
  paradigm, opportunities, and future directions,'' \emph{IEEE Internet of
  Things Magazine}, 2021.

\bibitem{li2022computation}
M.~Li, N.~Mao, X.~Zheng, and T.~R. Gadekallu, ``Computation offloading in edge
  computing based on deep reinforcement learning,'' in \emph{Proceedings of
  International Conference on Computing and Communication Networks: ICCCN
  2021}, vol. 394.\hskip 1em plus 0.5em minus 0.4em\relax Springer Nature,
  2022, p. 339.

\bibitem{liyanage2021driving}
M.~Liyanage, P.~Porambage, A.~Y. Ding, and A.~Kalla, ``Driving forces for
  multi-access edge computing (mec) iot integration in 5g,'' \emph{ICT
  Express}, 2021.

\bibitem{coronado2020enabling}
E.~Coronado, G.~Cebri{\'a}n-M{\'a}rquez, and R.~Riggio, ``Enabling autonomous
  and connected vehicles at the 5g network edge,'' in \emph{2020 6th IEEE
  Conference on Network Softwarization (NetSoft)}.\hskip 1em plus 0.5em minus
  0.4em\relax IEEE, 2020, pp. 350--352.

\bibitem{zhdanenko2019demonstration}
O.~Zhdanenko, J.~Liu, R.~Torre, S.~Mudriievskiy, H.~Salah, G.~T. Nguyen, and
  H.~F. Fitzek, ``Demonstration of mobile edge cloud for 5g connected cars,''
  in \emph{2019 16th IEEE Annual Consumer Communications \& Networking
  Conference (CCNC)}.\hskip 1em plus 0.5em minus 0.4em\relax IEEE, 2019, pp.
  1--2.

\bibitem{zhou2018mec}
S.~Zhou, P.~P. Netalkar, Y.~Chang, Y.~Xu, and J.~Chao, ``The mec-based
  architecture design for low-latency and fast hand-off vehicular networking,''
  in \emph{2018 IEEE 88th Vehicular Technology Conference (VTC-Fall)}.\hskip
  1em plus 0.5em minus 0.4em\relax IEEE, 2018, pp. 1--7.

\bibitem{giust2018multi}
F.~Giust, V.~Sciancalepore, D.~Sabella, M.~C. Filippou, S.~Mangiante,
  W.~Featherstone, and D.~Munaretto, ``Multi-access edge computing: The driver
  behind the wheel of 5g-connected cars,'' \emph{IEEE Communications Standards
  Magazine}, vol.~2, no.~3, pp. 66--73, 2018.

\bibitem{campolo2019mec}
C.~Campolo, A.~Iera, A.~Molinaro, and G.~Ruggeri, ``Mec support for 5g-v2x use
  cases through docker containers,'' in \emph{2019 IEEE Wireless Communications
  and Networking Conference (WCNC)}.\hskip 1em plus 0.5em minus 0.4em\relax
  IEEE, 2019, pp. 1--6.

\bibitem{barmpounakis2020collision}
S.~Barmpounakis, G.~Tsiatsios, M.~Papadakis, E.~Mitsianis, N.~Koursioumpas, and
  N.~Alonistioti, ``Collision avoidance in 5g using mec and nfv: The vulnerable
  road user safety use case,'' \emph{Computer Networks}, vol. 172, p. 107150,
  2020.

\bibitem{coronado2019addressing}
E.~Coronado, G.~Cebri{\'a}n-M{\'a}rquez, G.~Baggio, and R.~Riggio, ``Addressing
  bitrate and latency requirements for connected and autonomous vehicles,'' in
  \emph{IEEE INFOCOM 2019-IEEE Conference on Computer Communications Workshops
  (INFOCOM WKSHPS)}.\hskip 1em plus 0.5em minus 0.4em\relax IEEE, 2019, pp.
  961--962.

\bibitem{lee2020multiaccess}
J.~Lee, S.~Kang, J.~Jeon, and I.~Chun, ``Multiaccess edge computing-based
  simulation as a service for 5g mobile applications: a case study of tollgate
  selection for autonomous vehicles,'' \emph{Wireless Communications and Mobile
  Computing}, vol. 2020, 2020.

\bibitem{shi20185g}
Y.~Shi, Y.~Pan, Z.~Zhang, Y.~Li, and Y.~Xiao, ``A 5g-v2x based collaborative
  motion planning for autonomous industrial vehicles at road intersections,''
  in \emph{2018 IEEE International Conference on Systems, Man, and Cybernetics
  (SMC)}.\hskip 1em plus 0.5em minus 0.4em\relax IEEE, 2018, pp. 3744--3748.

\bibitem{rasheed2020application}
A.~Rasheed, A.~Anwar, K.~K. Sudheera, P.~H. Chong, W.~Liu, M.~Yaqub, and
  M.~Jafri, ``Application-aware hierarchical offloading for mec-enabled
  autonomous vehicle architecture,'' in \emph{2020 IEEE Globecom Workshops (GC
  Wkshps}.\hskip 1em plus 0.5em minus 0.4em\relax IEEE, 2020, pp. 1--6.

\bibitem{ma2020cooperative}
H.~Ma, S.~Li, E.~Zhang, Z.~Lv, J.~Hu, and X.~Wei, ``Cooperative autonomous
  driving oriented mec-aided 5g-v2x: Prototype system design, field tests and
  ai-based optimization tools,'' \emph{IEEE Access}, vol.~8, pp.
  54\,288--54\,302, 2020.

\bibitem{lian2020semantic}
Y.~Lian, L.~Qian, L.~Ding, F.~Yang, and Y.~Guan, ``Semantic fusion
  infrastructure for unmanned vehicle system based on cooperative 5g mec,'' in
  \emph{2020 IEEE/CIC International Conference on Communications in China
  (ICCC)}.\hskip 1em plus 0.5em minus 0.4em\relax IEEE, 2020, pp. 202--207.

\bibitem{sabella2020mec}
D.~Sabella, D.~Brevi, E.~Bonetto, A.~Ranjan, A.~Manzalini, and D.~Salerno,
  ``Mec-based infotainment services for smart roads in 5g environments,'' in
  \emph{2020 IEEE 91st Vehicular Technology Conference (VTC2020-Spring)}.\hskip
  1em plus 0.5em minus 0.4em\relax IEEE, 2020, pp. 1--6.

\bibitem{3gpp2017study}
3GPP, ``Study on management and orchestration of network slicing for next
  generation network,'' 2017.

\bibitem{alliance2016Description}
N.~Alliance, ``{Description of Network Slicing Concept},'' \emph{Ngmn}, pp.
  1--7, 2016.

\bibitem{wijethilaka2021survey}
S.~Wijethilaka and M.~Liyanage, ``Survey on network slicing for internet of
  things realization in 5g networks,'' \emph{IEEE Communications Surveys and
  Tutorials}, 2021.

\bibitem{huang20205g}
S.~Huang, B.~Guo, and Y.~Liu, ``5g-oriented optical underlay network slicing
  technology and challenges,'' \emph{IEEE Communications Magazine}, vol.~58,
  no.~2, pp. 13--19, 2020.

\bibitem{campolo20175g}
C.~Campolo, A.~Molinaro, A.~Iera, and F.~Menichella, ``5g network slicing for
  vehicle-to-everything services,'' \emph{IEEE Wireless Communications},
  vol.~24, no.~6, pp. 38--45, 2017.

\bibitem{khan2021network}
H.~Khan, P.~Luoto, S.~Samarakoon, M.~Bennis, and M.~Latva-Aho, ``Network
  slicing for vehicular communication,'' \emph{Transactions on Emerging
  Telecommunications Technologies}, vol.~32, no.~1, p. e3652, 2021.

\bibitem{khan2018application}
H.~Khan, P.~Luoto, M.~Bennis, and M.~Latva-aho, ``On the application of network
  slicing for 5g-v2x,'' in \emph{European Wireless 2018; 24th European Wireless
  Conference}.\hskip 1em plus 0.5em minus 0.4em\relax VDE, 2018, pp. 1--6.

\bibitem{sanchez2019empowering}
R.~Sanchez-Iborra, J.~Santa, J.~Gallego-Madrid, S.~Covaci, and A.~Skarmeta,
  ``Empowering the internet of vehicles with multi-rat 5g network slicing,''
  \emph{Sensors}, vol.~19, no.~14, p. 3107, 2019.

\bibitem{mlika2021network}
Z.~Mlika and S.~Cherkaoui, ``Network slicing with mec and deep reinforcement
  learning for the internet of vehicles,'' \emph{IEEE Network}, vol.~35, no.~3,
  pp. 132--138, 2021.

\bibitem{xiong2019smart}
K.~Xiong, S.~Leng, J.~Hu, X.~Chen, and K.~Yang, ``Smart network slicing for
  vehicular fog-rans,'' \emph{IEEE Transactions on Vehicular Technology},
  vol.~68, no.~4, pp. 3075--3085, 2019.

\bibitem{mahmood2019software}
A.~Mahmood, W.~E. Zhang, and Q.~Z. Sheng, ``Software-defined heterogeneous
  vehicular networking: The architectural design and open challenges,''
  \emph{Future Internet}, vol.~11, no.~3, p.~70, 2019.

\bibitem{goudarzi2020dynamic}
S.~Goudarzi, M.~H. Anisi, H.~Ahmadi, and L.~Musavian, ``Dynamic resource
  allocation model for distribution operations using sdn,'' \emph{IEEE Internet
  of Things Journal}, vol.~8, no.~2, pp. 976--988, 2020.

\bibitem{garg2019sdn}
S.~Garg, K.~Kaur, G.~Kaddoum, S.~H. Ahmed, and D.~N.~K. Jayakody, ``Sdn-based
  secure and privacy-preserving scheme for vehicular networks: A 5g
  perspective,'' \emph{IEEE Transactions on Vehicular Technology}, vol.~68,
  no.~9, pp. 8421--8434, 2019.

\bibitem{duan2017sdn}
X.~Duan, Y.~Liu, and X.~Wang, ``Sdn enabled 5g-vanet: Adaptive vehicle
  clustering and beamformed transmission for aggregated traffic,'' \emph{IEEE
  Communications Magazine}, vol.~55, no.~7, pp. 120--127, 2017.

\bibitem{garg2019mobqos}
S.~Garg, K.~Kaur, S.~H. Ahmed, A.~Bradai, G.~Kaddoum, and M.~Atiquzzaman,
  ``Mobqos: Mobility-aware and qos-driven sdn framework for autonomous
  vehicles,'' \emph{IEEE Wireless Communications}, vol.~26, no.~4, pp. 12--20,
  2019.

\bibitem{cheng2020vehicular}
X.~Cheng, Z.~Huang, and S.~Chen, ``Vehicular communication channel measurement,
  modelling, and application for beyond 5g and 6g,'' \emph{IET Communications},
  vol.~14, no.~19, pp. 3303--3311, 2020.

\bibitem{duan2020emerging}
W.~Duan, J.~Gu, M.~Wen, G.~Zhang, Y.~Ji, and S.~Mumtaz, ``Emerging technologies
  for 5g-iov networks: Applications, trends and opportunities,'' \emph{IEEE
  Network}, vol.~34, no.~5, pp. 283--289, 2020.

\bibitem{boban2017use}
M.~Boban, A.~Kousaridas, K.~Manolakis, J.~Eichinger, and W.~Xu, ``Use cases,
  requirements, and design considerations for 5g v2x,'' \emph{arXiv preprint
  arXiv:1712.01754}, 2017.

\bibitem{dahlman20205g}
E.~Dahlman, S.~Parkvall, and J.~Skold, \emph{5G NR: The next generation
  wireless access technology}.\hskip 1em plus 0.5em minus 0.4em\relax Academic
  Press, 2020.

\bibitem{bagheri20215g}
H.~Bagheri, M.~Noor-A-Rahim, Z.~Liu, H.~Lee, D.~Pesch, K.~Moessner, and
  P.~Xiao, ``5g nr-v2x: Toward connected and cooperative autonomous driving,''
  \emph{IEEE Communications Standards Magazine}, vol.~5, no.~1, pp. 48--54,
  2021.

\bibitem{albonda2019efficient}
H.~D.~R. Albonda and J.~P{\'e}rez-Romero, ``An efficient ran slicing strategy
  for a heterogeneous network with embb and v2x services,'' \emph{IEEE access},
  vol.~7, pp. 44\,771--44\,782, 2019.

\bibitem{pokhrel2020towards}
S.~R. Pokhrel, J.~Ding, J.~Park, O.-S. Park, and J.~Choi, ``Towards enabling
  critical mmtc: A review of urllc within mmtc,'' \emph{IEEE Access}, vol.~8,
  pp. 131\,796--131\,813, 2020.

\bibitem{ashraf2018dynamic}
M.~I. Ashraf, C.-F. Liu, M.~Bennis, W.~Saad, and C.~S. Hong, ``Dynamic resource
  allocation for optimized latency and reliability in vehicular networks,''
  \emph{IEEE Access}, vol.~6, pp. 63\,843--63\,858, 2018.

\bibitem{anwar2019physical}
W.~Anwar, N.~Franchi, and G.~Fettweis, ``Physical layer evaluation of v2x
  communications technologies: 5g nr-v2x, lte-v2x, ieee 802.11 bd, and ieee
  802.11 p,'' in \emph{2019 IEEE 90th Vehicular Technology Conference
  (VTC2019-Fall)}.\hskip 1em plus 0.5em minus 0.4em\relax IEEE, 2019, pp. 1--7.

\bibitem{mei2019intelligent}
J.~Mei, X.~Wang, and K.~Zheng, ``Intelligent network slicing for v2x services
  toward 5g,'' \emph{IEEE Network}, vol.~33, no.~6, pp. 196--204, 2019.

\bibitem{husain2019ultra}
S.~S. Husain, A.~Kunz, A.~Prasad, E.~Pateromichelakis, and K.~Samdanis,
  ``Ultra-high reliable 5g v2x communications,'' \emph{IEEE Communications
  Standards Magazine}, vol.~3, no.~2, pp. 46--52, 2019.

\bibitem{ashraf2020supporting}
S.~A. Ashraf, R.~Blasco, H.~Do, G.~Fodor, C.~Zhang, and W.~Sun, ``Supporting
  vehicle-to-everything services by 5g new radio release-16 systems,''
  \emph{IEEE Communications Standards Magazine}, vol.~4, no.~1, pp. 26--32,
  2020.

\bibitem{ganesan2020nr}
K.~Ganesan, J.~Lohr, P.~B. Mallick, A.~Kunz, and R.~Kuchibhotla, ``Nr sidelink
  design overview for advanced v2x service,'' \emph{IEEE Internet of Things
  Magazine}, vol.~3, no.~1, pp. 26--30, 2020.

\bibitem{lien20203gpp}
S.-Y. Lien, D.-J. Deng, C.-C. Lin, H.-L. Tsai, T.~Chen, C.~Guo, and S.-M.
  Cheng, ``3gpp nr sidelink transmissions toward 5g v2x,'' \emph{IEEE Access},
  vol.~8, pp. 35\,368--35\,382, 2020.

\bibitem{gyawali2020challenges}
S.~Gyawali, S.~Xu, Y.~Qian, and R.~Q. Hu, ``Challenges and solutions for
  cellular based v2x communications,'' \emph{IEEE Communications Surveys \&
  Tutorials}, 2020.

\bibitem{chen2017vehicle}
S.~Chen, J.~Hu, Y.~Shi, Y.~Peng, J.~Fang, R.~Zhao, and L.~Zhao,
  ``Vehicle-to-everything (v2x) services supported by lte-based systems and
  5g,'' \emph{IEEE Communications Standards Magazine}, vol.~1, no.~2, pp.
  70--76, 2017.

\bibitem{lu2020feasibility}
Y.~Lu, M.~Gerasimenko, R.~Kovalchukov, M.~Stusek, J.~Urama, J.~Hosek,
  M.~Valkama, and E.~S. Lohan, ``Feasibility of location-aware handover for
  autonomous vehicles in industrial multi-radio environments,'' \emph{Sensors},
  vol.~20, no.~21, p. 6290, 2020.

\bibitem{rasheed2021intelligent}
I.~Rasheed and F.~Hu, ``Intelligent super-fast vehicle-to-everything 5g
  communications with predictive switching between mmwave and thz links,''
  \emph{Vehicular Communications}, vol.~27, p. 100303, 2021.

\bibitem{kampert2020millimeter}
E.~Kampert, C.~Schettler, R.~Woodman, P.~A. Jennings, and M.~D. Higgins,
  ``Millimeter-wave communication for a last-mile autonomous transport
  vehicle,'' \emph{IEEE Access}, vol.~8, pp. 8386--8392, 2020.

\bibitem{manias2021making}
D.~M. Manias and A.~Shami, ``Making a case for federated learning in the
  internet of vehicles and intelligent transportation systems,'' \emph{IEEE
  Network}, vol.~35, no.~3, pp. 88--94, 2021.

\bibitem{savazzi2021opportunities}
S.~Savazzi, M.~Nicoli, M.~Bennis, S.~Kianoush, and L.~Barbieri, ``Opportunities
  of federated learning in connected, cooperative, and automated industrial
  systems,'' \emph{IEEE Communications Magazine}, vol.~59, no.~2, pp. 16--21,
  2021.

\bibitem{du2020federated}
Z.~Du, C.~Wu, T.~Yoshinaga, K.-L.~A. Yau, Y.~Ji, and J.~Li, ``Federated
  learning for vehicular internet of things: Recent advances and open issues,''
  \emph{IEEE Open Journal of the Computer Society}, vol.~1, pp. 45--61, 2020.

\bibitem{pokhrel2020improving}
S.~R. Pokhrel and J.~Choi, ``Improving tcp performance over wifi for internet
  of vehicles: A federated learning approach,'' \emph{IEEE Transactions on
  Vehicular Technology}, vol.~69, no.~6, pp. 6798--6802, 2020.

\bibitem{lu2020blockchain}
Y.~Lu, X.~Huang, K.~Zhang, S.~Maharjan, and Y.~Zhang, ``Blockchain empowered
  asynchronous federated learning for secure data sharing in internet of
  vehicles,'' \emph{IEEE Transactions on Vehicular Technology}, vol.~69, no.~4,
  pp. 4298--4311, 2020.

\bibitem{lu2020federated}
Y.~Lu, X.~Huang, Y.~Dai, S.~Maharjan, and Y.~Zhang, ``Federated learning for
  data privacy preservation in vehicular cyber-physical systems,'' \emph{IEEE
  Network}, vol.~34, no.~3, pp. 50--56, 2020.

\bibitem{kong2021fedvcp}
X.~Kong, H.~Gao, G.~Shen, G.~Duan, and S.~K. Das, ``Fedvcp: A
  federated-learning-based cooperative positioning scheme for social internet
  of vehicles,'' \emph{IEEE Transactions on Computational Social Systems},
  2021.

\bibitem{gadekallu2021federated}
T.~R. Gadekallu, Q.-V. Pham, T.~Huynh-The, S.~Bhattacharya, P.~K.~R.
  Maddikunta, and M.~Liyanage, ``Federated learning for big data: A survey on
  opportunities, applications, and future directions,'' \emph{arXiv preprint
  arXiv:2110.04160}, 2021.

\bibitem{gao2019blockchain}
J.~Gao, K.~O.-B.~O. Agyekum, E.~B. Sifah, K.~N. Acheampong, Q.~Xia, X.~Du,
  M.~Guizani, and H.~Xia, ``A blockchain-sdn-enabled internet of vehicles
  environment for fog computing and 5g networks,'' \emph{IEEE Internet of
  Things Journal}, vol.~7, no.~5, pp. 4278--4291, 2019.

\bibitem{xie2019blockchain}
L.~Xie, Y.~Ding, H.~Yang, and X.~Wang, ``Blockchain-based secure and
  trustworthy internet of things in sdn-enabled 5g-vanets,'' \emph{IEEE
  Access}, vol.~7, pp. 56\,656--56\,666, 2019.

\bibitem{bera2020blockchain}
B.~Bera, S.~Saha, A.~K. Das, N.~Kumar, P.~Lorenz, and M.~Alazab,
  ``Blockchain-envisioned secure data delivery and collection scheme for
  5g-based iot-enabled internet of drones environment,'' \emph{IEEE
  Transactions on Vehicular Technology}, vol.~69, no.~8, pp. 9097--9111, 2020.

\bibitem{aloqaily2021design}
M.~Aloqaily, O.~Bouachir, A.~Boukerche, and I.~Al~Ridhawi, ``Design guidelines
  for blockchain-assisted 5g-uav networks,'' \emph{IEEE Network}, vol.~35,
  no.~1, pp. 64--71, 2021.

\bibitem{gupta2021blockchain}
R.~Gupta, S.~Tanwar, and N.~Kumar, ``Blockchain and 5g integrated softwarized
  uav network management: Architecture, solutions, and challenges,''
  \emph{Physical Communication}, vol.~47, p. 101355, 2021.

\bibitem{jian2021blockchain}
X.~Jian, P.~Leng, Y.~Wang, M.~Alrashoud, and M.~S. Hossain,
  ``Blockchain-empowered trusted networking for unmanned aerial vehicles in the
  b5g era,'' \emph{IEEE Network}, vol.~35, no.~1, pp. 72--77, 2021.

\bibitem{feng2021efficient}
C.~Feng, K.~Yu, A.~K. Bashir, Y.~D. Al-Otaibi, Y.~Lu, S.~Chen, and D.~Zhang,
  ``Efficient and secure data sharing for 5g flying drones: a
  blockchain-enabled approach,'' \emph{IEEE Network}, vol.~35, no.~1, pp.
  130--137, 2021.

\bibitem{ghimire2021sharding}
B.~Ghimire, D.~B. Rawat, C.~Liu, and J.~Li, ``Sharding-enabled blockchain for
  software-defined internet of unmanned vehicles in the battlefield,''
  \emph{IEEE Network}, vol.~35, no.~1, pp. 101--107, 2021.

\bibitem{gumaei2021deep}
A.~Gumaei, M.~Al-Rakhami, M.~M. Hassan, P.~Pace, G.~Alai, K.~Lin, and
  G.~Fortino, ``Deep learning and blockchain with edge computing for 5g-enabled
  drone identification and flight mode detection,'' \emph{IEEE Network},
  vol.~35, no.~1, pp. 94--100, 2021.

\bibitem{nguyen2021federated}
D.~C. Nguyen, M.~Ding, Q.-V. Pham, P.~N. Pathirana, L.~B. Le, A.~Seneviratne,
  J.~Li, D.~Niyato, and H.~V. Poor, ``Federated learning meets blockchain in
  edge computing: Opportunities and challenges,'' \emph{IEEE Internet of Things
  Journal}, vol.~8, no.~16, pp. 12\,806--12\,825, 2021.

\bibitem{kumar2021ppsf}
P.~Kumar, R.~Kumar, G.~Srivastava, G.~P. Gupta, R.~Tripathi, T.~R. Gadekallu,
  and N.~Xiong, ``Ppsf: A privacy-preserving and secure framework using
  blockchain-based machine-learning for iot-driven smart cities,'' \emph{IEEE
  Transactions on Network Science and Engineering}, 2021.

\bibitem{bouraga2021taxonomy}
S.~Bouraga, ``A taxonomy of blockchain consensus protocols: A survey and
  classification framework,'' \emph{Expert Systems with Applications}, vol.
  168, p. 114384, 2021.

\bibitem{ortega2018trusted}
V.~Ortega, F.~Bouchmal, and J.~F. Monserrat, ``Trusted 5g vehicular networks:
  Blockchains and content-centric networking,'' \emph{IEEE Vehicular Technology
  Magazine}, vol.~13, no.~2, pp. 121--127, 2018.

\bibitem{shrestha2020evolution}
R.~Shrestha, S.~Y. Nam, R.~Bajracharya, and S.~Kim, ``Evolution of v2x
  communication and integration of blockchain for security enhancements,''
  \emph{Electronics}, vol.~9, no.~9, p. 1338, 2020.

\bibitem{rahmadika2019blockchain}
S.~Rahmadika, K.~Lee, and K.-H. Rhee, ``Blockchain-enabled 5g autonomous
  vehicular networks,'' in \emph{2019 International Conference on Sustainable
  Engineering and Creative Computing (ICSECC)}.\hskip 1em plus 0.5em minus
  0.4em\relax IEEE, 2019, pp. 275--280.

\bibitem{reebadiya2021blockchain}
D.~Reebadiya, T.~Rathod, R.~Gupta, S.~Tanwar, and N.~Kumar, ``Blockchain-based
  secure and intelligent sensing scheme for autonomous vehicles activity
  tracking beyond 5g networks,'' \emph{Peer-to-Peer Networking and
  Applications}, pp. 1--18, 2021.

\bibitem{nkenyereye2020secure}
L.~Nkenyereye, B.~Adhi~Tama, M.~K. Shahzad, and Y.-H. Choi, ``Secure and
  blockchain-based emergency driven message protocol for 5g enabled vehicular
  edge computing,'' \emph{Sensors}, vol.~20, no.~1, p. 154, 2020.

\bibitem{chen2021exploiting}
J.~Chen, W.~Wang, Y.~Zhou, S.~H. Ahmed, and W.~Wei, ``Exploiting 5g and
  blockchain for medical applications of drones,'' \emph{IEEE Network},
  vol.~35, no.~1, pp. 30--36, 2021.

\bibitem{kakkavas2021network}
G.~Kakkavas, A.~Stamou, V.~Karyotis, and S.~Papavassiliou, ``Network tomography
  for efficient monitoring in sdn-enabled 5g networks and beyond: Challenges
  and opportunities,'' \emph{IEEE Communications Magazine}, vol.~59, no.~3, pp.
  70--76, 2021.

\bibitem{benzaid2020ai}
C.~Benzaid and T.~Taleb, ``Ai-driven zero touch network and service management
  in 5g and beyond: Challenges and research directions,'' \emph{IEEE Network},
  vol.~34, no.~2, pp. 186--194, 2020.

\bibitem{liyanage2022survey}
M.~Liyanage, Q.-V. Pham, K.~Dev, S.~Bhattacharya, P.~K.~R. Maddikunta, T.~R.
  Gadekallu, and G.~Yenduri, ``A survey on zero touch network and service (zsm)
  management for 5g and beyond networks,'' \emph{Journal of Network and
  Computer Applications}, p. 103362, 2022.

\bibitem{cui2019review}
J.~Cui, L.~S. Liew, G.~Sabaliauskaite, and F.~Zhou, ``A review on safety
  failures, security attacks, and available countermeasures for autonomous
  vehicles,'' \emph{Ad Hoc Networks}, vol.~90, p. 101823, 2019.

\bibitem{malla2013security}
A.~M. Malla and R.~K. Sahu, ``Security attacks with an effective solution for
  dos attacks in vanet,'' \emph{International Journal of Computer
  Applications}, vol.~66, no.~22, 2013.

\bibitem{manvi2017survey}
S.~S. Manvi and S.~Tangade, ``A survey on authentication schemes in vanets for
  secured communication,'' \emph{Vehicular Communications}, vol.~9, pp. 19--30,
  2017.

\bibitem{othmane2015survey}
L.~B. Othmane, H.~Weffers, M.~M. Mohamad, and M.~Wolf, ``A survey of security
  and privacy in connected vehicles,'' in \emph{Wireless sensor and mobile
  ad-hoc networks}.\hskip 1em plus 0.5em minus 0.4em\relax Springer, 2015, pp.
  217--247.

\bibitem{zhang2017secure}
S.~Zhang, Y.~Lin, Q.~Liu, J.~Jiang, B.~Yin, and K.-K.~R. Choo, ``Secure hitch
  in location based social networks,'' \emph{Computer Communications}, vol.
  100, pp. 65--77, 2017.

\bibitem{raya2007securing}
M.~Raya and J.-P. Hubaux, ``Securing vehicular ad hoc networks,'' \emph{Journal
  of computer security}, vol.~15, no.~1, pp. 39--68, 2007.

\bibitem{petit2015remote}
J.~Petit, B.~Stottelaar, M.~Feiri, and F.~Kargl, ``Remote attacks on automated
  vehicles sensors: Experiments on camera and lidar,'' \emph{Black Hat Europe},
  vol.~11, no. 2015, p. 995, 2015.

\bibitem{stottelaar2015practical}
B.~G. Stottelaar, ``Practical cyber-attacks on autonomous vehicles,'' Master's
  thesis, University of Twente, 2015.

\bibitem{chauhan2014platform}
R.~Chauhan, \emph{A platform for false data injection in frequency modulated
  continuous wave radar}.\hskip 1em plus 0.5em minus 0.4em\relax Utah State
  University, 2014.

\bibitem{buehler2014airborne}
W.~E. Buehler, R.~M. Whitson, and M.~J. Lewis, ``Airborne radar jamming
  system,'' Sep.~9 2014, uS Patent 8,830,112.

\bibitem{fairfield2011traffic}
N.~Fairfield and C.~Urmson, ``Traffic light mapping and detection,'' in
  \emph{2011 IEEE International Conference on Robotics and Automation}.\hskip
  1em plus 0.5em minus 0.4em\relax IEEE, 2011, pp. 5421--5426.

\bibitem{hillel2014recent}
A.~B. Hillel, R.~Lerner, D.~Levi, and G.~Raz, ``Recent progress in road and
  lane detection: a survey,'' \emph{Machine vision and applications}, vol.~25,
  no.~3, pp. 727--745, 2014.

\bibitem{wang2019pseudo}
Y.~Wang, W.-L. Chao, D.~Garg, B.~Hariharan, M.~Campbell, and K.~Q. Weinberger,
  ``Pseudo-lidar from visual depth estimation: Bridging the gap in 3d object
  detection for autonomous driving,'' in \emph{Proceedings of the IEEE/CVF
  Conference on Computer Vision and Pattern Recognition}, 2019, pp. 8445--8453.

\bibitem{brown2017adversarial}
T.~B. Brown, D.~Man{\'e}, A.~Roy, M.~Abadi, and J.~Gilmer, ``Adversarial
  patch,'' \emph{arXiv preprint arXiv:1712.09665}, 2017.

\bibitem{kassakian1996automotive}
J.~G. Kassakian, H.-C. Wolf, J.~M. Miller, and C.~J. Hurton, ``Automotive
  electrical systems circa 2005,'' \emph{IEEE spectrum}, vol.~33, no.~8, pp.
  22--27, 1996.

\bibitem{jitpakdee2008neural}
R.~Jitpakdee and T.~Maneewarn, ``Neural networks terrain classification using
  inertial measurement unit for an autonomous vehicle,'' in \emph{2008 SICE
  Annual Conference}.\hskip 1em plus 0.5em minus 0.4em\relax IEEE, 2008, pp.
  554--558.

\bibitem{matsumoto2012method}
T.~Matsumoto, M.~Hata, M.~Tanabe, K.~Yoshioka, and K.~Oishi, ``A method of
  preventing unauthorized data transmission in controller area network,'' in
  \emph{2012 IEEE 75th Vehicular Technology Conference (VTC Spring)}.\hskip 1em
  plus 0.5em minus 0.4em\relax IEEE, 2012, pp. 1--5.

\bibitem{lin2012cyber}
C.-W. Lin and A.~Sangiovanni-Vincentelli, ``Cyber-security for the controller
  area network (can) communication protocol,'' in \emph{2012 International
  Conference on Cyber Security}.\hskip 1em plus 0.5em minus 0.4em\relax IEEE,
  2012, pp. 1--7.

\bibitem{checkoway2011comprehensive}
S.~Checkoway, D.~McCoy, B.~Kantor, D.~Anderson, H.~Shacham, S.~Savage,
  K.~Koscher, A.~Czeskis, F.~Roesner, T.~Kohno \emph{et~al.}, ``Comprehensive
  experimental analyses of automotive attack surfaces.'' in \emph{USENIX
  Security Symposium}, vol.~4, no. 447-462.\hskip 1em plus 0.5em minus
  0.4em\relax San Francisco, 2011, p. 2021.

\bibitem{koscher2010experimental}
K.~Koscher, S.~Savage, F.~Roesner, S.~Patel, T.~Kohno, A.~Czeskis, D.~McCoy,
  B.~Kantor, D.~Anderson, H.~Shacham \emph{et~al.}, ``Experimental security
  analysis of a modern automobile,'' in \emph{2010 IEEE Symposium on Security
  and Privacy}.\hskip 1em plus 0.5em minus 0.4em\relax IEEE Computer Society,
  2010, pp. 447--462.

\bibitem{kosmanos2020novel}
D.~Kosmanos, A.~Pappas, L.~Maglaras, S.~Moschoyiannis, F.~J. Aparicio-Navarro,
  A.~Argyriou, and H.~Janicke, ``A novel intrusion detection system against
  spoofing attacks in connected electric vehicles,'' \emph{Array}, vol.~5, p.
  100013, 2020.

\bibitem{hossain2021observer}
M.~M. Hossain and C.~Peng, ``Observer-based event triggering hlfc for
  multi-area power systems under dos attacks,'' \emph{Information Sciences},
  vol. 543, pp. 437--453, 2021.

\bibitem{fraiji2018cyber}
Y.~Fraiji, L.~B. Azzouz, W.~Trojet, and L.~A. Saidane, ``Cyber security issues
  of internet of electric vehicles,'' in \emph{2018 IEEE Wireless
  Communications and Networking Conference (WCNC)}.\hskip 1em plus 0.5em minus
  0.4em\relax IEEE, 2018, pp. 1--6.

\bibitem{ilahi2021challenges}
I.~Ilahi, M.~Usama, J.~Qadir, M.~U. Janjua, A.~Al-Fuqaha, D.~T. Hoang, and
  D.~Niyato, ``Challenges and countermeasures for adversarial attacks on deep
  reinforcement learning,'' \emph{IEEE Transactions on Artificial
  Intelligence}, vol.~3, no.~2, pp. 90--109, 2021.

\bibitem{pham2021detecting}
T.~N. Pham, A.~M.~T. Oo, and H.~Trinh, ``Detecting and isolating false data
  injection attacks on electric vehicles of smart grids using distributed
  functional observers,'' \emph{IET Generation, Transmission \& Distribution},
  vol.~15, no.~4, pp. 762--779, 2021.

\bibitem{strom2011medium}
E.~G. Strom, ``On medium access and physical layer standards for cooperative
  intelligent transport systems in europe,'' \emph{Proceedings of the IEEE},
  vol.~99, no.~7, pp. 1183--1188, 2011.

\bibitem{van2012impact}
M.~van Eenennaam, A.~van~de Venis, and G.~Karagiannis, ``Impact of ieee 1609.4
  channel switching on the ieee 802.11 p beaconing performance,'' in \emph{2012
  IFIP Wireless Days}.\hskip 1em plus 0.5em minus 0.4em\relax IEEE, 2012, pp.
  1--8.

\bibitem{parkvall2017nr}
S.~Parkvall, E.~Dahlman, A.~Furuskar, and M.~Frenne, ``Nr: The new 5g radio
  access technology,'' \emph{IEEE Communications Standards Magazine}, vol.~1,
  no.~4, pp. 24--30, 2017.

\bibitem{abboud2016interworking}
K.~Abboud, H.~A. Omar, and W.~Zhuang, ``Interworking of dsrc and cellular
  network technologies for v2x communications: A survey,'' \emph{IEEE
  transactions on vehicular technology}, vol.~65, no.~12, pp. 9457--9470, 2016.

\bibitem{5gdrive}
\BIBentryALTinterwordspacing
{``5G HarmoniseD Research and TrIals for serVice Evolution between EU and China
  (5G-DRIVE)"}. [Accessed on 29.07.2021]. [Online]. Available:
  \url{https://5g-drive.eu/}
\BIBentrySTDinterwordspacing

\bibitem{5gcarmen}
\BIBentryALTinterwordspacing
{``5G for Connected and Automated Road Mobility in the European UnioN
  (5G-Carmen)"}. [Accessed on 29.07.2021]. [Online]. Available:
  \url{https://5gcarmen.eu}
\BIBentrySTDinterwordspacing

\bibitem{5GCroco}
\BIBentryALTinterwordspacing
{``Fifth Generation Cross-Border Control (5GCroco)"}. [Accessed on 29.07.2021].
  [Online]. Available: \url{https://5gcroco.eu}
\BIBentrySTDinterwordspacing

\bibitem{5G-MOBIX}
\BIBentryALTinterwordspacing
{``5G for cooperative and connected automated MOBIility on X-border corridors
  (5G-MOBIX)"}. [Accessed on 29.07.2021]. [Online]. Available:
  \url{https://www.5g-mobix.com}
\BIBentrySTDinterwordspacing

\bibitem{ict4cart}
\BIBentryALTinterwordspacing
{``ICT Infrastructure for Connected and Automated Road Transport (ICT4CART)"}.
  [Accessed on 29.07.2021]. [Online]. Available: \url{https://www.ict4cart.eu}
\BIBentrySTDinterwordspacing

\bibitem{car2tera}
\BIBentryALTinterwordspacing
{``Terahertz sensors and networks for next generation smart automotive
  electronic systems (car2TERA)"}. [Accessed on 29.07.2021]. [Online].
  Available: \url{https://car2tera.eu}
\BIBentrySTDinterwordspacing

\bibitem{5GCar}
\BIBentryALTinterwordspacing
{``Fifth Generation Communication Automotive Research and innovation (5GCar)"}.
  [Accessed on 29.07.2021]. [Online]. Available: \url{https://5gcar.eu/}
\BIBentrySTDinterwordspacing

\bibitem{PAsCAL}
\BIBentryALTinterwordspacing
{``Enhance driver behaviour and Public Acceptance of Connected and Autonomous
  vehicLes (PAsCAL)"}. [Accessed on 29.07.2021]. [Online]. Available:
  \url{https://www.pascal-project.eu}
\BIBentrySTDinterwordspacing

\bibitem{CARAMEL}
\BIBentryALTinterwordspacing
{``Artificial Intelligence based cybersecurity for connected and automated
  vehicles (CARAMEL)"}. [Accessed on 29.07.2021]. [Online]. Available:
  \url{https://www.h2020caramel.eu}
\BIBentrySTDinterwordspacing

\bibitem{5gblueprint}
\BIBentryALTinterwordspacing
{``Next generation connectivity for enhanced, safe and efficient transport and
  logistics (5G-Blueprint)"}. [Accessed on 29.07.2021]. [Online]. Available:
  \url{https://www.5gblueprint.eu/}
\BIBentrySTDinterwordspacing

\bibitem{5GDrones}
\BIBentryALTinterwordspacing
{``Unmanned Aerial Vehicle Vertical Applications' Trials Leveraging Advanced 5G
  Facilities (5G!Drones)"}. [Accessed on 29.07.2021]. [Online]. Available:
  \url{https://5gdrones.eu/}
\BIBentrySTDinterwordspacing

\bibitem{EC1}
\BIBentryALTinterwordspacing
``{Certificate Policy for Deployment and Operation of European Cooperative
  Intelligent Transport Systems (C-ITS)},'' 2018, [Accessed on 29.07.2021].
  [Online]. Available:
  \url{https://ec.europa.eu/transport/sites/default/files/c-its_certificate_policy-v1.1.pdf}
\BIBentrySTDinterwordspacing

\bibitem{EC2}
\BIBentryALTinterwordspacing
``{C-ITS Point of Contact (CPOC) Protocol},'' 2019, [Accessed on 29.07.2021].
  [Online]. Available:
  \url{https://ec.europa.eu/transport/sites/default/files/c-its_certificate_policy-v1.1.pdf}
\BIBentrySTDinterwordspacing

\bibitem{EC3}
\BIBentryALTinterwordspacing
``{Security Policy and Governance Framework for Deployment and Operation of
  European Cooperative Intelligent Transport Systems (C-ITS)},'' 2017,
  [Accessed on 29.07.2021]. [Online]. Available:
  \url{https://ec.europa.eu/transport/sites/default/files/c-its_security_policy_release_1.pdf}
\BIBentrySTDinterwordspacing

\bibitem{EC4}
\BIBentryALTinterwordspacing
``{COM (2016) 766: A European strategy on Cooperative Intelligent Transport
  Systems, a milestone towards cooperative, connected and automated
  mobility},'' 2016, [Accessed on 29.07.2021]. [Online]. Available:
  \url{https://eur-lex.europa.eu/legal-content/EN/TXT/?uri=CELEX\%3A52016DC0766}
\BIBentrySTDinterwordspacing

\bibitem{5GAA1}
\BIBentryALTinterwordspacing
``{ Safety Treatment in V2X Applications},'' 2021, [Accessed on 29.07.2021].
  [Online]. Available:
  \url{https://5gaa.org/news/safety-treatment-in-v2x-applications/}
\BIBentrySTDinterwordspacing

\bibitem{5GAA2}
\BIBentryALTinterwordspacing
``{Cooperation Models enabling deployment and use of 5G infrastructures for CAM
  in Europe },'' 2021, [Accessed on 29.07.2021]. [Online]. Available:
  \url{https://5gaa.org/news/cooperation-models-enabling-deployment-and-use-of-5g-infrastructures-for-cam-in-europe/}
\BIBentrySTDinterwordspacing

\bibitem{5GAA3}
\BIBentryALTinterwordspacing
``{C-V2X Use Cases Volume II: Examples and Service Level Requirements },''
  2020, [Accessed on 29.07.2021]. [Online]. Available:
  \url{https://5gaa.org/news/c-v2x-use-cases-volume-ii-examples-and-service-level-requirements/}
\BIBentrySTDinterwordspacing

\bibitem{5GAA4}
\BIBentryALTinterwordspacing
``{A Visionary Roadmap for Advanced Driving Use Cases, Connectivity
  Technologies, and Radio Spectrum Needs },'' 2020, [Accessed on 29.07.2021].
  [Online]. Available:
  \url{https://5gaa.org/news/the-new-c-v2x-roadmap-for-automotive-connectivity/}
\BIBentrySTDinterwordspacing

\bibitem{5GAA5}
\BIBentryALTinterwordspacing
``{5GAA Efficient Security Provisioning System },'' 2020, [Accessed on
  29.07.2021]. [Online]. Available:
  \url{https://5gaa.org/news/5gaa-efficient-security-provisioning-system/}
\BIBentrySTDinterwordspacing

\bibitem{5GAA6}
\BIBentryALTinterwordspacing
``{Making 5G Proactive and Predictive for the Automotive Industry},'' 2020,
  [Accessed on 29.07.2021]. [Online]. Available:
  \url{https://5gaa.org/news/5gaa-releases-white-paper-on-making-5g-proactive-and-predictive-for-the-automotive-industry/}
\BIBentrySTDinterwordspacing

\bibitem{5GAA7}
\BIBentryALTinterwordspacing
``{C-V2X Use Cases: Methodology, Examples and Service Level Requirements},''
  2019, [Accessed on 29.07.2021]. [Online]. Available:
  \url{https://5gaa.org/news/5gaa-releases-white-paper-on-c-v2x-use-cases-methodology-examples-and-service-level-requirements/}
\BIBentrySTDinterwordspacing

\bibitem{5GAA8}
\BIBentryALTinterwordspacing
``{C-V2X Conclusions based on Evaluation of Available Architectural Options},''
  2019, [Accessed on 29.07.2021]. [Online]. Available:
  \url{https://5gaa.org/news/5gaa-releases-white-paper-on-c-v2x-conclusions-based-on-evaluation-of-available-architectural-options/}
\BIBentrySTDinterwordspacing

\bibitem{5GAA9}
\BIBentryALTinterwordspacing
``{Benefits of using existing cellular networks for the delivery of C-ITS},''
  2019, [Accessed on 29.07.2021]. [Online]. Available:
  \url{https://5gaa.org/news/5gaa-releases-white-paper-on-the-benefits-of-using-existing-cellular-networks-for-the-delivery-of-c-its/}
\BIBentrySTDinterwordspacing

\bibitem{5GAA10}
\BIBentryALTinterwordspacing
``{ Toward fully connected vehicles: Edge computing for advanced automotive
  communications},'' 2017, [Accessed on 29.07.2021]. [Online]. Available:
  \url{https://5gaa.org/news/toward-fully-connected-vehicles-edge-computing-for-advanced-automotive-communications/}
\BIBentrySTDinterwordspacing

\bibitem{5GAA11}
\BIBentryALTinterwordspacing
``{The cost-benefit analysis on cellular vehicle-to-everything (C-V2X)
  technology and its evolution to 5G-V2X },'' 2017, [Accessed on 29.07.2021].
  [Online]. Available:
  \url{https://5gaa.org/news/5gaa-study-the-cost-benefit-analysis-on-cellular-vehicle-to-everything-c-v2x-technology-and-its-evolution-to-5g-v2x/}
\BIBentrySTDinterwordspacing

\bibitem{5GAA12}
\BIBentryALTinterwordspacing
``{The Case for Cellular V2X for Safety and Cooperative Driving},'' 2016,
  [Accessed on 29.07.2021]. [Online]. Available:
  \url{https://5gaa.org/news/white-paper-placeholder-news-for-testing/}
\BIBentrySTDinterwordspacing

\bibitem{ESTI1}
\BIBentryALTinterwordspacing
``{TR 103 099: Intelligent Transport Systems (ITS); Architecture of conformance
  validation framework },'' 2020, [Accessed on 29.07.2021]. [Online].
  Available:
  \url{https://portal.etsi.org/webapp/WorkProgram/Report_WorkItem.asp?WKI_ID=59537}
\BIBentrySTDinterwordspacing

\bibitem{ESTI2}
\BIBentryALTinterwordspacing
``{TR 103 193: Intelligent Transport Systems (ITS); Testing; Interoperability
  test specifications for ITS V2X use cases; Architecture of ITS
  Interoperability Validation Framework},'' 2020, [Accessed on 29.07.2021].
  [Online]. Available:
  \url{https://portal.etsi.org/webapp/WorkProgram/Report_WorkItem.asp?WKI_ID=59547}
\BIBentrySTDinterwordspacing

\bibitem{ESTI3}
\BIBentryALTinterwordspacing
``{ TR 101 607: Intelligent Transport Systems (ITS); Cooperative ITS (C-ITS);
  Release 2},'' 2021, [Accessed on 29.07.2021]. [Online]. Available:
  \url{https://portal.etsi.org/webapp/WorkProgram/Report_WorkItem.asp?WKI_ID=62504}
\BIBentrySTDinterwordspacing

\bibitem{ESTI4}
\BIBentryALTinterwordspacing
``{TS 103 141: Intelligent Transport Systems (ITS); Facilities layer;
  Communication congestion control },'' 2021, [Accessed on 29.07.2021].
  [Online]. Available:
  \url{https://portal.etsi.org/webapp/WorkProgram/Report_WorkItem.asp?WKI_ID=37124}
\BIBentrySTDinterwordspacing

\bibitem{ESTI5}
\BIBentryALTinterwordspacing
``{TR 102 638: Intelligent Transport Systems (ITS); Use cases; Description },''
  2021, [Accessed on 29.07.2021]. [Online]. Available:
  \url{https://portal.etsi.org/webapp/WorkProgram/Report_WorkItem.asp?WKI_ID=46538}
\BIBentrySTDinterwordspacing

\bibitem{ESTI6}
\BIBentryALTinterwordspacing
``{ TS 103 324: Intelligent Transport Systems (ITS); Cooperative Perception
  Services},'' 2021, [Accessed on 29.07.2021]. [Online]. Available:
  \url{https://portal.etsi.org/webapp/WorkProgram/Report_WorkItem.asp?WKI_ID=46541}
\BIBentrySTDinterwordspacing

\bibitem{ESTI7}
\BIBentryALTinterwordspacing
``{TS 103 561: Intelligent Transport Systems (ITS); Vehicular Communications;
  Basic Set of Applications; Maneuver Coordination Service },'' 2018, [Accessed
  on 29.07.2021]. [Online]. Available:
  \url{https://portal.etsi.org/webapp/WorkProgram/Report_WorkItem.asp?WKI_ID=53496}
\BIBentrySTDinterwordspacing

\bibitem{ESTI8}
\BIBentryALTinterwordspacing
``{TR 103 578: Intelligent Transport Systems (ITS); Vehicular Communications;
  Informative report for the Maneuver Coordination Service},'' 2020, [Accessed
  on 29.07.2021]. [Online]. Available:
  \url{https://portal.etsi.org/webapp/WorkProgram/Report_WorkItem.asp?WKI_ID=53991}
\BIBentrySTDinterwordspacing

\bibitem{ESTI9}
\BIBentryALTinterwordspacing
``{ TS 102 894: Intelligent Transport Systems (ITS); Users and applications
  requirements; },'' 2019, [Accessed on 29.07.2021]. [Online]. Available:
  \url{https://portal.etsi.org/webapp/WorkProgram/Report_WorkItem.asp?WKI_ID=57406}
\BIBentrySTDinterwordspacing

\bibitem{ESTI10}
\BIBentryALTinterwordspacing
``{ TS 103 724: Intelligent Transport Systems (ITS); Facilities layer function;
  Interference Management Zone Message (IMZM); Release 2},'' 2021, [Accessed on
  29.07.2021]. [Online]. Available:
  \url{https://portal.etsi.org/webapp/WorkProgram/Report_WorkItem.asp?WKI_ID=58665}
\BIBentrySTDinterwordspacing

\bibitem{ESTI12}
\BIBentryALTinterwordspacing
``{TS 103 696: Intelligent Transport System (ITS); Communication Architecture
  for Multi-Channel Operation (MCO) },'' 2021, [Accessed on 29.07.2021].
  [Online]. Available:
  \url{https://portal.etsi.org/webapp/WorkProgram/Report_WorkItem.asp?WKI_ID=56810}
\BIBentrySTDinterwordspacing

\bibitem{ESTI13}
\BIBentryALTinterwordspacing
``{TR 102 962: Intelligent Transport Systems (ITS); Framework for Public Mobile
  Networks in Cooperative ITS (C-ITS); Release 2 },'' 2021, [Accessed on
  29.07.2021]. [Online]. Available:
  \url{https://portal.etsi.org/webapp/WorkProgram/Report_WorkItem.asp?WKI_ID=54678}
\BIBentrySTDinterwordspacing

\bibitem{ESTI14}
\BIBentryALTinterwordspacing
``{ TR 103 439: Intelligent Transport Systems (ITS); Multi Channel Operation
  study},'' 2021, [Accessed on 29.07.2021]. [Online]. Available:
  \url{https://portal.etsi.org/webapp/WorkProgram/Report_WorkItem.asp?WKI_ID=49984}
\BIBentrySTDinterwordspacing

\bibitem{EATA}
\BIBentryALTinterwordspacing
{``European Automotive and Telecoms Alliance"}. [Accessed on 29.07.2021].
  [Online]. Available: \url{https://eata.be/}
\BIBentrySTDinterwordspacing

\bibitem{car2car}
\BIBentryALTinterwordspacing
{``CAR 2 CAR Communication Consortium"}. [Accessed on 29.07.2021]. [Online].
  Available: \url{https://www.car-2-car.org/}
\BIBentrySTDinterwordspacing

\bibitem{5GAA}
\BIBentryALTinterwordspacing
{``5G Automotive Association"}. [Accessed on 29.07.2021]. [Online]. Available:
  \url{https://www.5gaa.org/}
\BIBentrySTDinterwordspacing

\bibitem{ETSI}
\BIBentryALTinterwordspacing
{``European Telecommunications Standards Institute"}. [Accessed on 29.07.2021].
  [Online]. Available: \url{https://www.etsi.org/}
\BIBentrySTDinterwordspacing

\bibitem{3GPP}
\BIBentryALTinterwordspacing
{``3rd Generation Partnership Project"}. [Accessed on 29.07.2021]. [Online].
  Available: \url{https://www.3gpp.org/}
\BIBentrySTDinterwordspacing

\bibitem{ITU}
\BIBentryALTinterwordspacing
{``International Telecommunication Union - Telecommunication (ITU-T)"}.
  [Accessed on 29.07.2021]. [Online]. Available: \url{https://www.itu.int/}
\BIBentrySTDinterwordspacing

\bibitem{ATIS}
\BIBentryALTinterwordspacing
{``Alliance for Telecommunications Industry Solutions"}. [Accessed on
  29.07.2021]. [Online]. Available: \url{https://www.atis.org/}
\BIBentrySTDinterwordspacing

\bibitem{AITS1}
\BIBentryALTinterwordspacing
``{ATIS-I-0000059 Improving Vehicle Cybersecurity: ICT Industry Experience and
  Perspectives },'' 2017, [Accessed on 29.07.2021]. [Online]. Available:
  \url{https://www.atis.org/connected-cars/}
\BIBentrySTDinterwordspacing

\bibitem{5GAmericas}
\BIBentryALTinterwordspacing
{``5G Americas"}. [Accessed on 29.07.2021]. [Online]. Available:
  \url{https://www.5gamericas.org/}
\BIBentrySTDinterwordspacing

\bibitem{5GAmericas1}
\BIBentryALTinterwordspacing
``{Cellular V2X Communications Towards 5G},'' 2018, [Accessed on 29.07.2021].
  [Online]. Available:
  \url{https://www.5gamericas.org/cellular-v2x-communications-towards-5g/}
\BIBentrySTDinterwordspacing

\bibitem{NGMN}
\BIBentryALTinterwordspacing
{``The Next Generation Mobile Networks"}. [Accessed on 29.07.2021]. [Online].
  Available: \url{https://www.ngmn.org/}
\BIBentrySTDinterwordspacing

\bibitem{NGMN1}
\BIBentryALTinterwordspacing
``{V2X White Paper v 1.0},'' 2018, [Accessed on 29.07.2021]. [Online].
  Available:
  \url{https://ngmn.org/wp-content/uploads/V2X_white_paper_v1_0-1.pdf}
\BIBentrySTDinterwordspacing

\bibitem{eucar}
\BIBentryALTinterwordspacing
{``The European Council for Automotive R\&D (EUCAR)"}. [Accessed on
  29.07.2021]. [Online]. Available: \url{https://www.eucar.be/}
\BIBentrySTDinterwordspacing

\bibitem{5GACIA}
\BIBentryALTinterwordspacing
{``5G Alliance for Connected Industries and Automation"}. [Accessed on
  29.07.2021]. [Online]. Available: \url{https://www.5g-acia.org/}
\BIBentrySTDinterwordspacing

\bibitem{shah20185g}
S.~A.~A. Shah, E.~Ahmed, M.~Imran, and S.~Zeadally, ``5g for vehicular
  communications,'' \emph{IEEE Communications Magazine}, vol.~56, no.~1, pp.
  111--117, 2018.

\bibitem{arrieta2020explainable}
A.~B. Arrieta, N.~D{\'\i}az-Rodr{\'\i}guez, J.~Del~Ser, A.~Bennetot, S.~Tabik,
  A.~Barbado, S.~Garc{\'\i}a, S.~Gil-L{\'o}pez, D.~Molina, R.~Benjamins
  \emph{et~al.}, ``Explainable artificial intelligence (xai): Concepts,
  taxonomies, opportunities and challenges toward responsible ai,''
  \emph{Information Fusion}, vol.~58, pp. 82--115, 2020.

\bibitem{wang2021explainable}
S.~Wang, M.~A. Qureshi, L.~Miralles-Pechua{\'a}n, T.~Huynh-The, T.~R.
  Gadekallu, and M.~Liyanage, ``Explainable ai for b5g/6g: Technical aspects,
  use cases, and research challenges,'' \emph{arXiv preprint arXiv:2112.04698},
  2021.

\bibitem{srivastava2022xai}
G.~Srivastava, R.~H. Jhaveri, S.~Bhattacharya, S.~Pandya, P.~K.~R. Maddikunta,
  G.~Yenduri, J.~G. Hall, M.~Alazab, T.~R. Gadekallu \emph{et~al.}, ``Xai for
  cybersecurity: State of the art, challenges, open issues and future
  directions,'' \emph{arXiv preprint arXiv:2206.03585}, 2022.

\bibitem{kaur2021machine}
J.~Kaur, M.~A. Khan, M.~Iftikhar, M.~Imran, and Q.~E.~U. Haq, ``Machine
  learning techniques for 5g and beyond,'' \emph{IEEE Access}, vol.~9, pp.
  23\,472--23\,488, 2021.

\bibitem{reddy2020analysis}
G.~T. Reddy, M.~P.~K. Reddy, K.~Lakshmanna, R.~Kaluri, D.~S. Rajput,
  G.~Srivastava, and T.~Baker, ``Analysis of dimensionality reduction
  techniques on big data,'' \emph{IEEE Access}, vol.~8, pp. 54\,776--54\,788,
  2020.

\bibitem{hewavithana2022overcoming}
T.~Hewavithana, A.~Chopra, B.~Mondal, S.~Wong, A.~Davydov, and M.~Majmundar,
  ``Overcoming channel aging in massive mimo basestations with open ran
  fronthaul,'' in \emph{2022 IEEE Wireless Communications and Networking
  Conference (WCNC)}.\hskip 1em plus 0.5em minus 0.4em\relax IEEE, 2022, pp.
  2577--2582.

\end{thebibliography}
\end{document}